\documentclass[twocolumn,preprint]{aastex62}

\usepackage{amsmath,amstext, amssymb}
\usepackage[T1]{fontenc}
\usepackage{apjfonts} 
\usepackage{graphicx}
\usepackage{color}
\usepackage{tablefootnote}
\usepackage{subfigure}
\usepackage{natbib}
\usepackage{bm}

\begin{document}

\title{\large \bf The Super Eight Galaxies: Properties of a Sample of Very Bright Galaxies at $7<z<8$}

\author[0000-0002-8584-1903]{Joanna S. Bridge}
\email{joanna.bridge@louisville.edu}
\affiliation{Department of Physics and Astronomy, 102 Natural Science Building, University of Louisville, Louisville, KY 40292}

\author[0000-0002-4884-6756]{Benne W. Holwerda}
\affiliation{Department of Physics and Astronomy, 102 Natural Science Building, University of Louisville, Louisville, KY 40292}

\author[0000-0001-7768-5309]{Mauro Stefanon}
\affiliation{Leiden Observatory, Leiden University, NL-2300 RA Leiden, Netherlands}

\author[0000-0002-4989-2471]{Rychard J. Bouwens}
\affiliation{Leiden Observatory, Leiden University, NL-2300 RA Leiden, Netherlands}

\author[0000-0001-5851-6649]{Pascal A. Oesch}
\affiliation{Department of Astronomy, University of Geneva, Ch. des Maillettes 51, 1290 Versoix, Switzerland}

\author[0000-0001-9391-305X]{Michele Trenti}
\affiliation{School of Physics, The University of Melbourne, Parkville VIC 3010, Australia}
\affiliation{ARC Centre of Excellence for All Sky Astrophysics in 3 Dimensions (ASTRO 3D)}

\author[0000-0003-0956-0728]{Stephanie R. Bernard}
\affiliation{School of Physics, The University of Melbourne, Parkville VIC 3010, Australia}
\affiliation{ARC Centre of Excellence for All-Sky Astrophysics (CAASTRO)}

\author[0000-0002-7908-9284]{Larry D. Bradley}
\affiliation{Space Telescope Science Institute, 3700 San Martin Drive Baltimore, MD 21218}

\author[0000-0002-8096-2837]{Garth D. Illingworth}
\affiliation{Department of Astronomy and Astrophysics, University of California, Santa Cruz, CA 95064}

\author[0000-0002-0761-1985]{Samir Kusmic}
\affiliation{Department of Physics and Astronomy, 102 Natural Science Building, University of Louisville, Louisville, KY 40292}

\author{Dan Magee}
\affiliation{UCO/Lick Observatory, University of California, Santa Cruz, CA 95064}

\author[0000-0002-8512-1404]{Takahiro Morishita}
\affiliation{Space Telescope Science Institute, 3700 San Martin Drive, Baltimore, MD 21218, USA}

\author[0000-0002-4140-1367]{Guido W. Roberts-Borsani}
\affiliation{University College London, Gower Street, London WC1E 6BT, UK}

\author[0000-0001-8034-7802]{Renske Smit}
\affiliation{Cavendish Laboratory, University of Cambridge, 19 JJ Thomson
Avenue, Cambridge CB3 0HE, UK}
\affiliation{Kavli Institute for Cosmology, University of Cambridge, Madingley Road, Cambridge CB3 0HA}

\author[0000-0001-9537-5814]{Rebecca L. Steele}
\affiliation{Department of Physics and Astronomy, 102 Natural Science Building, University of Louisville, Louisville, KY 40292}

\begin{abstract}
We present the Super Eight galaxies - a set of very luminous, high-redshift ($7.1<z<8.0$) galaxy candidates found in Brightest of Reionizing Galaxies (BoRG) Survey fields. The original sample includes eight galaxies that are $Y$-band dropout objects with $H$-band magnitudes of $m_H<25.5$. Four of these objects were originally reported in \cite{calvi2016}. Combining new \emph{Hubble Space Telescope} (\emph{HST}) WFC3/F814W imaging and \emph{Spitzer} IRAC data with archival imaging from BoRG and other surveys, we explore the properties of these galaxies. Photometric redshift fitting places six of these galaxies in the redshift range of $7.1<z<8.0$, resulting in three new high-redshift galaxies and confirming three of the four high-redshift galaxy candidates from \cite{calvi2016}. We calculate the half-light radii of the Super Eight galaxies using the \emph{HST} F160W filter and find that the Super Eight sizes are in line with typical evolution of size with redshift. The Super Eights have a mean mass of log(M$_*$/M$_\sun$) $\sim10$, which is typical for sources in this luminosity range. Finally, we place our sample on the UV $z\sim8$ luminosity function and find that the Super Eight number density is consistent with other surveys in this magnitude and redshift range.
\\
\end{abstract}

\section{Introduction}
Two of the most compelling questions when studying the high-redshift universe are when and how the intergalactic medium (IGM) transitioned from neutral, opaque hydrogen to transparent and ionized. Recent observations from {\em Planck} \& studies of $z\sim 6$ quasars \citep[e.g.,][]{fan2006} place strong constraints on the basic time line of cosmic reionization: Reionization appears to have started around $z\sim 20$ and to be complete by $z\sim6$, with a mid-point around $z\sim8.8\pm1.3$ \citep{planck2016}.

 Remarkable progress has been made in extending observations to this earliest period: many candidate galaxies have been identified at $7<z<8$  \citep[e.g.,][]{mclure2013, bradley2014, bowler2014,bouwens2015,finkelstein2015,schmidt2014, bowler2017,stefanon2017,ono2018} and a dozen candidate galaxies have been found as far back as $9<z<11$ \citep[e.g.,][]{bouwens2011, coe2013, ellis2013, mclure2013,oesch2013, oesch2014, oesch2016, hashimoto2018}.

Probing the ionization state of the Universe is usually done through observations of the Ly$\alpha$ emission line in $z>6$ galaxies. The suppression of Ly$\alpha$ emission from galaxies is to be expected in an increasingly neutral Universe due to the resonant nature of the Ly$\alpha$ line. Indeed, the overall prevalence of Ly$\alpha$ emission in galaxies becomes increasingly rare towards higher redshift, beginning at $z\sim 6$ \citep{santos2004,malhotra2004, mcquinn2007, mesinger2008, stark2010, stark2011, fontana2010, dijkstra2012, ono2012,treu2013, caruana2012, caruana2014, tilvi2013}. By $z \sim 7$, only $\sim$ 10\% of galaxies show Ly$\alpha$ emission, a number that decreases to less than 3\% for $z>7$  \citep{fontana2010, pentericci2011, pentericci2014, finkelstein2013, schenker2012, schenker2014, ono2012, tilvi2013,oesch2015,zitrin2015, robertsborsani2016, song2016,tilvi2016, stark2017, larson2018}. Determining exactly when this transition from a neutral to an ionized IGM occurs is hampered by the low Ly$\alpha$ detection rate at these redshifts \citep[$<$3-10\%;][]{stark2011,stark2013,schenker2014}.

In order to bypass this detection deficit, bright emission line sources at $7<z<9$ can be also detected using the \emph{Spitzer}/IRAC 3.6-$\mu$m and 4.5-$\mu$m bands \citep[e.g.,][]{robertsborsani2016}. Strong nebular emission lines such as H$\alpha$ and [O~III]+H$\beta$ appear in these bands, causing a color gradient that is highly redshift dependent \citep{schaerer2009, labbe2013, stark2013, bowler2014, laporte2014, smit2014, smit2015}. High luminosity sources ($H_{160} < 25.5$) with red IRAC colors ($[3.6]-[4.5] > 0.5$) can therefore be identified reliably as long as care is taken to rule out lower redshift interlopers such as brown dwarfs. 

This method has been successfully used to identify four bright sources in the CANDELS \citep{grogin2011} Extended Groth Strip (EGS) \citep{davis2007} and COSMOS \citep{scoville2007} fields. All four of these sources have consequently been spectroscopically confirmed via their Ly$\alpha$ emission \citep[$z$ = 7.73, 8.68, 7.48, 7.15;][]{oesch2015,zitrin2015,robertsborsani2016,stark2017}. Previous work by \cite{ono2012} and \cite{finkelstein2013} identified very bright Ly$\alpha$ emitters using \emph{Spitzer}/IRAC photometry and found that two further $z>7$ Ly$\alpha$ emitters have IRAC colors of $[4.5] - [3.6] > 0.5$. Taken together, these results indicate that luminous ($M_{UV} < -21$) galaxies with red IRAC colors are strong candidates in which to look for Ly$\alpha$ emission at this redshift.

The Ly$\alpha$ confirmation of the EGS galaxies shows that these very bright galaxies do exist at this epoch. However, it does not establish if all photometrically-selected candidates are indeed $z\sim8$ galaxies, nor does it yet set strong constraints on the bright end of the $z\sim8$ luminosity function.

The luminosity function (LF) is usually described using a type of gamma distribution known as the Schechter function \citep{schechter1976}, with a characteristic luminosity that indicates where the slope of the LF begins to steepen towards the bright end. The LF of galaxies exhibits a smooth evolution with cosmic time, with some studies showing little evolution in the characteristic luminosity of star-forming galaxies over the redshift range $4<z<10$ \citep[e.g.][but see \citealt{lorenzoni2011, mclure2013, bowler2015}]{vanderburg2010, jaacks2012, atek2015, bouwens2015, finkelstein2015, mason2015, livermore2017, ishigaki2018, ono2018}. This limited evolution is perhaps counterintuitive; one might expect that the biggest and brightest galaxies will continually grow in UV luminosity as they build up cool gas reservoirs and stars. However, if galaxies reach sufficiently high masses, metal production and dust extinction help set an upper limit on their UV luminosities \citep{reddy2009, bouwens2009}. Feedback from active galactic nuclei also plays a role in this process \citep{granato2004, croton2006}. How and when these physical processes influence star formation in early galaxies are yet to be definitively determined, causing the nature of the bright end of the LF to be uncertain. Several studies of the LF at $4<z<7$ indicate that the bright end may flatten with redshift \citep{bowler2014, bowler2015, calvi2016, salmon2017, stefanon2017, ono2018}, while modelling suggests a Schecter-like LF even at high redshift \citep{trenti2010, jaacks2012}.

\begin{deluxetable*}{cccc}
\tablecolumns{4}
\tablecaption{Candidate Super Eight Sample\label{basic}}
\tablehead{\colhead{Super8 ID} &\colhead{Other ID(s)} & \colhead{R.A. (J2000)} & \colhead{Dec. (J2000)} }
\startdata
Super8-1 & - - & 23:50:34.66 & -43:32:32.50 \\[3pt]
Super8-2 & - - & 08:35:13.10 & +24:55:38.10\\[3pt]
Super8-3 & - - & 22:02:50.00 & +18:51:00.20 \\[3pt]
Super8-4 & borg\_0116+1425\_747\tablenotemark{a, b}& 01:16:08.93 & +14:24:24.50 \\[3pt]
Super8-5 & borg\_0853+0310\_145\tablenotemark{a}& 08:52:44.52 & +03:08:48.10 \\[3pt] 
Super8-6 & borg\_1152+3402\_912\tablenotemark{a}& 11:51:37.85 & +34:02:22.90 \\[3pt]
Super8-7 & borg\_2229-0945\_548\tablenotemark{a}& 22:28:45.72 & -09:44:56.51 \\[3pt]
Super8-8 & - - & 15:57:05.62 & -37:46:55.75
\enddata
\tablenotetext{a}{From \cite{calvi2016}.
\tablenotetext{b}{See \cite{livermore2018}}}
\end{deluxetable*}

Obtaining additional statistics on the brightest $z\sim8$ galaxies is important to draw robust conclusions about how the bright end of the LF evolves with cosmic time. These studies will provide us with insight into how bright or massive galaxies might become at very earliest times in the Universe and what physical processes drive the first star-formation.

At $z\sim8$, there is substantial field-to-field variance in the spatial density of the most luminous (and hence rarest) galaxies. To overcome this variance, several wide-area searches have been conducted. \cite{mclure2013} examined several hundred arcmin$^2$ using \emph{Hubble} Ultra Deep Field (HUDF) observations from HUDF09 and HUDF12, as well as a wealth of publicly available imaging to characterize the LF at $z\sim7-8$ using a catalog of $\sim600$ galaxies, and provided the first estimate of the LF at $z\sim9$. \cite{bouwens2015} compiled publicly available data from CANDELS, HUDF09, HUDF12 \citep{ellis2013}, ERS \citep{windhorst2011}, and BoRG \citep{trenti2011} to provide a consistent characterization of the luminosity function from $z\sim4-10$ using $\sim 10,000$ galaxies over $\sim1000$ arcmin$^2$. The UltraVISTA survey \citep{stefanon2017} originally found sixteen Lyman Break Galaxies at $z\sim8-9$ over a very large survey area of 1.6 degrees$^2$, three of which have robust high photometric redshifts determined using follow-up \emph{HST} observations.

The BoRG Survey is a pure-parallel \emph{HST} survey that is comprised of Wide Field Camera 3 (WFC3) observations that were simultaneously observed with COS spectroscopy occurring elsewhere. BoRG now consists of the BoRG[z8] \citep{bradley2012,trenti2012, schmidt2014}, BoRG[z9-10] (\citealt{calvi2016, bernard2016, morishita2018}, HIPPIES (GO 12286, PI H. Yan), and other GTO HST/WFC3 programs. These consist of single \emph{HST}/WFC3 pointings that are distributed almost randomly throughout the sky. The benefit of this strategy is that it is very efficient at providing a fair probe of the brightest sources at this redshift \citep{bradley2012, brammer2012, schmidt2014, calvi2016}. The drawback of this approach is that with only four or five \emph{HST} filters, redshift identifications are less certain, and we are tantalizingly left with bright, yet tentative $z\sim8$ candidate galaxies.

The set of eight galaxies selected for follow-up study here were previously identified in the BoRG survey, and were chosen as a result of their high-redshift candidacy and bright $H$-band magnitudes. In this work, we expand upon the available \emph{HST} photometry with new data using the \emph{HST} WFC3 F814W filter and the \emph{Spitzer} IRAC filters in order to explore the properties of these luminous Super Eight galaxies that were found in the BoRG survey. In Section 2, we discuss how the candidates were originally selected. With the candidates in hand, we detail the data (both archival and new) and the photometry for the galaxy candidates in Section 3. Section 4 explains the fitting procedure for determining the galaxies' photometric redshifts, and Section 5 discusses further properties of the sample. The volume density of the sample is discussed in Section 6, and in Section 7, we place our results in the larger context of high-redshift galaxies studies, and consider future work.

We assume a $\Lambda$CDM cosmology with $H_0 = 70$ km s$^{-1}$ Mpc$^{-1}$, $\Omega_m = 0.3$, and $\Omega_\Lambda = 0.7$. All magnitudes are in the AB system \citep{oke1983}.

\section{Candidate Selection}

Four of the Super Eight galaxies were previously identified as high-redshift candidates in the BoRG fields by \cite{calvi2016} (Super8-4-7). That search covered an area of $\sim130$ arcmin$^2$. The high-redshift candidates were selected via the Lyman-break technique \citep{steidel1996} and used external persistence maps to exclude spurious detections. Super8-7 was selected using the $z\sim9$ $Y$-dropout criteria:
\begin{gather*}
S/N_{350} < 1.5\\
S/N_{140} \geq 6\\
S/N_{160} \geq 4\\
Y_{105} - JH_{140} > 1.5\\
Y_{105} - JH_{140} > 5.33\times(JH_{140} - H_{160})+0.7\\ 
JH_{140} - H_{160} < 0.3
\end{gather*}
where $Y_{105}$, $JH_{140}$, and $H_{160}$ represent the F105W, F140W, and F160W \emph{HST} WFC3 filters. \cite{calvi2016} identified three further objects at $z\sim7-8$ using:
\begin{gather*}
S/N_{350} < 1.5\\
S/N_{125} \geq 6\\
S/N_{140} \geq 6\\
S/N_{160} \geq 4\\
Y_{105} - J_{125} > 0.45\\
Y_{105} - JH_{125} > 1.5\times(J_{125} - H_{160})+0.45\\
J_{125} - H_{160} < 0.5
\end{gather*}
where $J_{125}$ represents the F125W filter.

The remaining Super Eights (Super8-1-3, 8) were identified as high-redshift candidates in the BoRG survey fields by \cite{bouwens2015}. Super8-2 and Super8-3 were found in the original search that covered 218 arcmin$^2$. They were excluded from the catalog due the requirement that objects not be possibly associated with bright foreground sources. Super8-1 and Super8-8 were discovered in COS-GTO observations (IDs 11528 and 12036, PI J.\ Green) that were not included in the \cite{bouwens2015} search but were identified in additional data using the same criteria. This extended search added an additional 83.9 arcmin$^2$ in survey area, for a total of $\sim300$ arcmin$^2$. 

These four galaxies were selected using the criteria:
\begin{gather*}
S/N_{606/600} < 1.5\\
Y_{098}-J_{125}>1.3\\
J_{125}-H_{160} < 0.5\\
Y_{098}-J_{125}>0.75\times(J_{125}-H_{160})+1.3
\end{gather*}
where $Y_{098}$ represents the F098M band. The final candidate sample is given in Table~\ref{basic}.

\section{Data}
With the galaxy sample in hand, we reprocessed all of the data available for these objects in order to obtain accurate photometric redshifts. We therefore reacquired all of the imaging available in the \emph{Hubble} Legacy Archive\footnote{https://hla.stsci.edu/hlaview.html} (HLA) for the Super Eight galaxies. Additionally, we performed a new \emph{HST}+\emph{Spitzer} observing campaign to collect further imaging in filters not yet available.

\begin{deluxetable}{ccccccc}
\tablecaption{Candidate Super Eight \emph{HST} F160W and \emph{Spitzer} IRAC Exposure Times and Depths\label{f160w_depths}}
\tablehead{\colhead{ID} & \multicolumn{3}{c}{Exposure Time (s)} & \multicolumn{3}{c}{5$\sigma$ Depth\tablenotemark{a}} \\
             \colhead{}  & \colhead{F160W} & \colhead{[3.6]} & \colhead{[4.5]} &
              \colhead{F160W} & \colhead{[3.6]} & \colhead{[4.5]} }
\startdata
Super8-1                 & 2806 & 2372  & 650 & 26.23 & 24.8 & 23.9\\
Super8-2                 & 2009 & 2326  & 716 & 26.40 & 24.8 & 24.0\\
Super8-3                 & 1403 & 2336  & 828 & 26.99 & 24.6 & 24.1\\
Super8-4                 & 2409 & 2026  & 595 & 26.97 & 24.5 & 23.9\\
Super8-5                 & 1709 & 2225  & 902 & 26.98 & 24.8 & 24.2\\
Super8-6\tablenotemark{b}& 1606 &  - -  & - - & 26.46 & - -  & - - \\
Super8-7                 & 1759 & 3079  & 814 & 26.57 & 24.8 & 24.0\\
Super8-8                 & 453  & 2416  & 415 & 23.83 & 24.8 & 23.7   
\enddata
\tablenotetext{a}{The depth was determined by calculating the average rms noise in $0\farcs5$-diameter ($0\farcs9$-diameter) empty apertures placed throughout the \emph{HST} (IRAC) images.}
\tablenotetext{b}{There are no IRAC observations for Super8-6}
\end{deluxetable}

\begin{deluxetable*}{ccccccccc}
\tablecolumns{9}
\tablecaption{Candidate Super Eight \emph{HST} Magnitudes \label{photo}}
\tablehead{\colhead{Filter} & \colhead{Super8-1} & \colhead{Super8-2} & \colhead{Super8-3} & \colhead{Super8-4} & \colhead{Super8-5} & \colhead{Super8-6} & \colhead{Super8-7} & \colhead{Super8-8}}
\startdata
F350LP & - - 			     & - - 			  & - - 			  & $>28.19$      & $>27.96$		   & $>27.89$      & $29.03\pm0.98$      & - -			\\[3pt]
F475X  & - - 			     & - - 			  & $>26.70$    & - - 		        & - - 			       & - - 			     & - - 		           & - -			\\[3pt]
F475W  & $>26.80$ 	     & - - 			  & - - 			  & - - 		        & - -  			       & - - 			     & - - 		           & - -			\\[3pt]
F600LP & $>26.53$	     & - - 			  & $>26.93$ 	  & - - 		        & - - 			       & - -   		         & - - 		           & $>25.67$ \\[3pt]
F606W  & - - 		         & $>27.09$  & - - 	          & - - 		        & - - 			       & - - 			     & - - 		           & - -	        \\[3pt]
F814W  & $>28.02$ 	     & $>27.94$  & $>27.99$    & $>27.95$       & $>27.92$ 		       & $>27.24$      & $>28.00$ 	   & $26.03\pm0.37$ \\[3pt]
F098M  & $27.25\pm0.67$ & $26.50\pm0.69$ & $26.11\pm0.29$    & $>26.95$        & - - 			       & - - 			     & - - 		           & $24.23\pm0.24$ \\[3pt]
F105W  & - - 			     & - - 			  & - - 			  & $26.61\pm0.60$ & $25.78\pm0.27$  & $26.33\pm0.44$ & $>26.92$       & - -			\\[3pt]
F125W  & $25.64\pm0.15$ & $25.33\pm0.18$ & $25.42\pm0.21$    & $25.55\pm0.24$ & $25.16\pm0.16$  & $25.72\pm0.27$ & $25.86\pm0.29$ & $23.61\pm0.05$ \\[3pt]
F140W  & - - 			     & - - 			  & - - 			  & $25.51\pm0.21$ & $25.14\pm0.14$  & $25.61\pm0.22$ & $25.64\pm0.21$ & - -			\\[3pt]
F160W  & $25.42\pm0.20$ & $25.30\pm0.24$ & $25.15\pm0.18$    & $25.36\pm0.24$ & $25.26\pm0.21$  & $25.44\pm0.31$ & $25.46\pm0.24$ & $23.15\pm0.13$  
\enddata
\tablecomments{The 1$\sigma$ uncertainties are quoted for the non-detection upper limits.}
\end{deluxetable*}

\begin{deluxetable}{ccc}
\tablecolumns{3}
\tablecaption{Candidate Super Eight \emph{Spitzer} IRAC Magnitudes\label{IRAC}}
\tablehead{\colhead{\hspace{.8cm}ID\tablenotemark{a}}\hspace{.8cm} & \colhead{\hspace{.8cm}[3.6]}\hspace{.8cm}& \colhead{\hspace{.8cm}[4.5]}\hspace{.8cm}}
\startdata
Super8-1 & $24.8\pm0.5$  & $>24.7$  \\[3pt]
Super8-2 & $>25.5$       & $>24.7$  \\[3pt]
Super8-3 & $24.4\pm0.4$  & $24.0\pm0.5$\\[3pt]
Super8-4 & $24.2\pm0.4$  & $>24.7$  \\[3pt]
Super8-5 & $25.4\pm0.9$  & $>24.9$  \\[3pt]
Super8-7 & $23.8\pm0.2$  & $24.6\pm0.9$\\[3pt]
Super8-8 & $23.2\pm0.1$  & $23.0\pm 0.3$
\enddata
\tablenotetext{\footnotesize a}{There are no IRAC observations for Super8-6}
\tablecomments{The 1$\sigma$ uncertainties are quoted for the non-detection upper limits.}
\end{deluxetable}

\subsection{The BoRG Data}
The $z\sim8$ galaxies presented here are found in the BoRG fields that are discussed in \cite{bradley2012} and \cite{schmidt2014}, based on the 2009, 2012, and 2013 BoRG survey campaigns. The pure-parallel nature of the survey ensures that the survey area is divided into many independent lines-of-sight on the sky, reducing sample (or cosmic) variance below the level of statistical noise \citep{trenti2008}. Here, we use the standard multi-drizzle reduction of these undithered WFC3 data that are available on the \emph{Hubble} Legacy Archive. The exposure times and $5\sigma$ limiting depths in the $H$-band for Super Eights are given in Table~\ref{f160w_depths}.

\subsection{HST/WFC3 F814W and Spitzer/IRAC Data}
The bands provided by the BoRG survey are important to identify $Y$-band dropout candidates. In order to provide more robust photometric redshift fitting, we obtained \emph{HST}/WFC3 F814W observations of the eight Super Eight galaxy candidates (GO 14652, PI B. Holwerda), which also included \emph{Spitzer} photometry in both the 3.6-$\mu$m and 4.5-$\mu$m bands for five of the eight objects. \emph{Spitzer} IRAC observations were also obtained for two other objects via another program (ID 13103, PI S. Bernard). In total, there are IRAC observations for seven of the eight galaxies in the sample. The combination the F814W and IRAC data is critical for the reliable selection of galaxies at this epoch. 

Each object was observed with the \emph{HST} F814W filter for $\sim5200$ seconds, reaching an average $5\sigma$ limiting depth in $0\farcs4$-diameter empty apertures of $m = 27.4$. The images were reduced with the standard STScI \texttt{AstroDrizzle} package \citep{gonzaga2012}, which removes the geometric distortion, corrects for the sky background, flags cosmic-rays, and combines images with subsampling using the drizzle algorithm.

The \emph{Spitzer} IRAC observations consist of observations of varying exposure times and limiting magnitudes for each observation, given in Table~\ref{f160w_depths}. The IRAC data reduction was carried out using the \texttt{mophongo} pipeline developed by \cite{labbe2015}. Each image frame is corrected for background, cosmic rays, persistence from bright stars, and other artifacts. The corrected frames are successively registered to the reference frame (the \emph{HST} image) and median combined. For each mosaic, the pipeline generates spatially varying empirical point spread functions (PSF), applying the weights and rotations of each frame to a high signal-to-noise template PSF reconstructed from the deepest archival observations.

\subsection{Other Data}
In addition to the BoRG survey data and the \emph{HST/Spitzer} F814W campaign, we trawled the Hubble Legacy Archive for any other available observations that exist for these objects. This resulted in shorter wavelength filters of F475W (GO 11528, PI J. Green) and F475X (GO 11534, PI J. Green) for Super8-1 and Super8-3, respectively. The use of these data does not change the final results, but we include them for completeness. We also include F098M filter data (GO 14701, PI M. Trenti) for Super8-4. For these data, we used the standard drizzled data products from the \emph{HST} archive. 
\newpage
\subsection{Source Extraction and Aperture Photometry}
Once we gathered the imaging data for all of the objects, they were aligned using the \texttt{AstroPy} reprojection\footnote{https://reproject.readthedocs.io/} algorithm. To determine the \emph{HST} photometry, we ran \texttt{Source Extractor} \citep{bertin1996,holwerda2005} in dual-image mode using the F160W images as the detection image. We used MAG\_AUTO and associated errors as the total magnitude in each band. To correctly determine the photometric errors, we re-scaled the original weight maps for each image to correctly account for correlated noise \citep{casertano2000} following the method of \cite{trenti2011}. We verified that each source had a S/N $> 0.5$ with at least nine contiguous pixels in the detection image. Table~\ref{photo} gives the \emph{HST} photometry for our sample, which have been corrected for Galactic extinction using the dust maps of \cite{schlafly2011}\footnote{http://irsa.ipac.caltech.edu/applications/DUST/}.

Photometry in the IRAC 3.6-$\mu$m and 4.5-$\mu$m bands was performed with the code \texttt{Mophongo} \citep{labbe2005, labbe2006, labbe2010a, labbe2010b, labbe2013, labbe2015}. Briefly, the code reconstructs the light profile of all neighboring sources in $12\farcs0$ radius fitting to the IRAC frame a higher resolution image (the \emph{HST} detection mosaic) convolved with a kernel reconstructed from the empirical IRAC PSFs and the PSF of the prior image. Aperture photometry (with a diameter of $1\farcs8$) is then performed after removing the models of the neighboring objects. A correction based on the PSF and on the light profile of the source under analysis is finally applied to obtain total fluxes. Table~\ref{IRAC} gives the observed IRAC fluxes. As with the \emph{HST} data, the photometry has been corrected for Galactic extinction \citep{schlafly2011}, although this extinction does not affect the measurements given the lack of precision in the IRAC data.

The IRAC uncertainties were measured in the same manner as described in \cite{stefanon2017}. Briefly, the rms was calculated for the pixels in the apertures using the residual frames (science frame with \emph{all} objects subtracted.) Systematic errors from the kernel reconstruction are also taken 

\begin{figure*}[!h]
\begin{tabular}{cc}
  	\begin{minipage}{0.95\textwidth}
    	\centering
    	\scalebox{1}
    	{\includegraphics[width=1\textwidth]{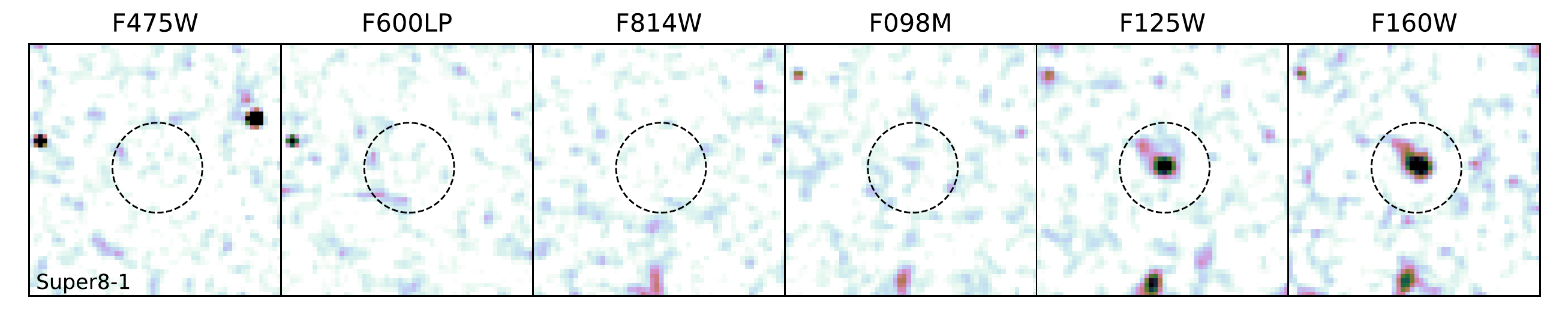}}
    \end{minipage}\\
    \begin{minipage}{0.95\textwidth} 
    	\scalebox{0.77}
    	{\includegraphics[width=0.45\textwidth]{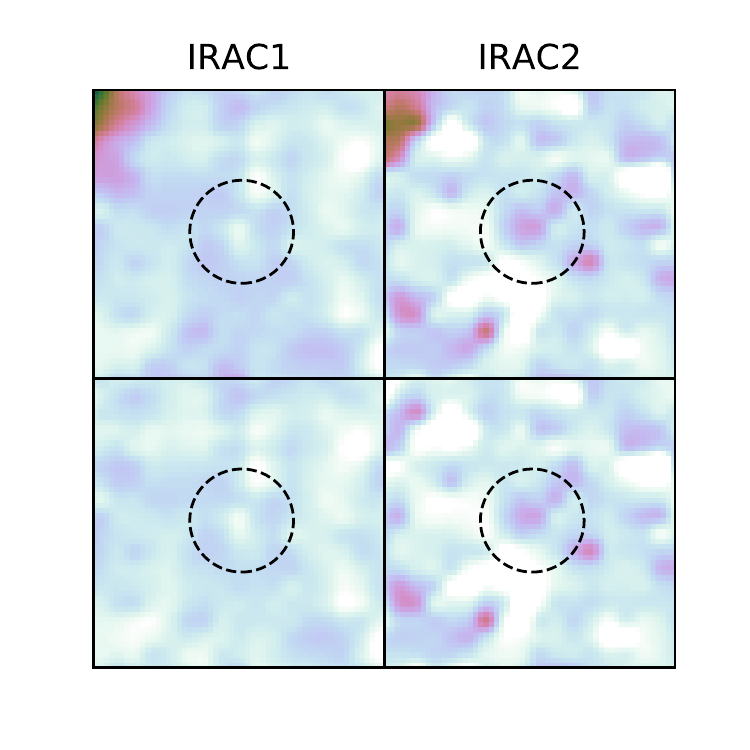}}
    	\scalebox{1.4}
    	{\includegraphics[width=0.45\textwidth]{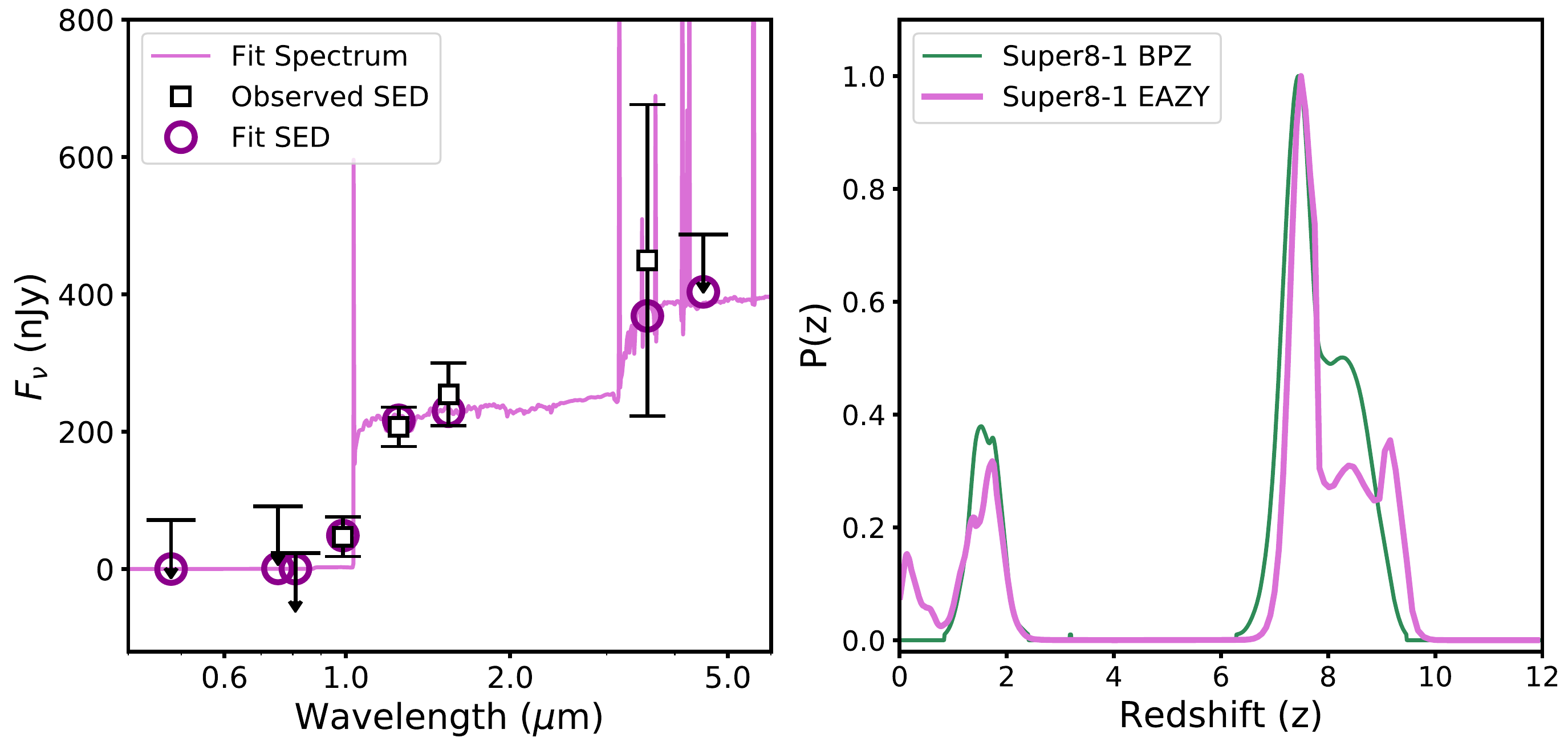}}
    \end{minipage}
\end{tabular}
\caption{\emph{HST} and \emph{Spitzer} cutouts for Super Eight 1. The cutouts are 5" $\times$ 5" and centered on the candidate galaxy. The central circles have a radius of 0\farcs9, matching the IRAC aperture, and are included to guide the eye. For the IRAC cutouts, both the original images (top) and images after neighbor subtraction (bottom) are shown. The bottom right figure gives the photometry and fit SED for Super Eight 1. The observed photometry is shown with black open squares, while the fit is shown with dark pink open circles. The fit spectrum is shown with a solid pink line. The upper limits indicate the $1\sigma$ uncertainties. The probability distributions from the redshift fitting performed using \texttt{EAZY} (pink) and \texttt{BPZ} (green) are given in the lower rightmost panel.}
\label{cutout1}
\end{figure*}

\begin{figure*}[!h]
\begin{tabular}{cc}
  	\begin{minipage}{0.95\textwidth}
    	\centering
    	\scalebox{0.8}
    	{\includegraphics[width=1\textwidth]{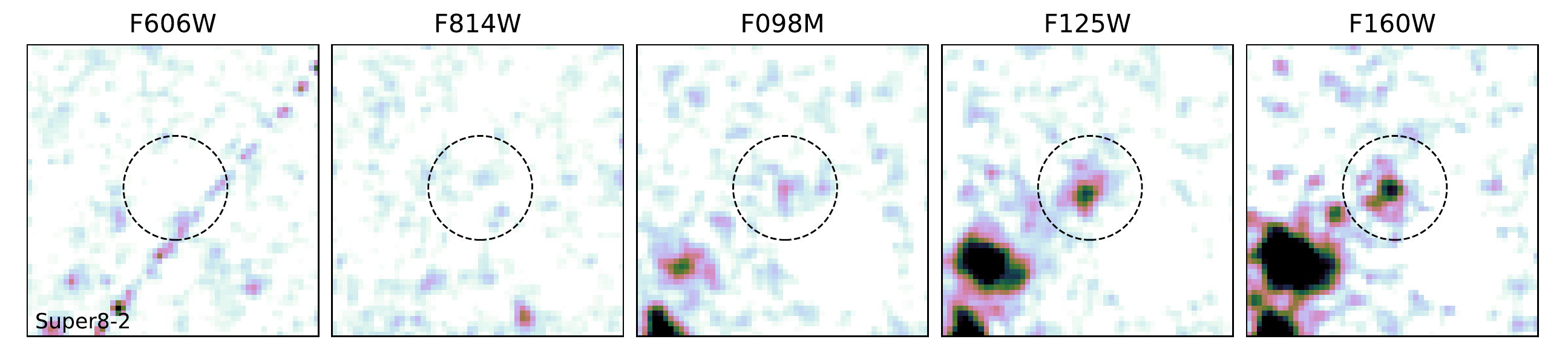}}
    \end{minipage}\\
    \begin{minipage}{0.95\textwidth} 
    	\scalebox{0.77}
    	{\includegraphics[width=0.45\textwidth]{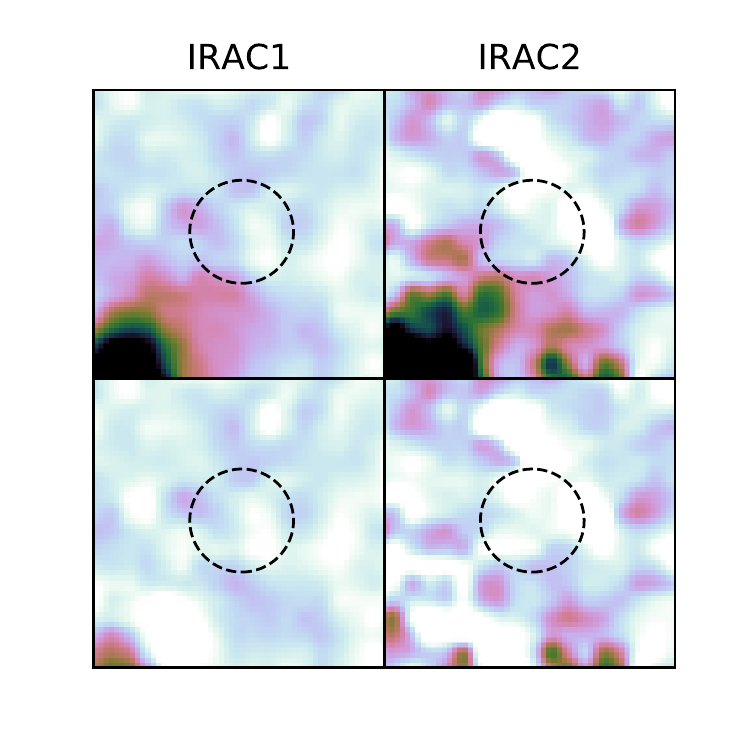}}
    	\scalebox{1.4}
    	{\includegraphics[width=0.45\textwidth]{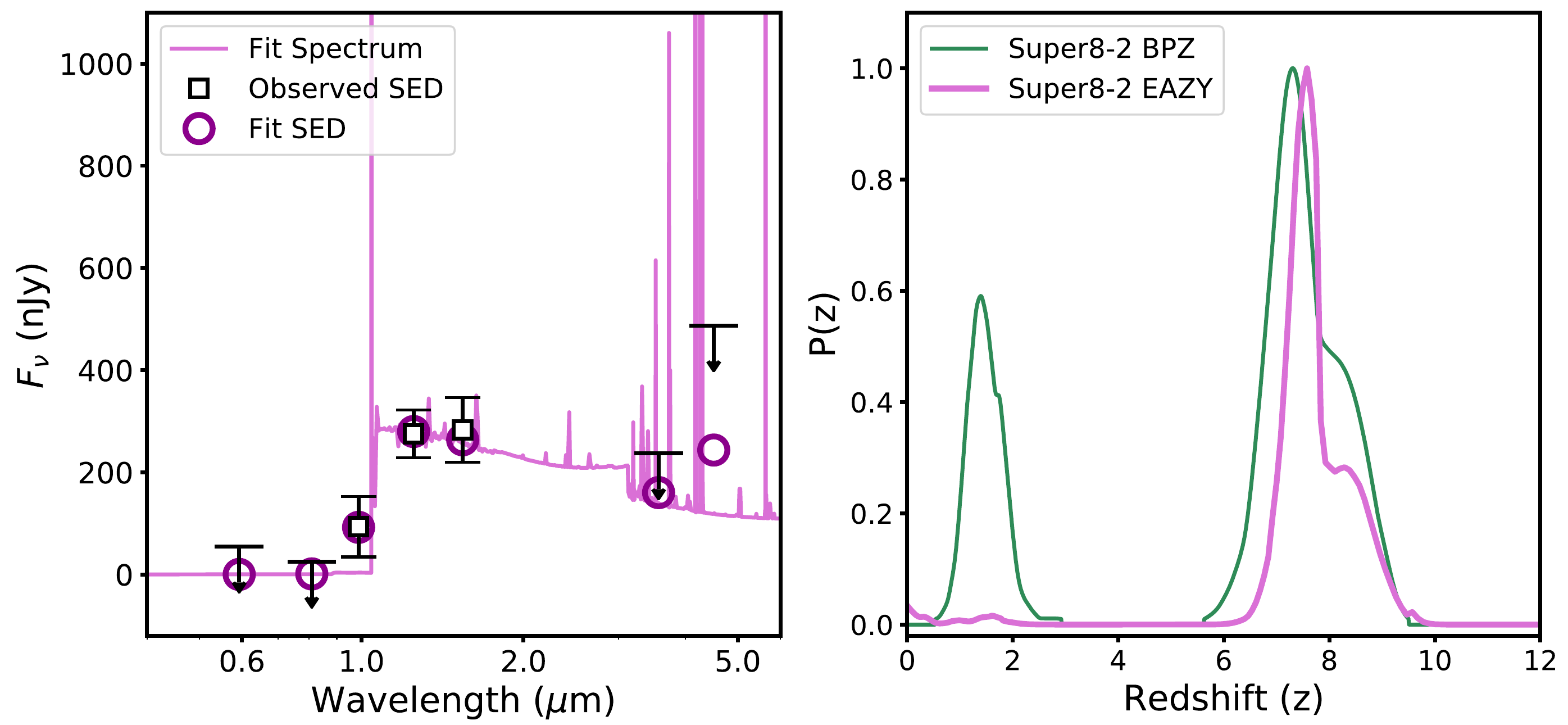}}
    \end{minipage}
\end{tabular}
\caption{Same as Figure~\ref{cutout1} but for Super Eight 2.}
\label{cutout2}
\end{figure*}

\begin{figure*}[!h]
\begin{tabular}{cc}
  	\begin{minipage}{0.95\textwidth}
    	\centering
    	\scalebox{1}
    	{\includegraphics[width=1\textwidth]{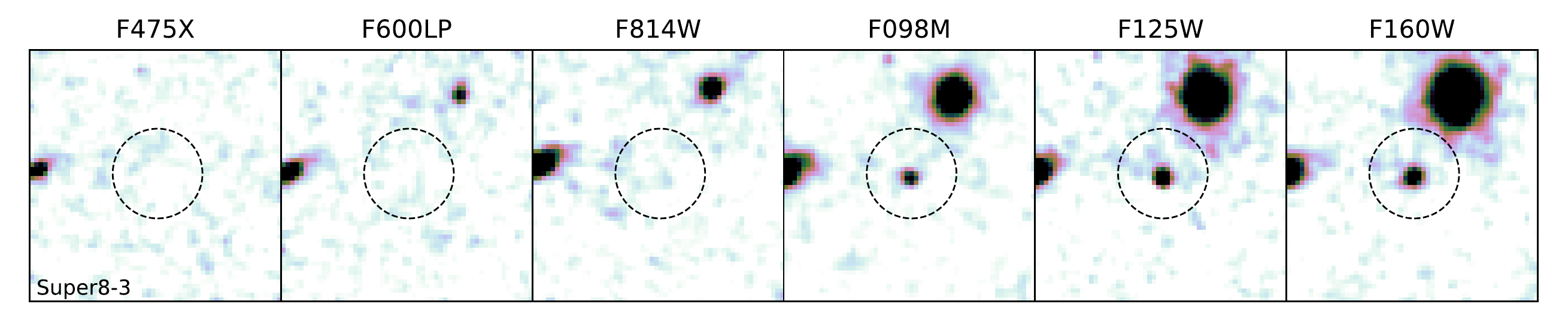}}
    \end{minipage}\\
    \begin{minipage}{0.95\textwidth} 
    	\scalebox{0.77}
    	{\includegraphics[width=0.45\textwidth]{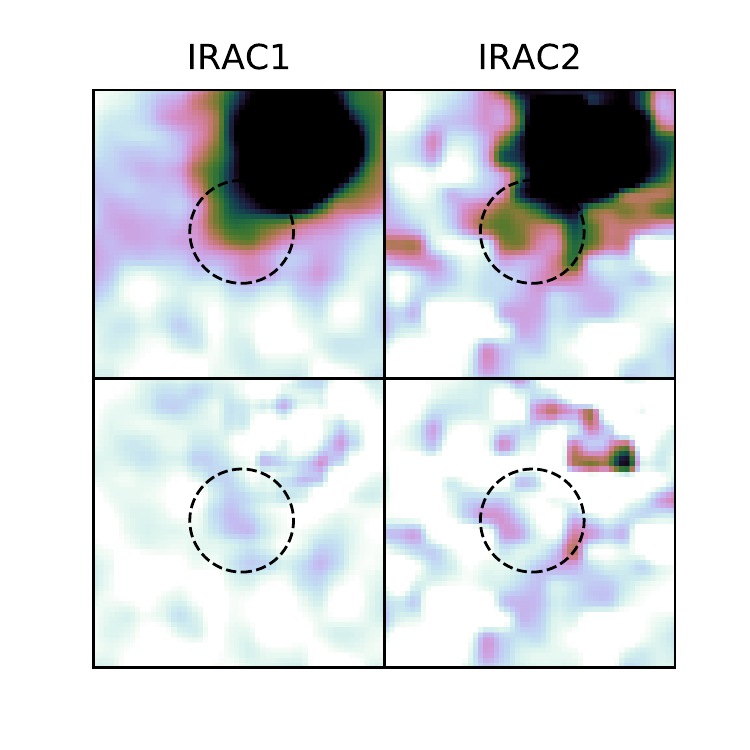}}
    	\scalebox{1.4}
    	{\includegraphics[width=0.45\textwidth]{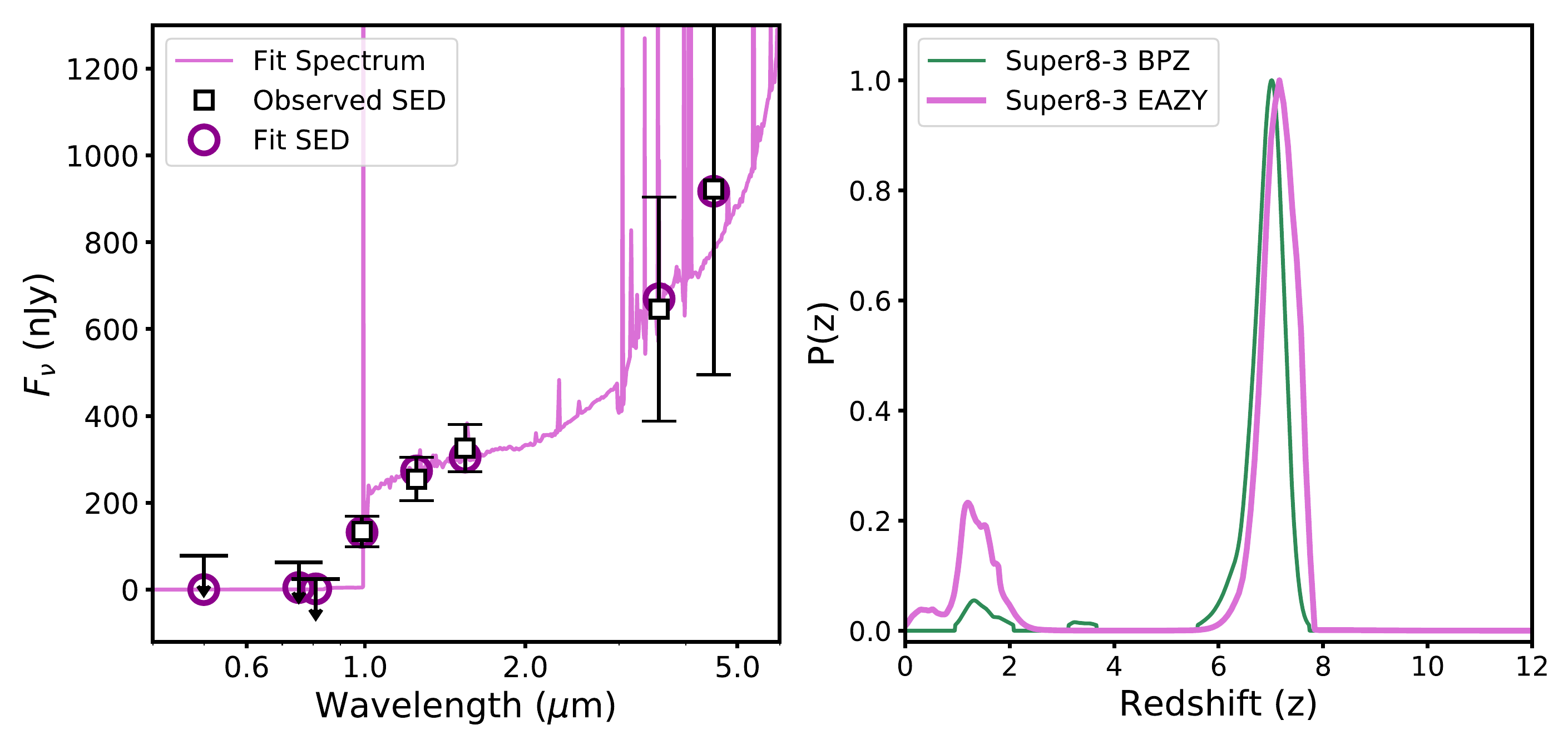}}
    \end{minipage}
\end{tabular}
\caption{Same as Figure~\ref{cutout1} but for Super Eight 3.}
\label{cutout3}
\end{figure*}

\begin{figure*}[!h]
\begin{tabular}{cc}
  	\begin{minipage}{0.95\textwidth}
    \centering
    	\scalebox{1}
    	{\includegraphics[width=1\textwidth]{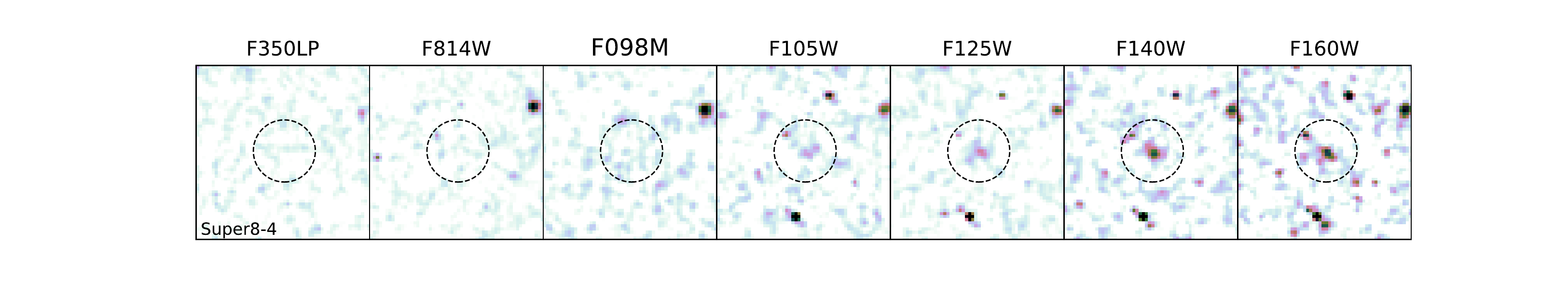}}
    \end{minipage}\\
    \begin{minipage}{0.95\textwidth} 
    	\scalebox{0.77}
    	{\includegraphics[width=0.45\textwidth]{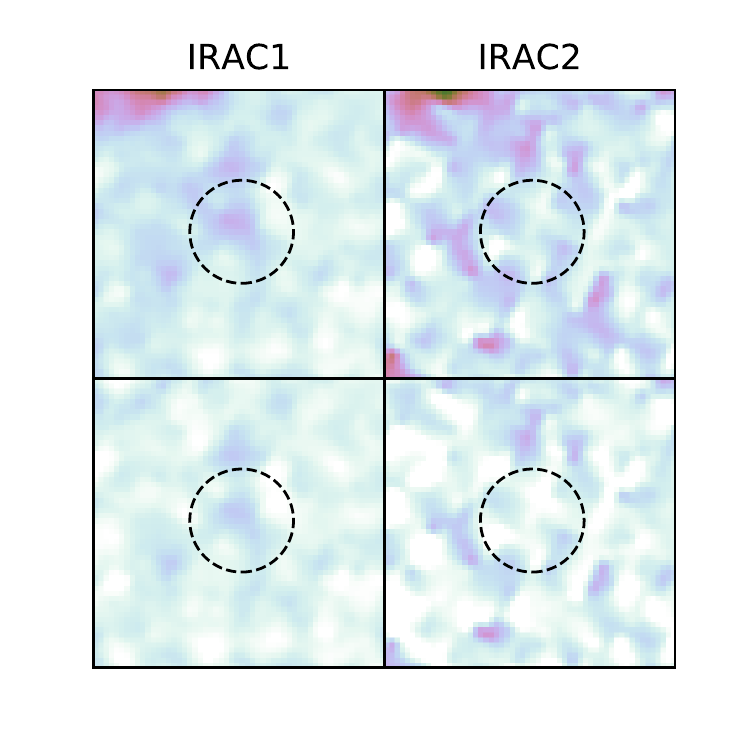}}
    	\scalebox{1.4}
    	{\includegraphics[width=0.45\textwidth]{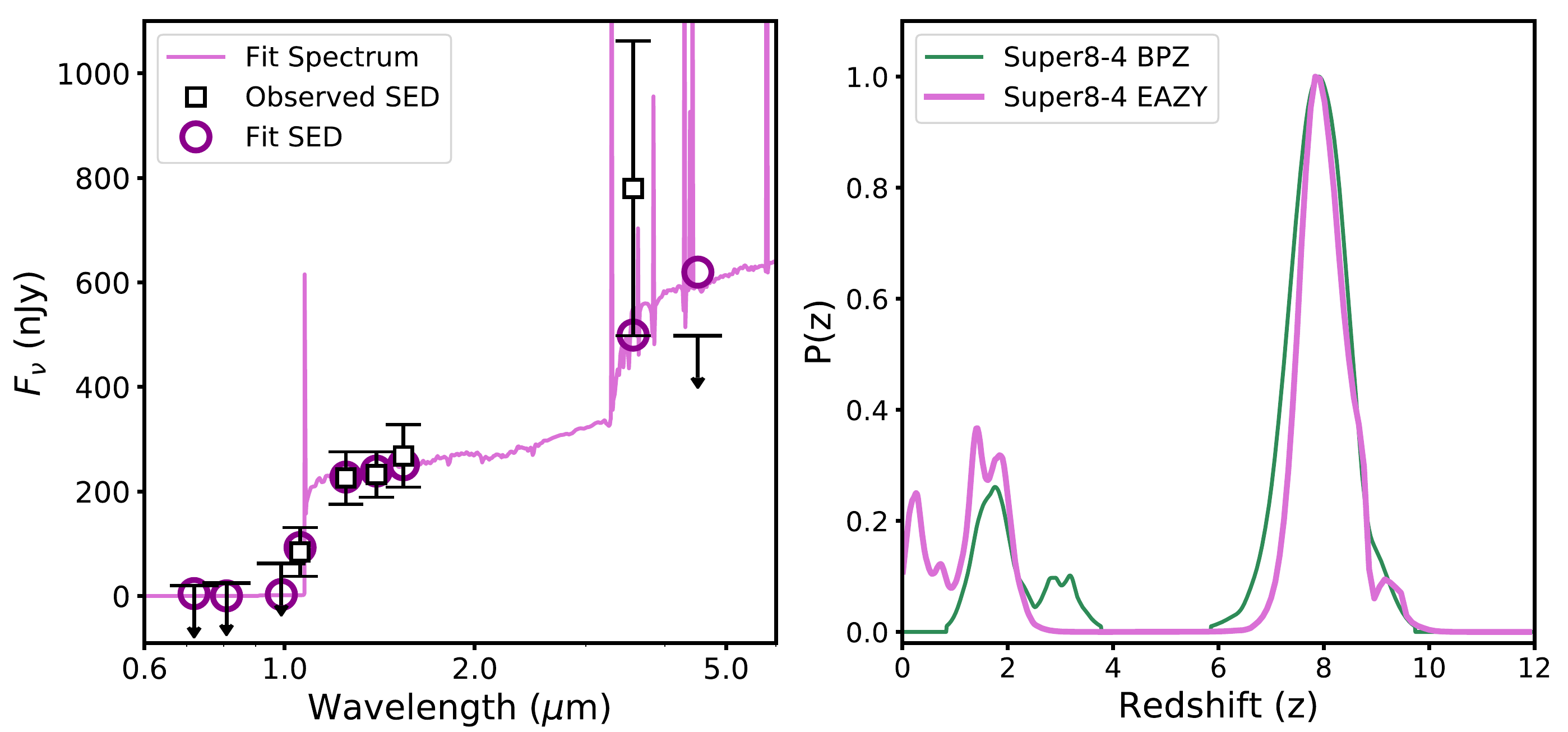}}
    \end{minipage}
\end{tabular}
\caption{Same as Figure~\ref{cutout1} but for Super Eight 4.}
\label{cutout4}
\end{figure*}

\begin{figure*}[!h]
\begin{tabular}{cc}
  	\begin{minipage}{0.95\textwidth}
    \centering
    	\scalebox{1}
    	{\includegraphics[width=1\textwidth]{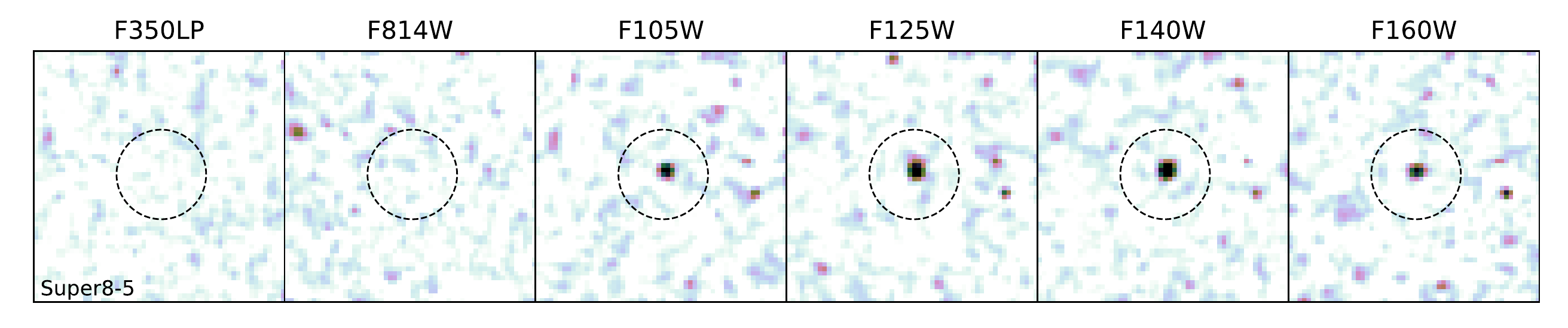}}
    \end{minipage}\\
    \begin{minipage}{0.95\textwidth} 
    	\scalebox{0.77}
    	{\includegraphics[width=0.45\textwidth]{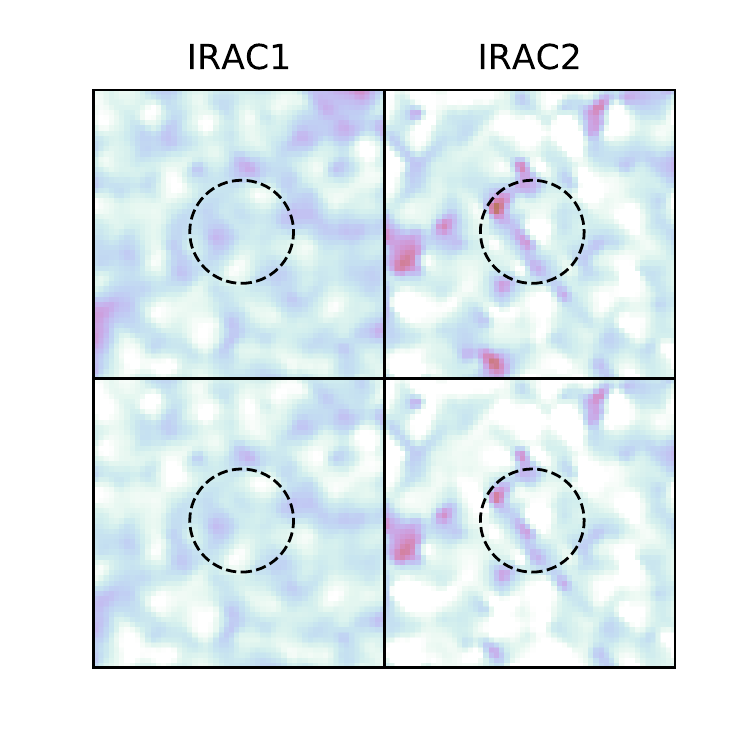}}
    	\scalebox{1.4}
    	{\includegraphics[width=0.45\textwidth]{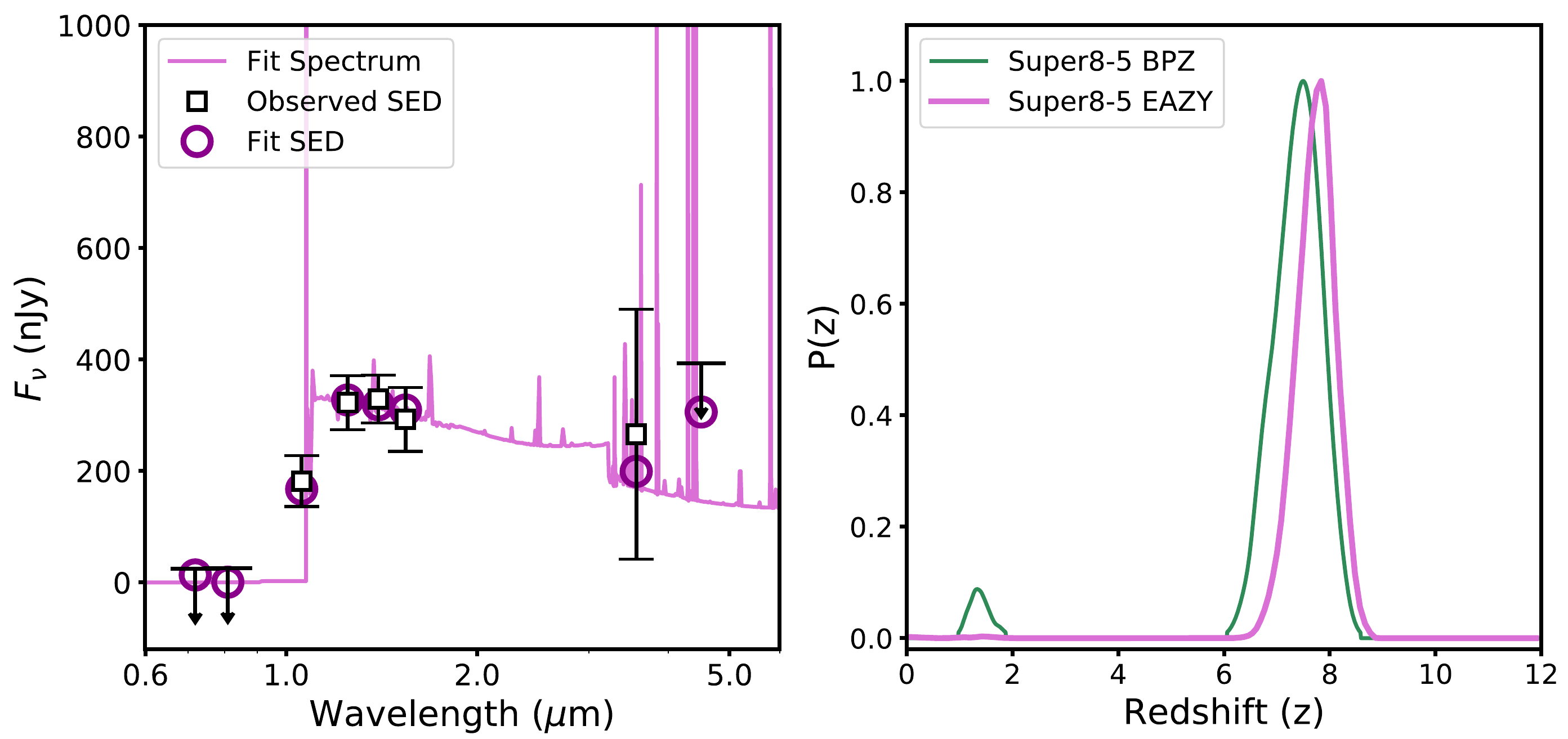}}
    \end{minipage}
\end{tabular}
\caption{Same as Figure~\ref{cutout1} but for Super Eight 5.}
\label{cutout5}
\end{figure*}

\begin{figure*}
\centering
\begin{subfigure}
	\centering
	\scalebox{1}
	{\includegraphics[width=1\textwidth]{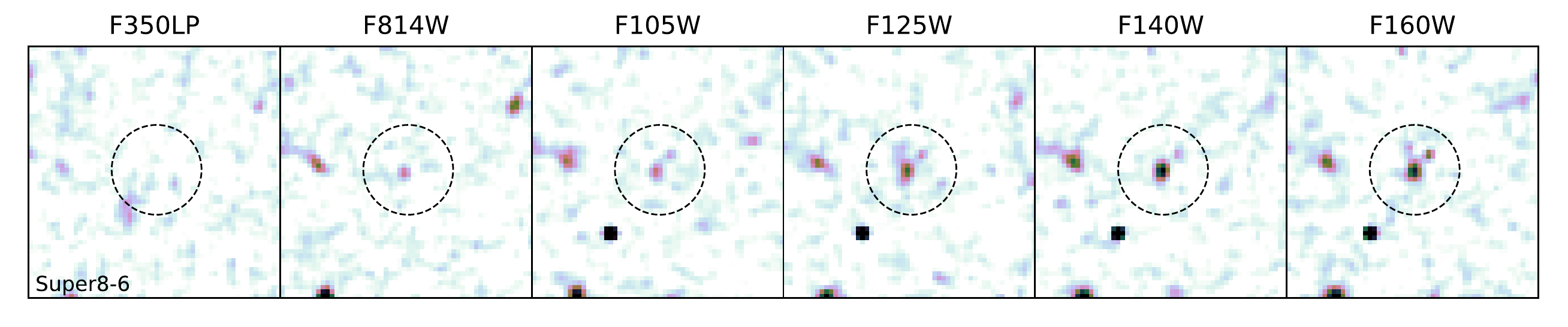}}
\end{subfigure}
\begin{subfigure}
	\centering
	\scalebox{0.35}
	{\includegraphics{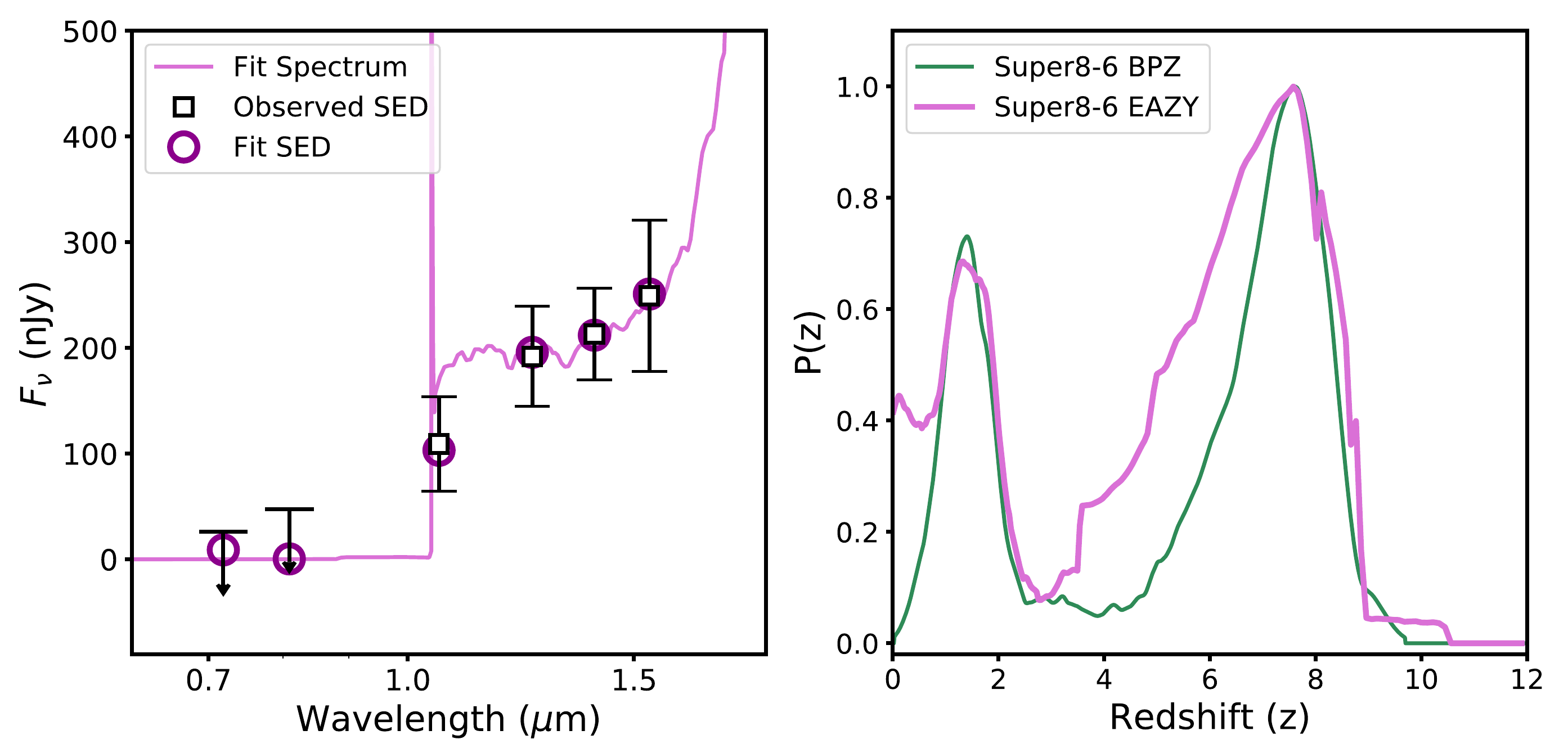}}
\end{subfigure}
\caption{Same as Figure~\ref{cutout1} but for Super Eight 6.}
\label{cutout6}
\end{figure*}

\begin{figure*}[!h]
\begin{tabular}{cc}
  	\begin{minipage}{0.95\textwidth}
    \centering
    	\scalebox{1}
    	{\includegraphics[width=1\textwidth]{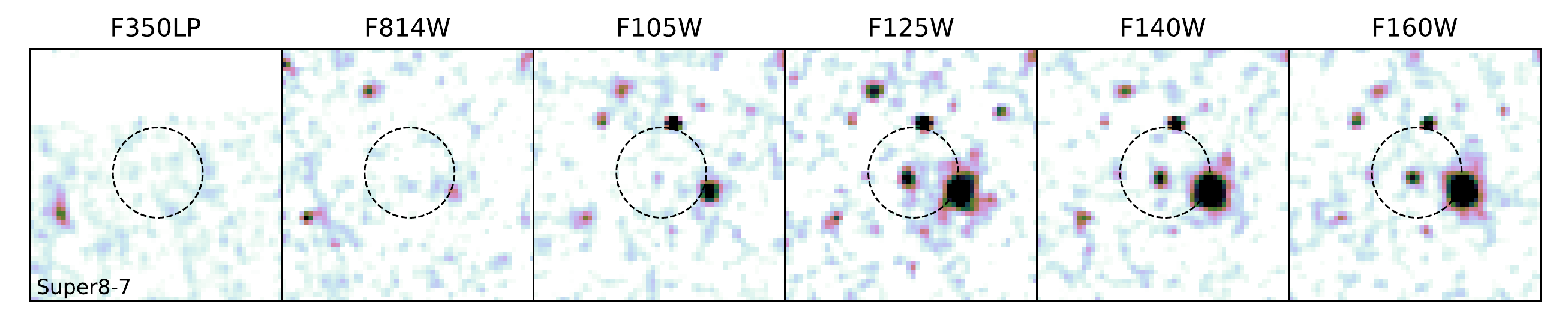}}
    \end{minipage}\\
    \begin{minipage}{0.95\textwidth} 
    	\scalebox{0.77}
    	{\includegraphics[width=0.45\textwidth]{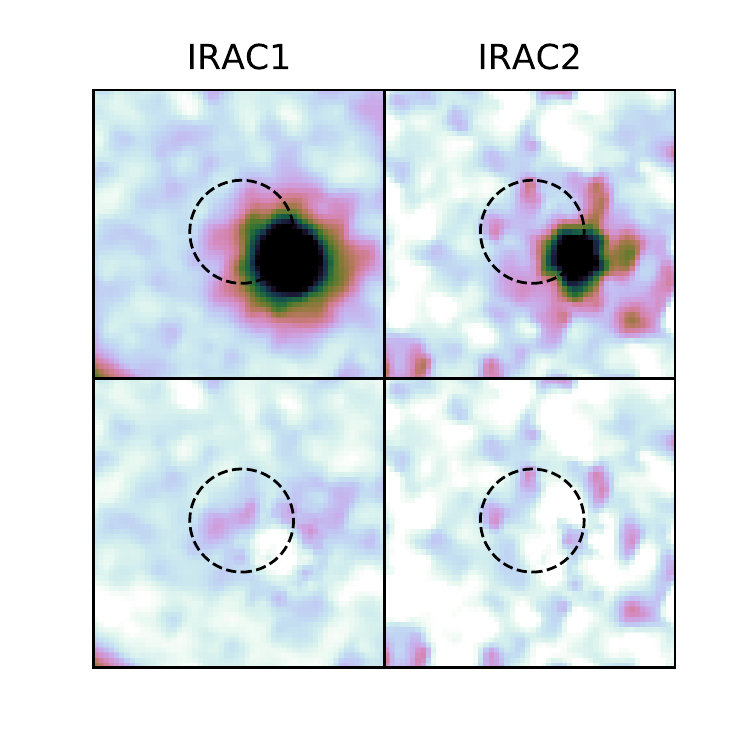}}
    	\scalebox{1.4}
    	{\includegraphics[width=0.45\textwidth]{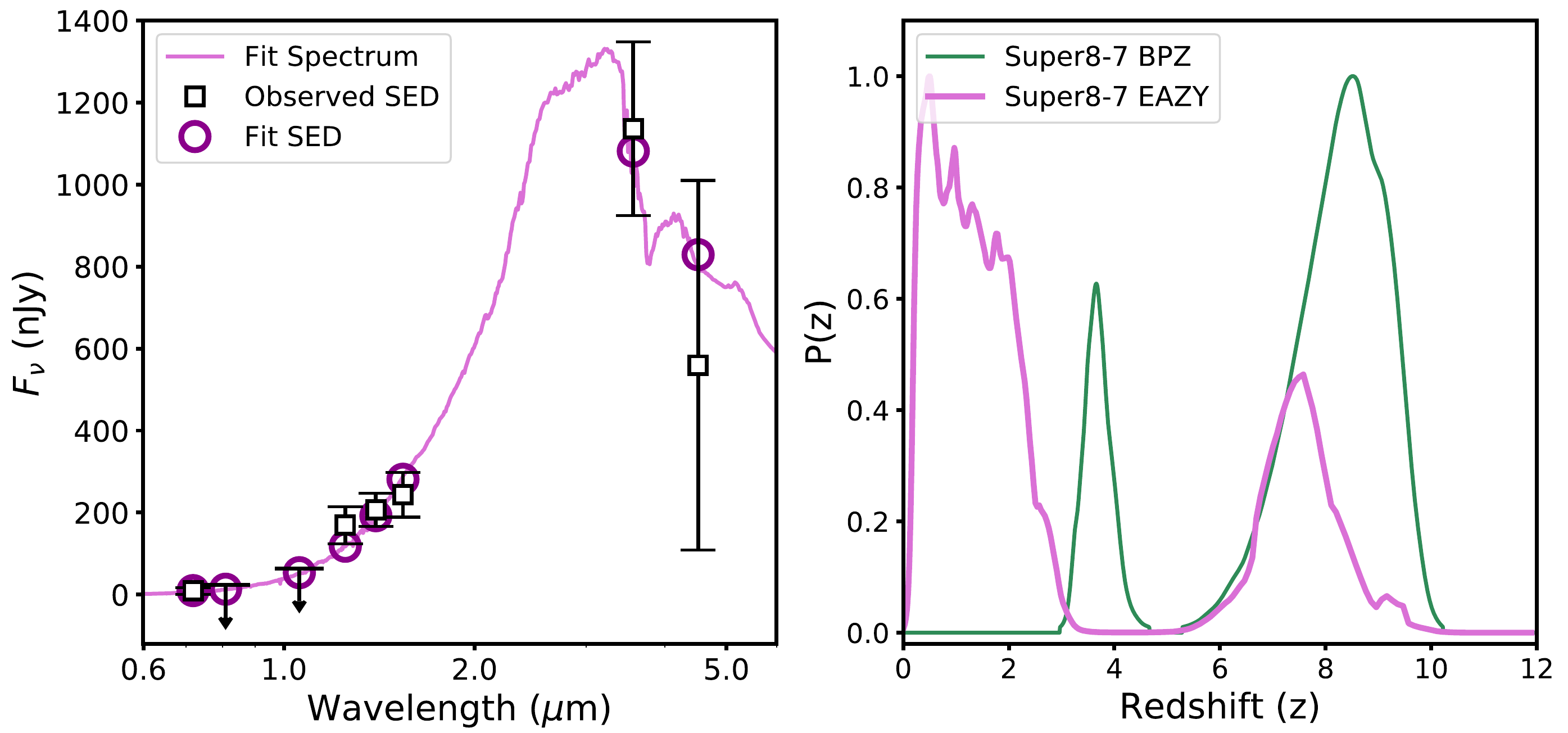}}
    \end{minipage}
\end{tabular}
\caption{Same as Figure~\ref{cutout1} but for Super Eight 7.}
\label{cutout7}
\end{figure*}

\begin{figure*}[!h]
\begin{tabular}{cc}
  	\begin{minipage}{0.95\textwidth}
    \centering
    	\scalebox{0.8}
    	{\includegraphics[width=1\textwidth]{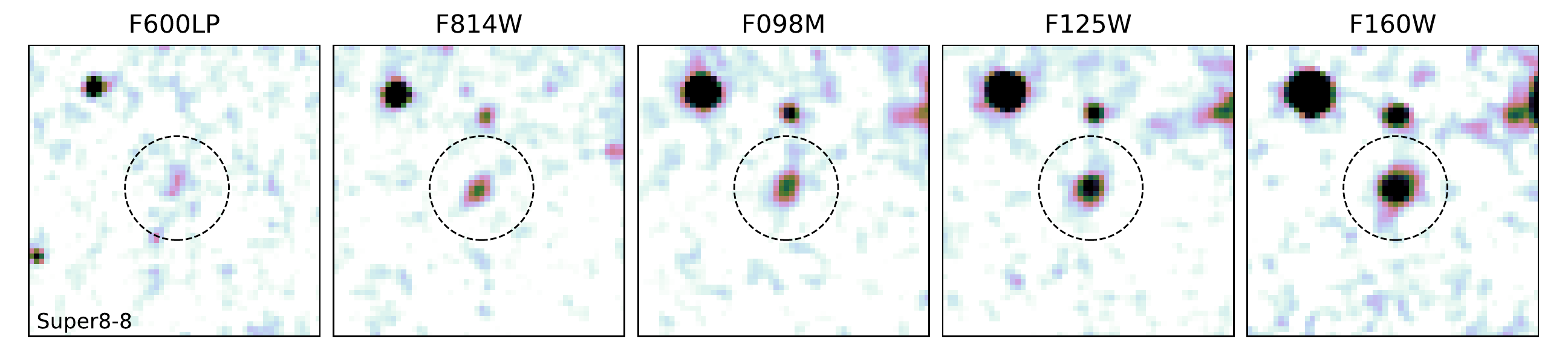}}
    \end{minipage}\\
    \begin{minipage}{0.95\textwidth} 
    	\scalebox{0.77}
    	{\includegraphics[width=0.45\textwidth]{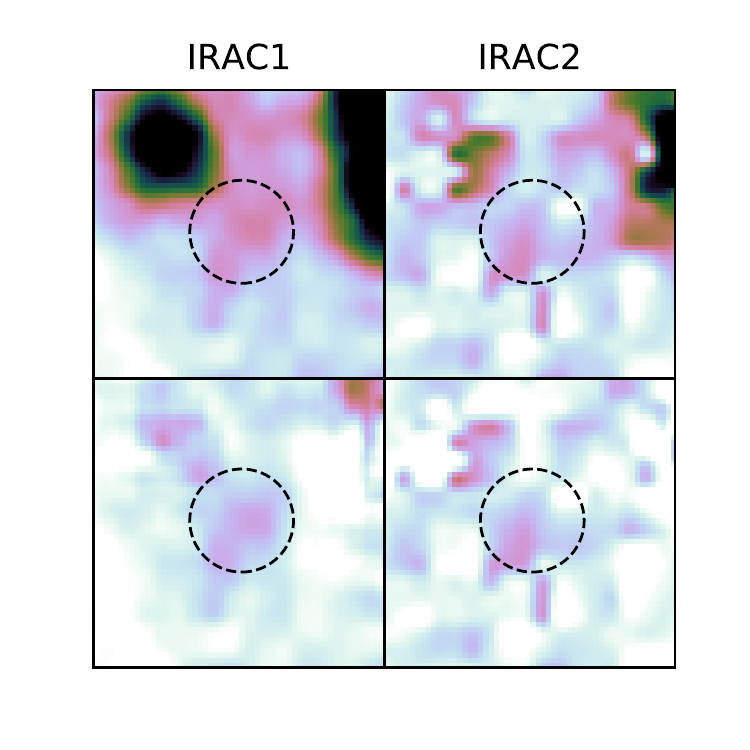}}
    	\scalebox{1.4}
    	{\includegraphics[width=0.45\textwidth]{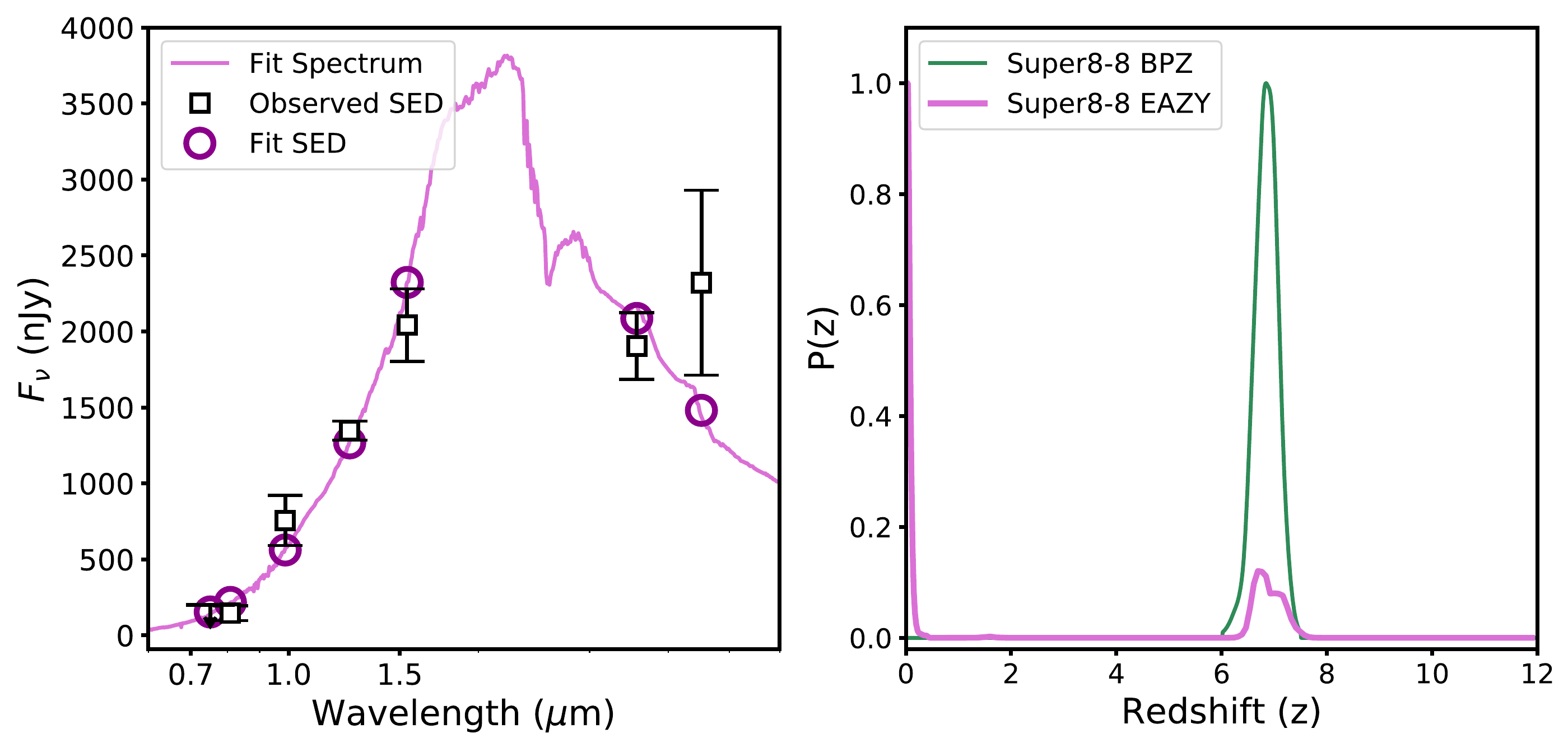}}
    \end{minipage}
\end{tabular}
\caption{Same as Figure~\ref{cutout1} but for Super Eight 8.}
\label{cutout8}
\end{figure*}
\clearpage

\noindent into account and the final result was scaled using the aperture correction.

We present the cutouts from the available \emph{HST} bands in Figures~\ref{cutout1}-\ref{cutout8} as well as the \emph{Spitzer} IRAC cutouts for the objects with available data.

\section{Photometric Redshift Fitting}
We performed photometric redshift fitting using the \texttt{EAZY} fitting code \citep{brammer2008}. This algorithm can quickly fit linear combinations of spectral templates that are based on semianalytic models (SAM), rather than spectroscopic samples. \cite{brinchmann2017} determined that the least-biased \texttt{EAZY} template for high-redshift galaxy fitting is the \texttt{eazy\_v1.3} template. This template is based on the original \texttt{EAZY} template that was developed from synthetic galaxy photometry using SAMs, but also includes several additional SEDs \citep{bruzual2003, maraston2005, erb2010} and emission lines \citep{ilbert2009}. In addition, we have included three lower-redshift templates of dusty, passively evolving galaxies with that represent a 2.5 Gyr-old population \citep{bruzual2003} with reddenings of $A_v=2, 5$ and 8 magnitudes using \cite{calzetti2000} extinction curves. We have also included four BPASS templates of young ($<10^{8.5}$ yr) systems with strong nebular lines and nebular continuum emission, as well as some dust extinction \citep{eldridge2017}. The inclusion of BPASS templates is particularly important because they include binary systems, which can elongate the lifetime of the youngest, most massive stars, resulting in strong emission, even in older galaxy populations. We include no magnitude priors in the fitting. The use of Bayesian priors is helpful when there are degeneracies between template colors and redshifts in fairly featureless spectra; the inclusion of the Lyman break as well as IRAC photometry in our SEDs precludes these degeneracies in our sample (see \citealt{brammer2008} for more detail.)  

In order to verify the \texttt{EAZY} photometric redshifts, we also used the Bayesian redshift fitting code \texttt{BPZ} \citep{benitez2000} using the same template SEDs as \cite{livermore2018}. We note that \texttt{BPZ} does not easily allow for the addition new galaxy templates, making it more difficult to compare the results directly.

\section{Galaxy Properties}
The photometric redshift fitting results for the Super Eight galaxies are given in Figures~\ref{cutout1} through~\ref{cutout8}. The addition of the F814W filter and the IRAC band(s) secure the photometric redshifts of five out of eight of the Super Eight galaxies candidates (Super8-1 through 5) as $7.1<z<8.0$. Super8-6 has a probability distribution that indicates that it is a high-redshift object, but the result is tenuous due to the lack of IRAC data for this galaxy. Super8-7 and Super8-8 are both likely low-redshift galaxies, although the higher redshift probability for Super8-7 is not small. The probability distributions for all objects are shown in the same series of figures. The redshifts of the six likely Super Eight galaxies are tabulated in Table~\ref{props}. The redshift errors were generated by determining the $1\sigma$ errors of the highest peak in the redshift probability distribution

\begin{figure*}
\centering
\scalebox{0.9}
{\includegraphics{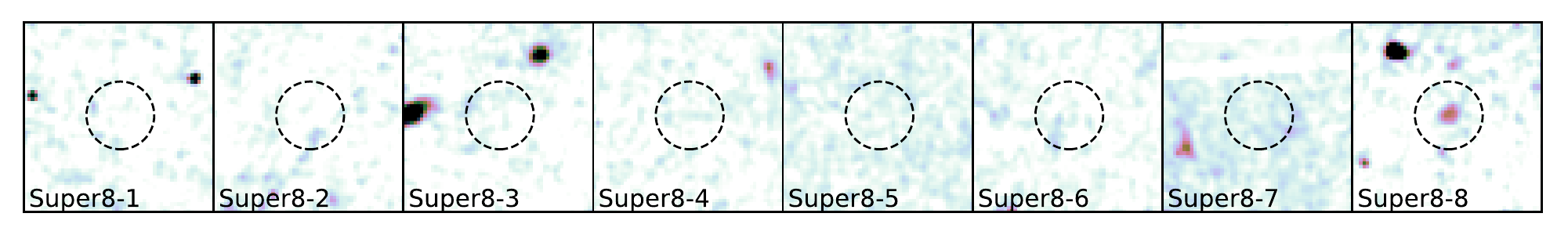}}
\caption{A stack of images blueward of the expected Lyman break for each of the Super Eights. All stacks include the \emph{HST} F814W filter and bluer.}
\label{blue}
\end{figure*}

As a check, we stacked the images blueward of the Lyman break (up to and including the \emph{HST} F814W filter) for all of the objects to verify that there was no flux in the blue bands, and we show the results in Figure~\ref{blue}. The ow-redshift nature of Super8-8 can be clearly seen in the stack.

\begin{figure}
\centering
\scalebox{0.4}
{\includegraphics{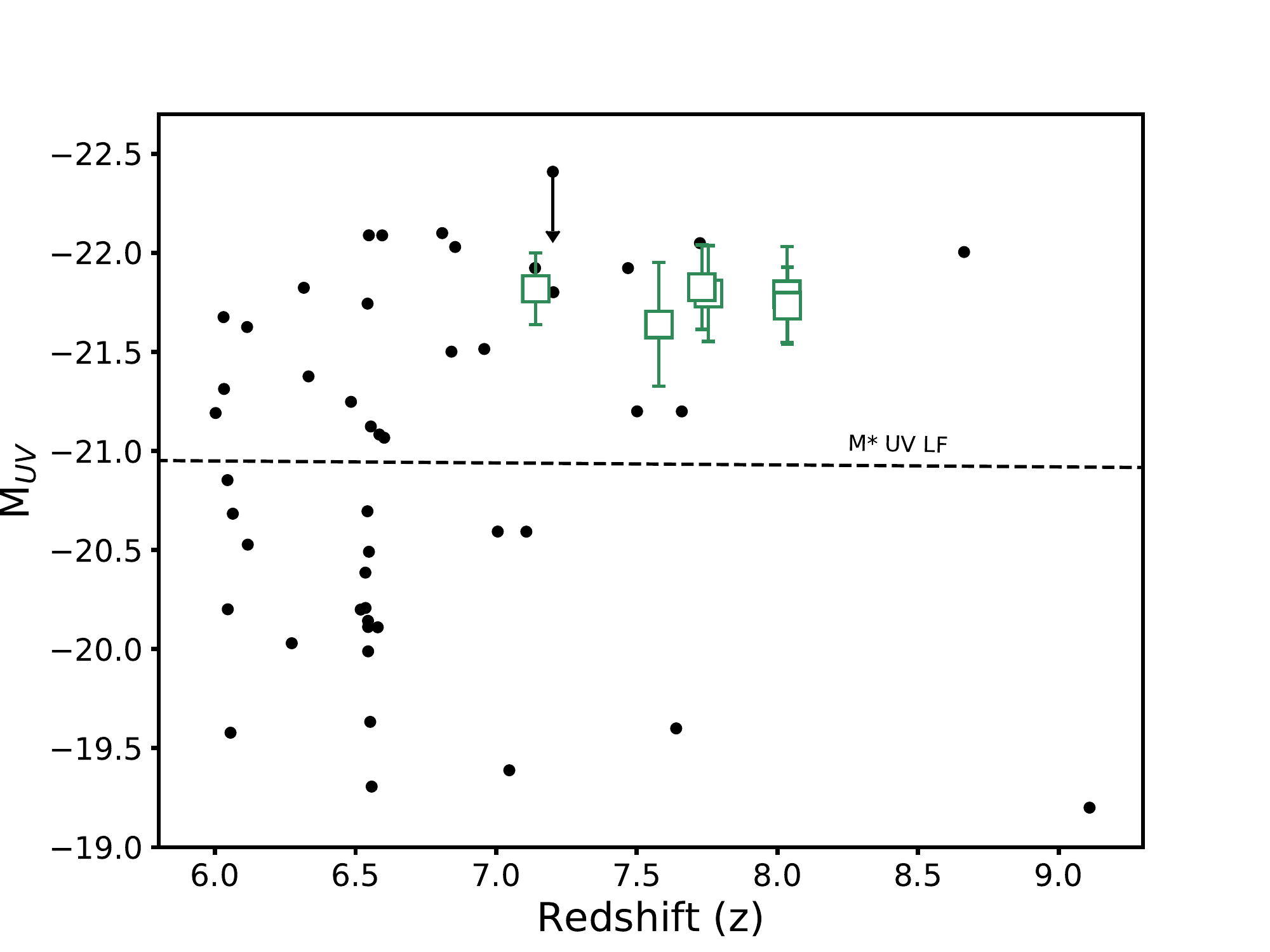}}
\caption{The rest-frame absolute UV magnitude vs. photometric redshift of the Super Eight galaxies (red squares). Also shown are $z>6$ galaxies with confirmed spectroscopic redshifts (black circles) \citep{vanzella2011, ono2012, shibuya2012, finkelstein2013, jiang2013, robertsborsani2016, oesch2015, zitrin2015, huang2016, song2016, hoag2017, stark2017, smit2018, hashimoto2018}. The dashed line traces the $z\sim8$ $M^*_{UV}$ luminosity function of \cite{bouwens2015}.}
\label{red_mag}
\end{figure}

In the appendix, we show the photometric redshift solutions both with and without the addition of the \emph{HST} F814W filter and the two IRAC bands. In most cases, the redshift solution is significantly affected by the addition of these filters.

Figure~\ref{red_mag} shows the rest-frame UV absolute magnitudes of the Super Eight galaxies evolution with redshift. Similar to the bright, high-redshift galaxies of \cite{robertsborsani2016}, these galaxies lie well above the $M^*_{UV}$ of the $z=8$ luminosity function of \cite{bouwens2015} and are some of the brightest known $z>7$ candidate galaxies.  

\subsection{$[3.6]-[4.5]$ IRAC Color}
There has been much discussion as to whether the $[3.6]-[4.5]$ IRAC color can be used to identify high-redshift galaxies \citep{schaerer2009, shim2011, labbe2013, stark2013, debarros2014, smit2015,stark2017}. Strong nebular line emission (such as that from H$\alpha$ and [O~III]+H$\beta$) can have a significant effect on \emph{Spitzer}/IRAC bands; in particular, $z\sim8$ galaxies could have 4.5-$\mu$m band emission from the [O~III]+H$\beta$ lines. \cite{robertsborsani2016} searched for $z>7$ candidates in the CANDELS fields with very red $[3.6]-[4.5]$ IRAC colors, and identified four candidates. These galaxies have all been spectroscopically confirmed as high-redshift Ly$\alpha$ emitters \citep{oesch2015,zitrin2015,stark2017}.

We give the $[3.6]-[4.5]$ IRAC colors for the Super Eights in Table~\ref{props}. Only one of the galaxies is detected at $1\sigma$ in both IRAC bands. While the upper limits (and $1\sigma$ detections) are useful for more precisely measuring the redshift of these objects, the IRAC colors are not well determined, but we include them here for completeness. Deeper IRAC observations will be needed to robustly determine the infrared colors of these objects.

\subsection{Sizes}
\begin{figure}
\centering
\scalebox{0.4}
{\includegraphics{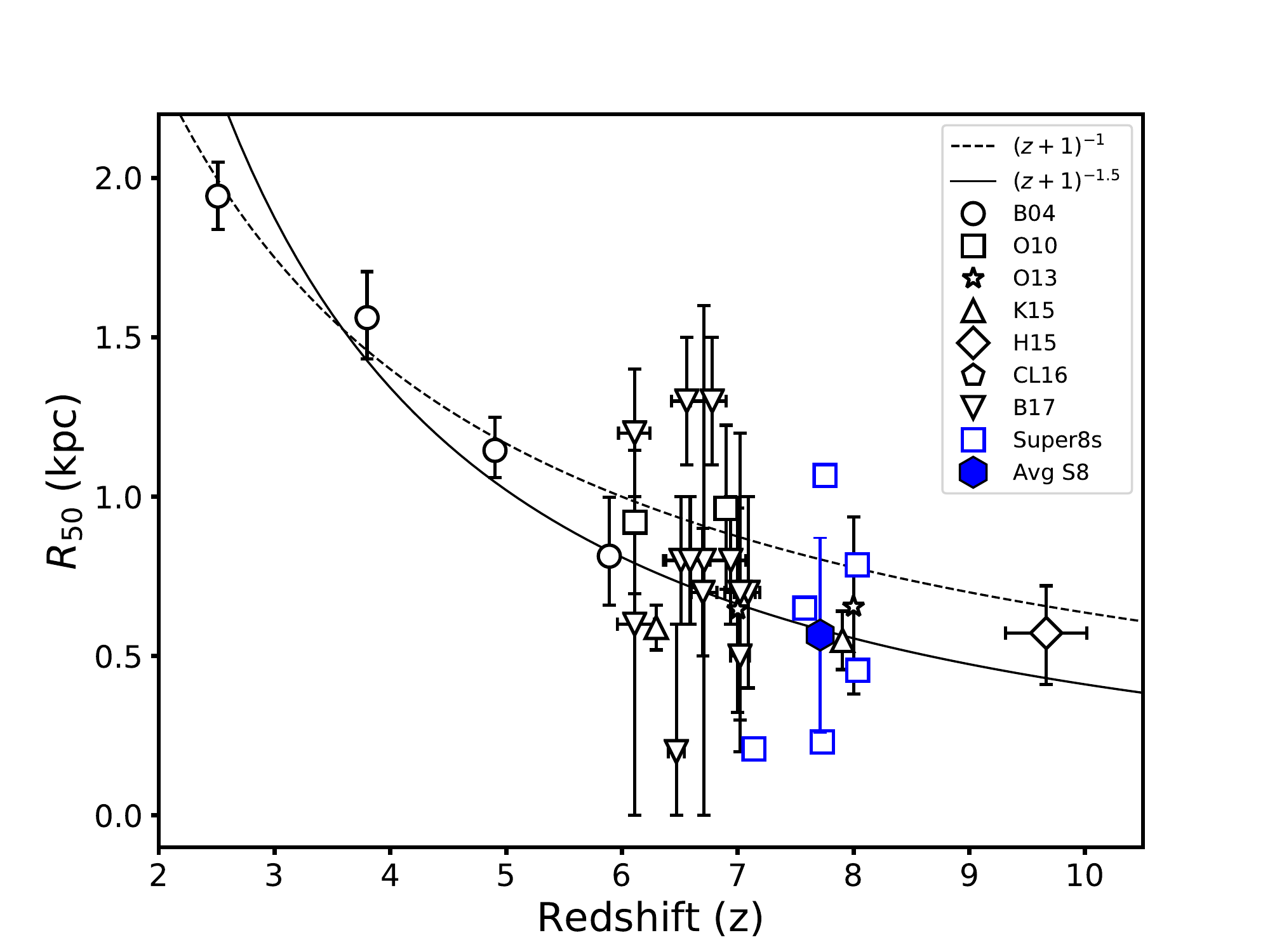}}
\caption{Physical sizes of the Super Eight sample versus redshift, shown with empty blue squares. Also shown are the measurements from \cite{bouwens2004} (B04), \cite{oesch2010} (O10), \cite{ono2013} (O13), \cite{holwerda2015} (H15), \cite{kawamata2015} (K15), \cite{curtislake2016} (CL16), and the single-component galaxies of \cite{bowler2017} (B17). Typical relations of $(1+z)^{-1}$ \citep[e.g.,][]{bouwens2004} and $(1+z)^{-1.5}$ \citep{ferguson2004} (normalized arbitrarily) are also shown for reference with dashed and solid lines, respectively.}
\label{size}
\end{figure}

\begin{figure}
\centering
\scalebox{0.8}
{\includegraphics{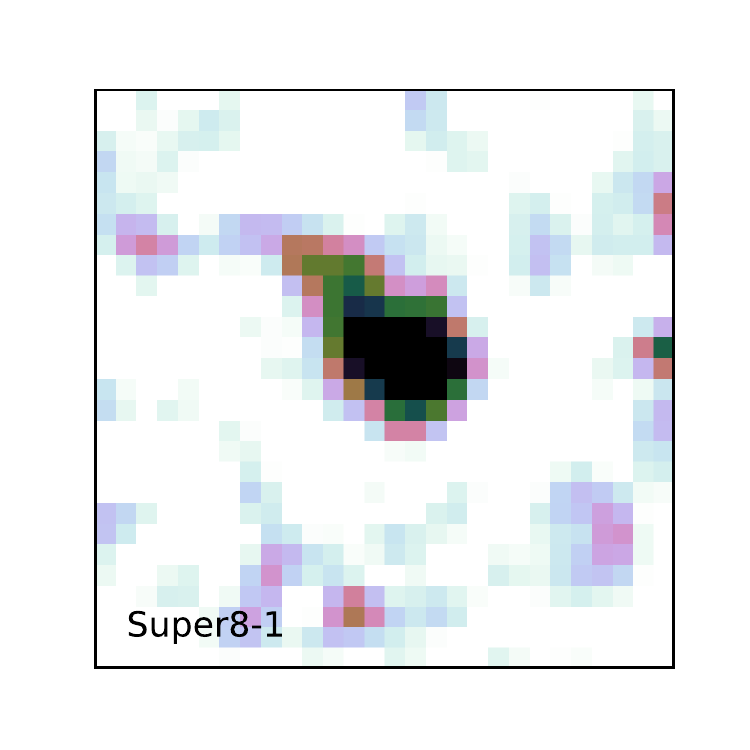}}
\caption{A 2\farcs5 $\times$ 2\farcs5 cutout of the F160W filter of Super8-1. The galaxy is fairly extended and asymmetric.}
\label{single_cut}
\end{figure}

We determined the rest-frame UV half-light radii of the Super Eight sample using \texttt{Source Extractor} on the F160W images. We converted the measured effective radii to arcseconds using the respective pixel sizes and corrected the measured sizes for the PSF using $R_{50} = \sqrt{R_{SE}^2 - R_{PSF}^2}$, where $R_{PSF}=0\farcs14$ for the F160W filter on WFC3. Figure~\ref{size} shows the Super Eight galaxy physical size evolution with redshift, and the sizes are given in Table~\ref{props}. Super8-2 is the most extended object in the sample with a size of $R_{50} = 0\farcs21$. The mean physical size of the resolved objects is ($R_{50}=0.6\pm0.3$ kpc), which is typical of what has been found for other high-redshift samples; \cite{bowler2017} found a range of 0.5-3 kpc for a sample of $z\sim7$ galaxies. Figure~\ref{single_cut} shows a 2\farcs5$\times$2\farcs5 cutout of the F160W filter of Super8-1 to show typical size exhibited by some of the Super Eight galaxies.

Size estimates can be affected by the choice of measurement technique \cite[e.g.,][]{huang2013, curtislake2016}. We chose the \texttt{Source Extractor} sizes for several reasons: (1) \texttt{GALFIT} \citep{peng2002, peng2010} can only perform a decomposition if the data is sufficiently sampled; however, the BoRG data lacks the resampling from a intentional dither strategy, (2) \texttt{Source Extractor} sizes have proven been shown to be accurate enough for size estimates (see \citealt{grazian2012,holwerda2014}), and (3) \texttt{Source Extractor} is the standard tool for size measurements at lower redshift. While \texttt{Source Extractor} sizes may be underestimates in the case of very extended, low-surface brightness galaxies \citep{grazian2012, ono2013}, this is an issue only for objects more extended and with lower surface brightness than those presented here. 

\begin{deluxetable*}{cccccc}[!t]
\tablecolumns{6}
\tablecaption{Super Eight Properties\label{props}}
\tablehead{\colhead{ID} & \colhead{$z_{\textrm{phot}}$\tablenotemark{a}} & \colhead{log(M$_*$) (M$_\sun$)} & $M_{UV}$ & \colhead{$R_{50}$ (kpc)} & \colhead{$[3.6]-[4.5]$}}
\startdata
Super8-1 &  $8.0^{+1.0}_{-0.7}$  & $10.1$ & $-21.73\pm0.31$ & $0.45$ & $-1.2\pm3.5$  \\[3pt]
Super8-2 &  $7.8^{+0.7}_{-0.6}$  & $10.4$ & $-21.80\pm0.24$ & $1.07$ & $1.0\pm1.8$  \\[3pt]
Super8-3 &  $7.1^{+0.4}_{-0.3}$  & $10.1$ & $-21.82\pm0.18$ & $0.21$ & $0.4\pm0.7$  \\[3pt]
Super8-4 &  $8.0^{+0.5}_{-0.5}$  & $10.4$ & $-21.79\pm0.24$ & $0.79$ & - - \\[3pt]
Super8-5 &  $7.7^{+0.4}_{-0.4}$  & $9.8$  & $-21.83\pm0.21$ & $0.23$ & $0.4\pm1.5$  \\[3pt] 
Super8-6 &  $7.6^{+0.5}_{-2.6}$  & $10.6$  & $-21.64\pm0.31$ & $0.66$ &    - -     
\enddata
\centering
\tablenotetext{\tiny a}{The photometric redshift output by \texttt{EAZY}. The errors are $\pm1\sigma$ \\of the highest peak in the p$p(z)$ probability distribution.}
\end{deluxetable*}

Previous explorations of the size of high-redshift galaxies have been done either using a few faint galaxies in the \emph{Hubble} Ultra Deep Field or the eXtremely Deep Field \citep[e.g.,][]{oesch2010, ono2012, hathi2012}, or with larger samples of galaxies at $z\sim9$ in the CANDELS fields \citep{oesch2014,holwerda2015}. \cite{oesch2014} found little evolution from $7<z<8$ in their sample, noting that there were no bright, extended objects at $z\sim8$, while \cite{bowler2017} found more extended and merging systems at $z\sim7$. There is significant scatter in the sizes of the Super Eight galaxies, but there is room to state that the most luminous of these early galaxies may have more extended sizes, tentatively extending the trend found by \cite{bowler2017} to the $z\sim8$ regime.

The mean galaxy size has been shown to decrease steadily with increasing redshift, although this relationship has significant scatter. Analytical models from \cite{fall1980} and \cite{mo1998} predict effective radii should scale somewhere in between $(1 + z)^{-1}$ for galaxies in halos of fixed mass and $(1 + z)^{-1.5}$ for a fixed circular velocity. More recent theoretical work \citep[e.g.,][]{somerville2008, wyithe2011, stringer2014} retains this global redshift relation for massive galaxies. Observational evidence from earlier samples point to a relation somewhere in between with some disagreement over the precise scaling relation, where some studies indicate $(1 + z)^{-1}$ \citep{bouwens2004, bouwens2006, oesch2010, holwerda2014}, other studies preferring $(1 + z)^{-1.5}$ \citep{ferguson2004}, and yet others lying between these two relations \citep{hathi2008, ono2013}. 

Both \cite{shibuya2015} and \cite{curtislake2016} have pointed out, however, that the arithmetic mean may not be the optimal description for the growth of galaxy sizes with redshift in this epoch, given the existence of a ``tail'' of extended galaxies, which would skew the mean and introduce a selection effect due to surface brightness dimming. The mean of the Super Eight sizes, shown in Figure~\ref{size} as a filled blue hexagon, does not indicate such an effect, and shows an agreement with the $(1 + z)^{-1.5}$ relation.

\cite{holwerda2015} examined the sizes of a set of candidate $z\sim9-10$ galaxies from CANDELS \citep{oesch2013} and found that the dusty low-redshift interlopers for that sample had a mean effective radius of $\langle r_e\rangle =0\farcs59$. These potential contaminants are significantly larger than any of the Super Eight galaxies.

\subsection{Masses}
We determine the masses of the Super Eight galaxies by performing SED fits using the \texttt{MAGPHYS} algorithm \citep{dacunha2008}. \texttt{MAGPHYS} a Bayesian algorithm that uses panchromatic galaxy model spectra from \cite{bruzual2003} and \cite{charlot2000} to derive likelihood estimates on various galaxy properties. Specifically, we use the high-redshift extension as described in \cite{dacunha2015}, which includes new star formation histories that are appropriate for galaxies whose star formation is still increasing, and improved prescriptions for dust attenuation and effects of the intergalactic medium on UV photons.

To perform the SED fitting, we employ a \cite{chabrier2003} initial mass function (IMF) and a star formation history (SFH) that increases linearly then decreases exponentially with time (see \cite{dacunha2015} for more details.) The code uses a power law to describe the dust content \citep{charlot2000}. We specify the redshift determined by the photometric redshift fitting in Section 3. The resulting mass estimates are reported in Table~\ref{props}.

\begin{figure*}
\centering
\scalebox{0.7}
{\includegraphics{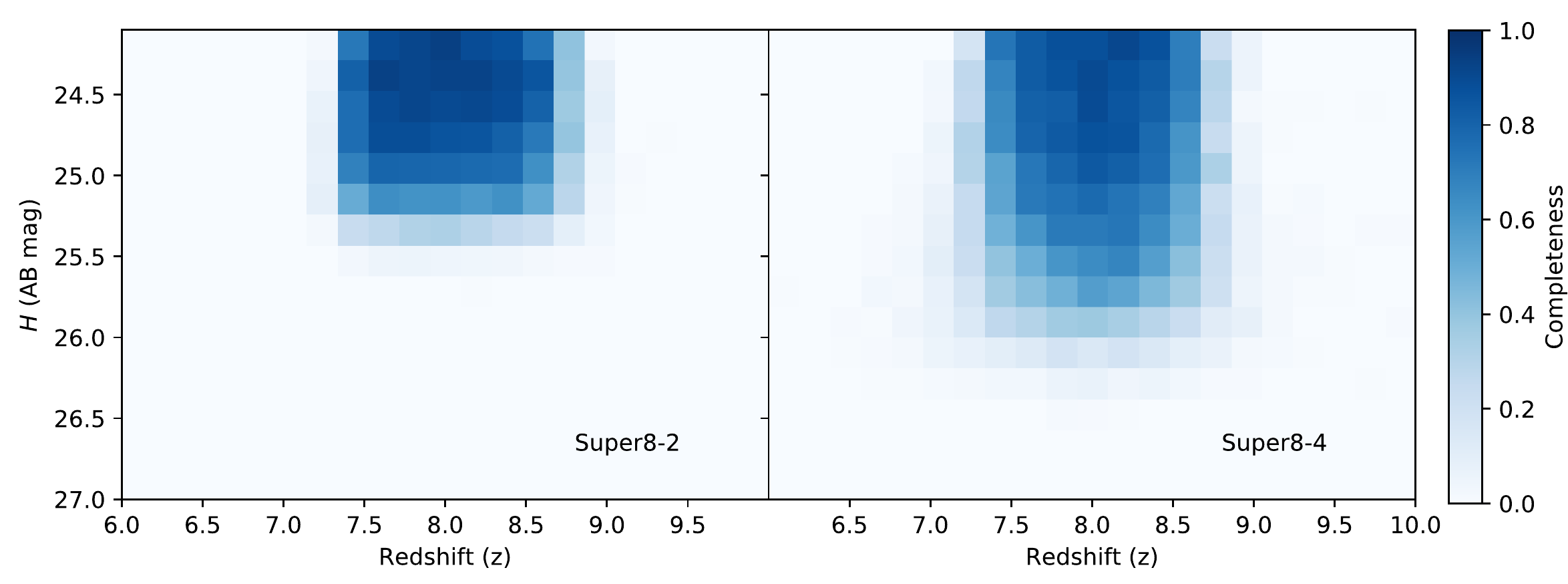}}
\caption{The selection probability $S(z,m)\; C(m)$ for Super8-2 and 4. Super8-2 is given as an example of the three galaxies found in the extended BoRG search, while Super8-4 is representative of the four galaxies originally found by \cite{calvi2016}. The selection function was used to determine the total effective volume probed in the search for the Super Eights galaxies. }
\label{sel_func}
\end{figure*}

The masses derived do not explicitly take into account the fact that at high-redshift, the 4.5-$\mu$m band may be contaminated with strong emission lines. We therefore performed the SED fitting without the [4.5] band, and found that the results are well within $1\sigma$ of the original results.

As a check, we also determined the masses using the SED fitting code \texttt{FAST} \citep{kriek2009}. Using a \cite{chabrier2003} IMF, the \cite{calzetti2000} dust law, and an exponentially decreasing SFH, the code uses the Flexible Stellar Population Synthesis (FSPS) of \cite{conroy2009} to perform chi-squared fitting. The \texttt{FAST} skewed lower than the \texttt{MAGPHYS} results by about $2\sigma$ for two out of six objects. Given that we cannot pin down the infrared emission from these galaxies at this redshift, this is not surprising. We report the \texttt{MAGPHYS} results as that code, in particular the star formation history, has been optimized for high-redshift galaxy analysis.

The masses of the Super Eight galaxies are comparable to what has been found previously for very bright galaxies at this redshift. \cite{oesch2015} report the mass for a spectroscopically-confirmed $z=7.73$ galaxy with $m_H = 25.0$ as log(M$_*$/M$_\sun$) = $9.9\pm0.2$. This is in line with the masses of the Super Eights.

\section{Volume Density of the Super Eights}
In order to place the Super Eight galaxies on the bright end of the luminosity function, we must first determine the total volume sampled. The three galaxies originally identified by \cite{calvi2016} (Super Eights 4-6) were found over an effective area of $\sim130$ arcmin$^2$. Due to the pure-parallel nature of the BoRG survey, each field has a distinct limiting magnitude, which are reported in Table 1 of \cite{calvi2016}, along with the effective area of each field. For the remaining three Super Eights galaxies, the survey area is $\sim300$ arcmin$^2$ with a limiting magnitude of $m_{H}=26.3$, as reported by \cite{bouwens2015}.

In order to derive the completeness of our sample, we performed simulations to recover artificial sources in the fields \citep{oesch2007, oesch2009}. These simulations determine a completeness function, $C(m)$, and a redshift selection function $S(z,m)$ at $z\sim8$ for each field, resulting in an overall selection probability. The effective volume as a function of magnitude is
\begin{equation}
V_{\textrm{eff}}(m) = \int_0^{\infty} S(z,m)\; C(m) \frac{dV}{dz} dz
\end{equation}
The artificial sources cover a range of SEDs, magnitudes, sizes, and redshifts. Using the same selection criteria used for the candidate selection, we then determined how well we were able to recover these artificial sources, resulting in the source selection and completeness functions. These simulations therefore allowed us to account for volume loss due to foreground sources or limiting magnitudes of the fields. The selection function for Super8-2 and 4 are shown in Figure~\ref{sel_func}. The BoRG search that resulted in the four galaxies originally identified by \cite{calvi2016}, of which Super8-4 is an example, reached fainter F160W magnitudes than the other three galaxies whose identity as high-redshift galaxies was less certain (see Section 2.)

\begin{figure}
\centering
\scalebox{0.45}
{\includegraphics{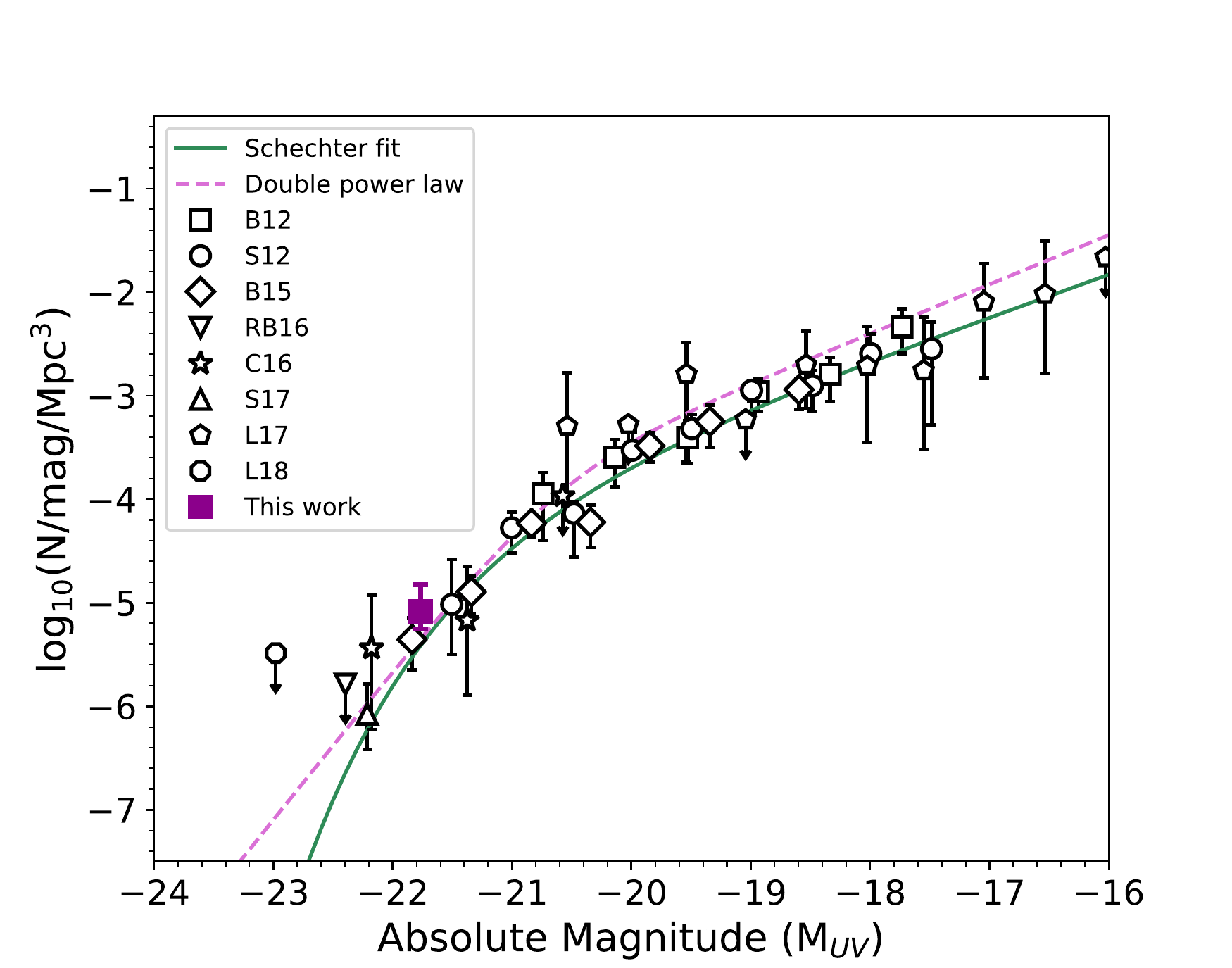}}
\caption{\textit{Left}: The UV luminosity functions for $z\sim8$ galaxies. The dark magenta indicates the Super Eight sample. The error bars are dominated by the Poisson error. The solid green line gives the $z\sim8$ Schechter function fit to the UV luminosity function from \cite{bouwens2015}. The double power law of \cite{bowler2014} is shown with a pink dashed line. Previous LF results from \cite{bradley2012} (B12), \cite{schenker2012} (S12), \cite{bouwens2015} (B15), \cite{robertsborsani2016} (RB16), \cite{calvi2016} (C16), \cite{stefanon2017} (S17), \cite{livermore2017} (L17), and \cite{livermore2018} (L18) are also shown. }
\label{LF}
\end{figure}

We calculate the $z\sim8$ LF using this effective volume, including only the five objects with $z>7.5$ (Super8-3 has $z = 7.1$). The resulting volume density is $8.24^{+4.90}_{-3.26} \times 10^{-6}$ mag$^{-1}$ Mpc$^{-3}$ for $25 < m_H< 25.5$. The errors are dominated by Poisson uncertainties. Figure~\ref{LF} shows this result, along with various volume densities from the literature. The Super Eights lie slightly above the $z\sim8$ Schechter luminosity function determined by \cite{bouwens2015}. It has been posited that the bright end of the LF may be better described by a double power law \citep[e.g.,][]{bowler2014}; the Super Eights cannot conclusive rule out either a Schechter function or double power law.

\section{Discussion}
We have presented here the Super Eight galaxies, a set of six very luminous, $7.1<z<8.0$ galaxy candidates. The properties of these galaxies are similar to the high $H$-band magnitude galaxies that have been found in the CANDELS EGS and COSMOS fields by \cite{robertsborsani2016}.

Finding such bright objects at $z\sim8$ challenges our understanding of the evolution of the luminosity function at the bright end. It has been suggested in recent studies that the bright end of the high-redshift luminosity function may in fact not follow a Schechter fit, and instead is better characterized by a double power-law \citep[e.g., ][]{bowler2014, bowler2015, stefanon2017, ono2018}. The fact that the existence of very bright $z\sim8$ galaxies is becoming much more commonplace may support this idea. Both \cite{livermore2018} and \cite{morishita2018} found that the $z\sim9$ luminosity function does not favor a power-law form at the bright end, but also note that it is not ruled out for the BoRG data. According to the derived luminosity functions of \cite{mason2015}, \cite{bouwens2015}, and \cite{finkelstein2015}, finding galaxies of these UV magnitudes at this redshift is very unlikely. The Super Eight sample of galaxies presented here favors a double power law interpretation, but a Schechter form for the luminosity function at this redshift is not ruled out.

The fact that Ly$\alpha$ has been observed in galaxies of comparable luminosity to these particular galaxies, while Ly$\alpha$ is largely attenuated in lower luminosity galaxies that are also in the heart of the reionization epoch, makes them unique targets to study to better understand the details regarding reionization in the early universe. It has been posited that these very bright objects could be ionizing the neutral hydrogen around them more efficiently than their less luminous counterparts \citep{zitrin2015,stark2017}.

The Super Eight galaxies make ideal candidates in which to search for Ly$\alpha$. If Ly$\alpha$ is not detected in these galaxies, this may signify that the EGS field, with its 100\% success rate in detecting Ly$\alpha$, is a unique line-of-sight. If we are successfully able to detect Ly$\alpha$ in the Super Eight targets, it will be an indication that bright galaxies play a significant role in the reionization process, as it is expected that the IGM is largely neutral from $7<z<9$.

The detection of Ly$\alpha$ coupled with red IRAC colors (which can be a sign of strong [O~III]+H$\beta$ emission) indicates a young stellar population in these galaxies. As noted by \cite{stark2017} and backed up by modeling by \cite{mason2018}, the hard ionizing radiation from these galaxies may be preferentially ionizing the IGM surrounding them, creating large ionized bubbles within the neutral hydrogen. While the IRAC colors of the Super Eights are inconclusive, deeper \emph{Spitzer} data may reveal the emission lines properties of these sources.

\section{Conclusions}
In this paper we have presented a sample of eight very bright ($m_H < 25.5$) $z\sim8$ galaxy candidats from the BoRG survey called the Super Eights. We began with a set of eight $Y$-band dropout candidates, four of which had previously been presented in \cite{calvi2016}, while the remaining four were found in the BoRG fields using the selection criteria of \cite{bouwens2015}. In this work, six of the eight galaxies have been determined to be high-redshift objects.

The properties of this sample are as follows:
\begin{itemize}
\item Using the photometric redshift codes \texttt{EAZY} \citep{brammer2008} and \texttt{BPZ} \citep{benitez2000}, we fit all available photometry to determine if the galaxies are at high redshift. We employed new \emph{HST} F814W and \emph{Spitzer} IRAC data (GO 14652, PI B. Holwerda) in addition to archival \emph{HST} data and found that seven of the original eight candidate galaxies are high-redshift objects with $7.1<z<8.0$. This confirms the redshift fitting of \citep{calvi2016}. We found one source to be a likely low-redshift interloper.
\item The IRAC data help constrain the photometric redshift solutions for the Super Eight galaxies. However, due the the shallow nature of our \emph{Spitzer} IRAC observations, we cannot conclusively state that these galaxies have red IRAC colors, such as those reported by \cite{robertsborsani2016}. Deeper \emph{Spitzer} data are needed to show if an infrared excess in the 4.5-$\mu$m band due to nebular emission lines such as [O~III]+H$\beta$ exists. 
\item The sizes of the Super Eight galaxies as determined from their rest-frame UV half-light radii are on par with those measured at similar redshifts. The mean physical size is $R_{50} = 0.6\pm0.3$ kpc, in line with the $(1+z)^{-1.5}$ relationship, but there is significant scatter in the size-redshift relationship.
\item We performed SED fitting of the Super Eights using both \texttt{MAGPHYS} \citep{dacunha2008} and \texttt{FAST} \citep{kriek2009} and found that these galaxies have a mean mass of log(M$_*$/M$_\sun$) $\sim10$.
\item The volume density of this sample of galaxies is $8.24^{+4.90}_{-3.26} \times 10^{-6}$ mag$^{-1}$ Mpc$^{-3}$ for $25 < m_H< 25.5$ at $z\sim8$. This is a slightly higher density than predicted at the bright end of the luminosity function using a Schechter fit \citep[e.g., ][]{bouwens2015}. Although this could be an indication that the the luminosity function is better modeled using, for example, a double power law function, rather than the Schechter function, it could be simply be result of small number statistics.
\end{itemize}
The Super Eight galaxies join a growing sample of very bright high-redshift objects that are motivating our understanding of the history of reionization. In the future, they will make excellent early candidates to follow up with \emph{JWST}. 
\\
\\
\indent We thank the referee for their useful feedback. We acknowledge the support of NASA/STScI grant HST-GO-14652. R.\ S.\ acknowledges a Rubicon program with project number 680-50-1518. This work makes use of NASA's Astrophysics Data System.
\\
\facilities{\textit{HST} (ACS, WFC3), \textit{Spitzer} (IRAC)}
\software{\texttt{AstroPy} \citep{astropy2013, astropy2018}, \texttt{FAST} \citep{kriek2009}, \texttt{MAGPHYS} \citep{dacunha2008sc}, \texttt{Matplotlib} \citep{hunter2007}, \texttt{NumPy} \citep{vanderwalt2011}, \texttt{SciPy} \citep{jones2001}, \texttt{EAZY} \citep{brammer2008}}

\appendix
\section{Photometric redshift comparison}
We performed photometric redshift fitting on the eight candidate Super Eight galaxies using \texttt{EAZY} \citep{brammer2008}, both with and without the \emph{HST} F814W and \emph{Spitzer} IRAC bands. The results are presented in Figures~15-17. In many cases, these additional data made a significant difference in the redshift fit. For several other galaxies, the probability distribution of the redshift fit was bettered significantly. It should be noted the our results without the F814W and IRAC bands differ significantly from the \texttt{BPZ} \citep{benitez2000} redshifts reported by \cite{calvi2016}. There are several likely reasons for this, the foremost being that \texttt{BPZ} does not easily allow for the addition new galaxy templates, make the results less robust. Additionally, variation in the measured photometry can affect the results.

Finally, it should be noted that \texttt{EAZY} fits the templates via the linear combination of all of the input templates simultaneously, stepping through the redshift grid to find the best-fitting template spectrum via chi squared minimization. The result is that in some cases, the best-fit template may be inappropriate for the output redshift (e.g., the bottom panel of Figure~\ref{sed_compare1}.) This underlines the importance of having the F814W and IRAC filters - they are necessary to omit unphysical results.

\begin{figure*}[!h]
\centering
\begin{subfigure}
    \centering
	\scalebox{0.45}
	{\includegraphics{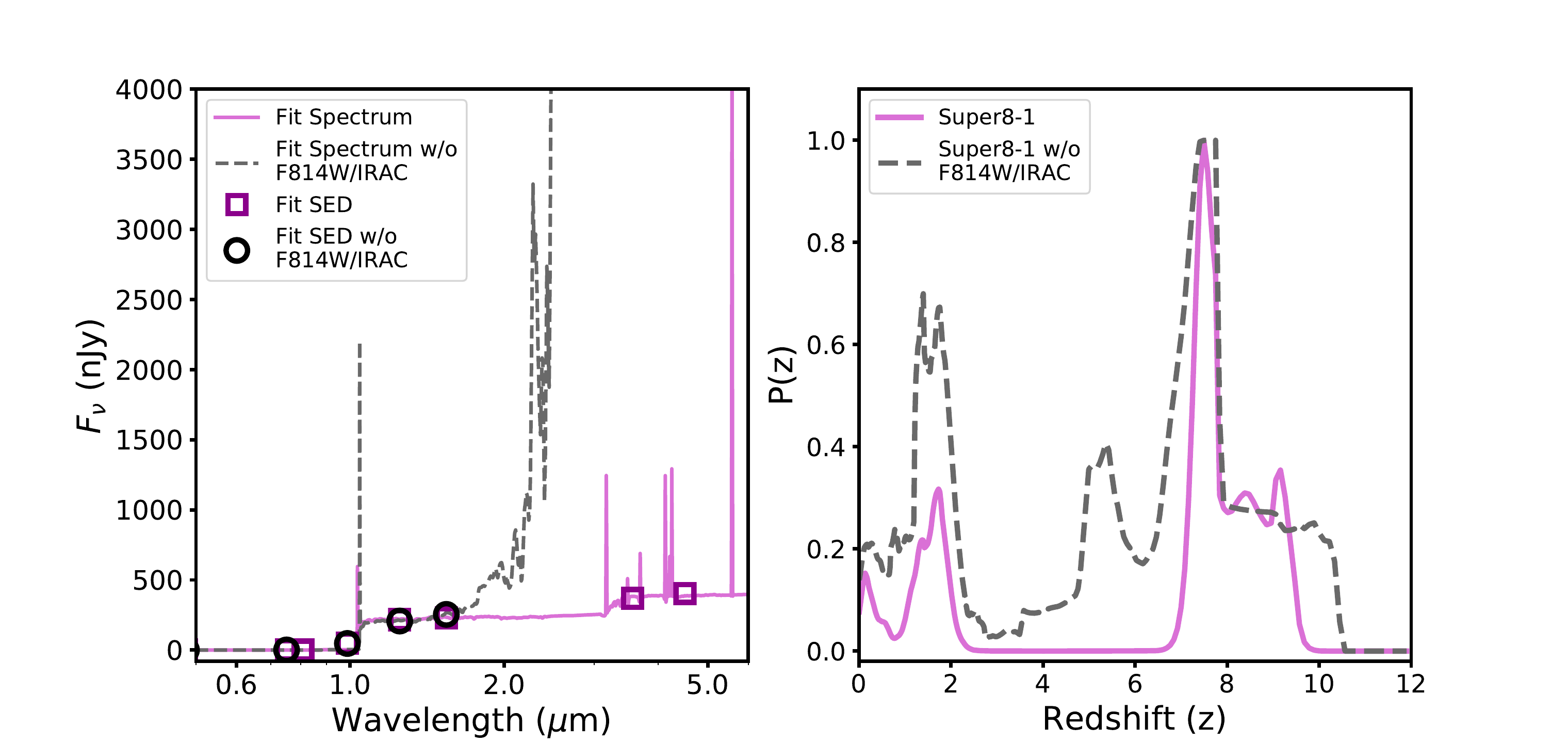}}
\end{subfigure}
\begin{subfigure}
	\centering
	\scalebox{0.45}
	{\includegraphics{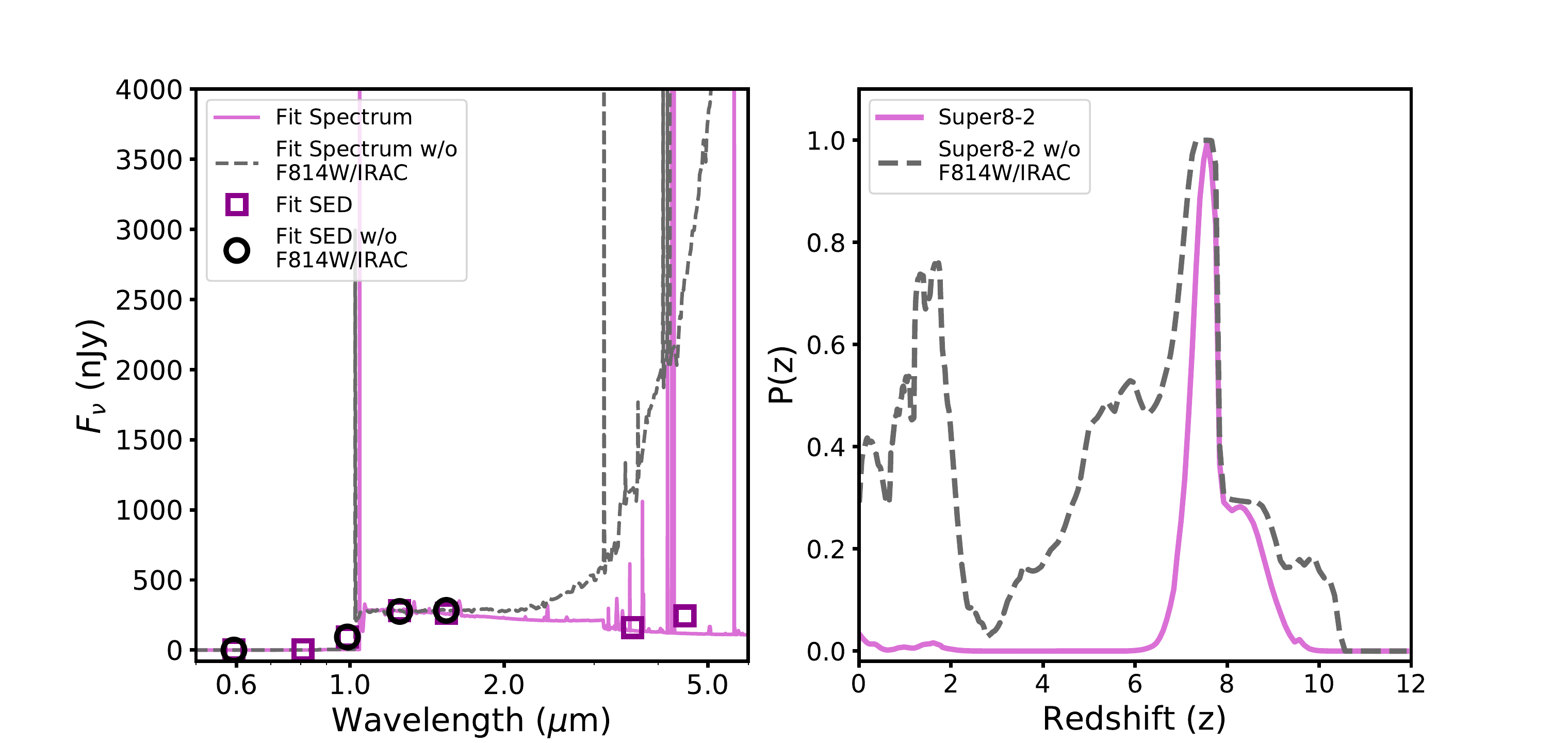}}
\end{subfigure}
\begin{subfigure}
	\centering
	\scalebox{0.45}
	{\includegraphics{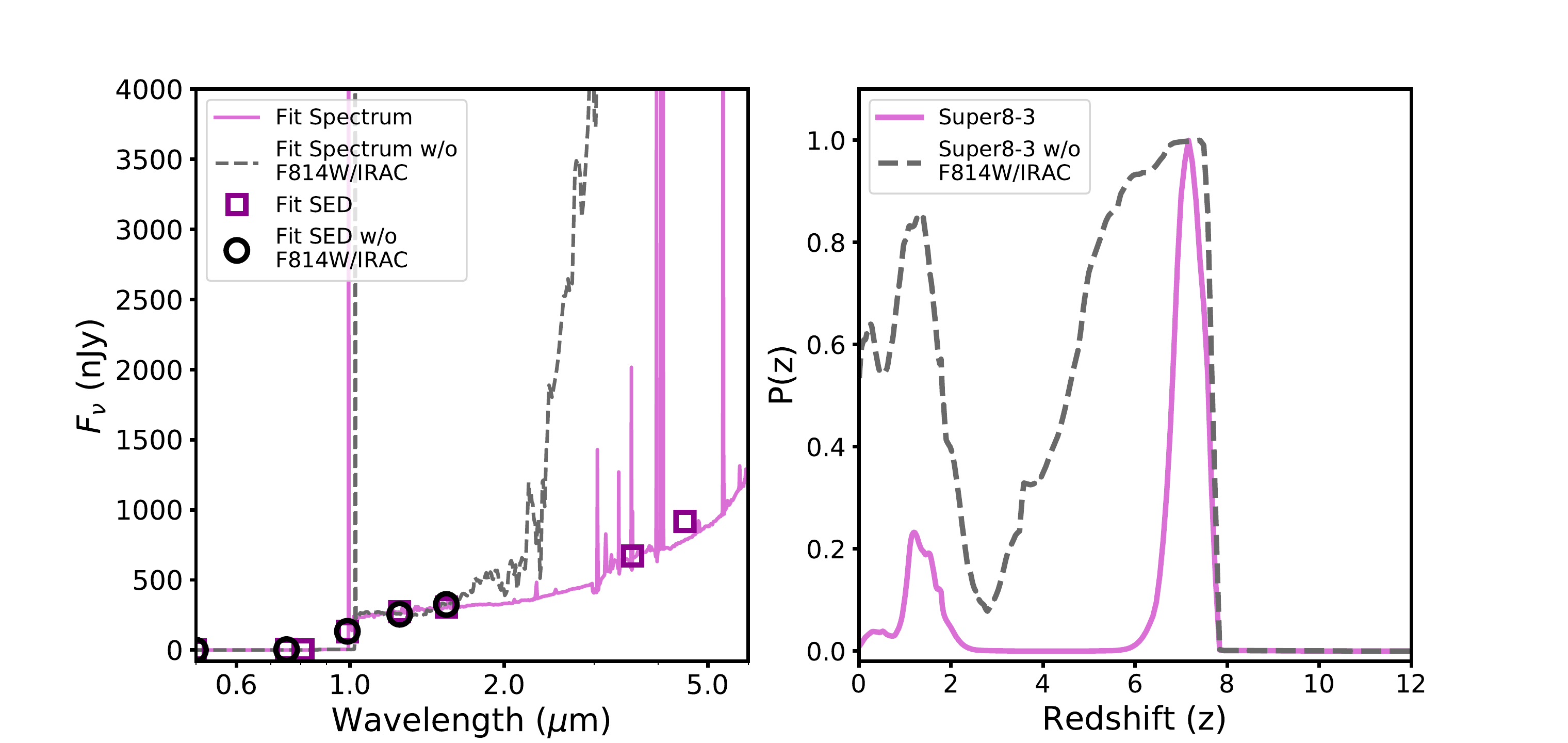}}
\end{subfigure}
\label{sed_compare}
\caption{\texttt{EAZY} photometric redshift fits for Super Eights 1-3, both with and without the \emph{HST} F814W and IRAC bands. The pink colors correspond to the fits using all the bands, while the cyan colors do not include the F814W and IRAC data.}
\end{figure*}
\clearpage

\begin{figure*}[!h]
\centering
\begin{subfigure}
	\centering
	\scalebox{0.45}
	{\includegraphics{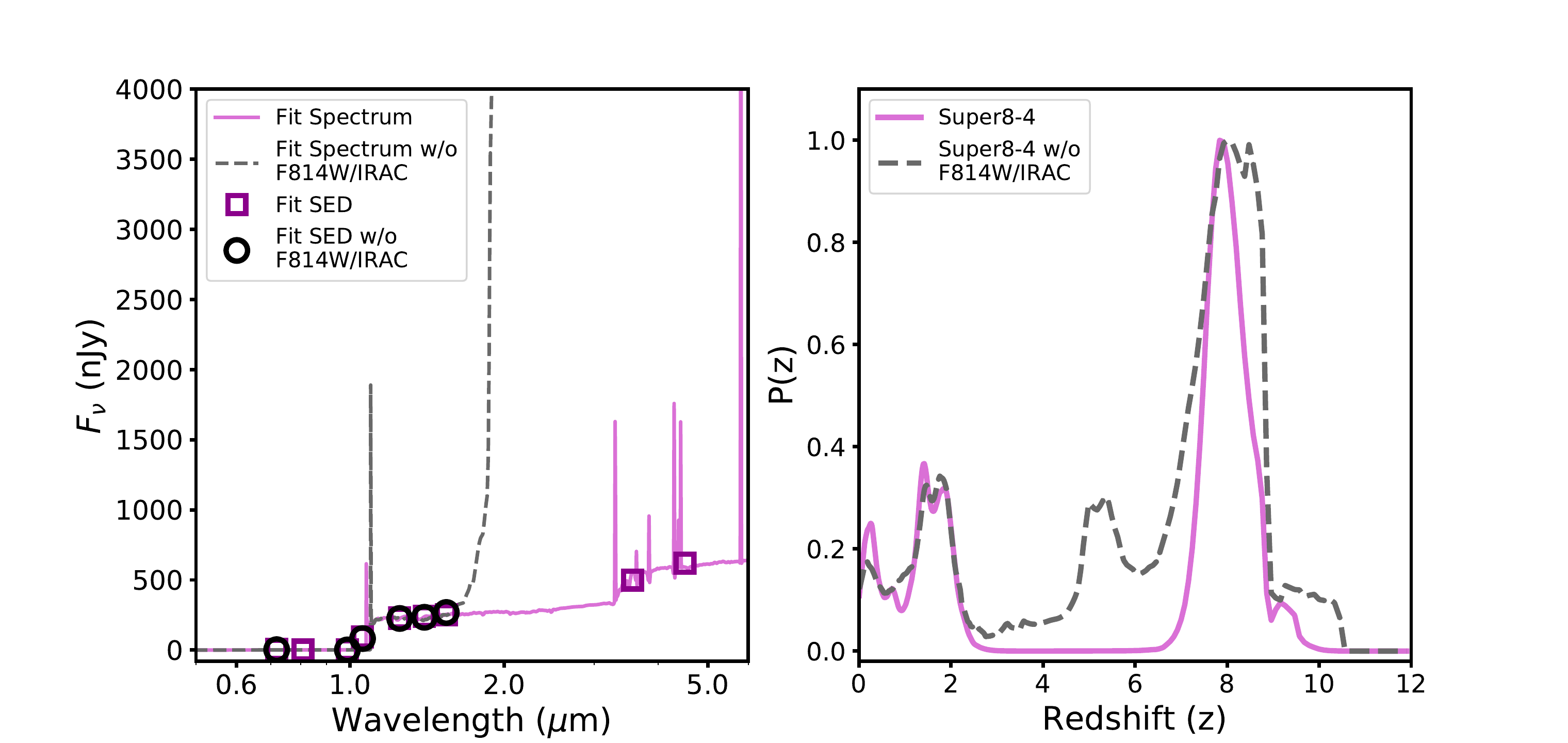}}
\end{subfigure}
\begin{subfigure}
	\centering
	\scalebox{0.45}
	{\includegraphics{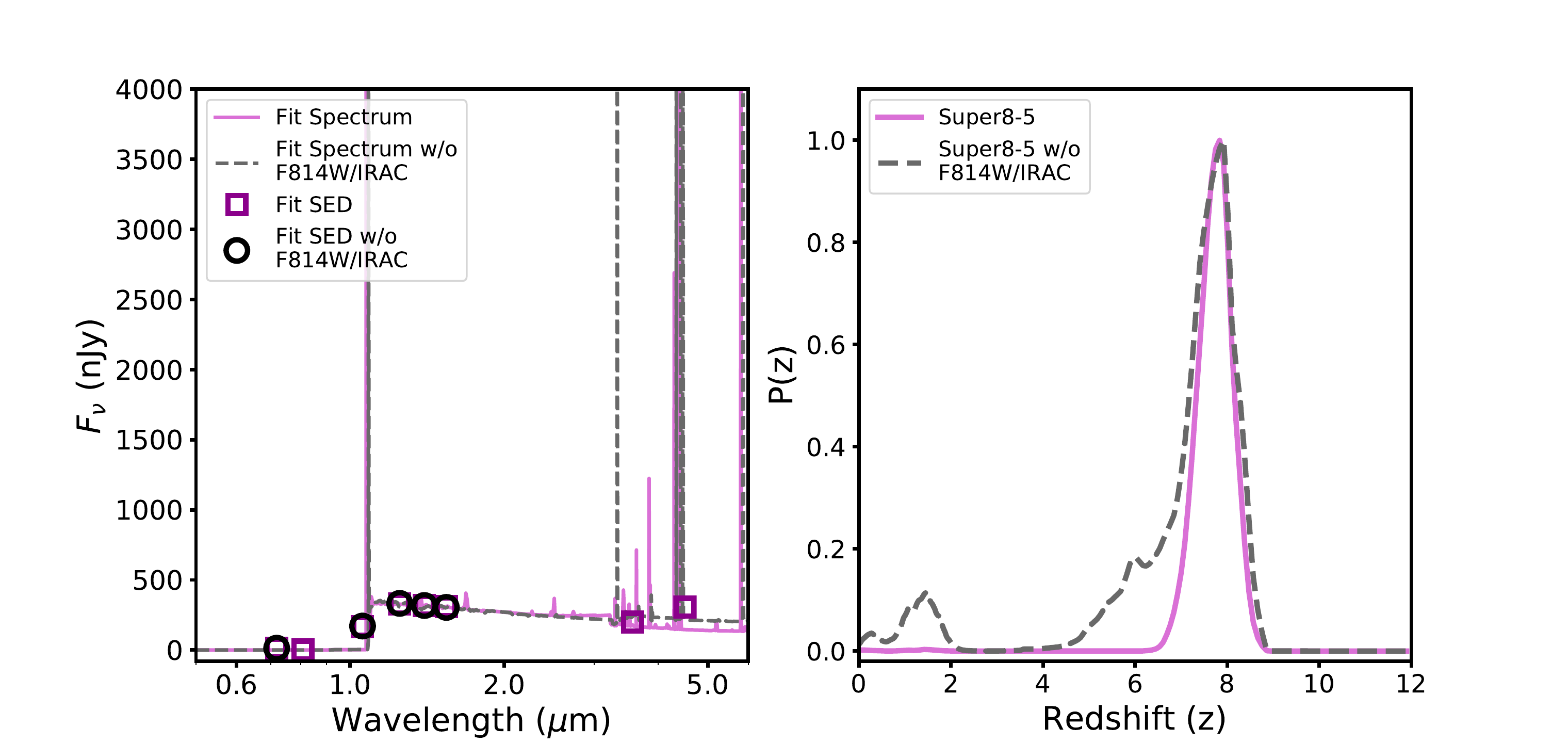}}
\end{subfigure}
\begin{subfigure}
	\centering
	\scalebox{0.45}
	{\includegraphics{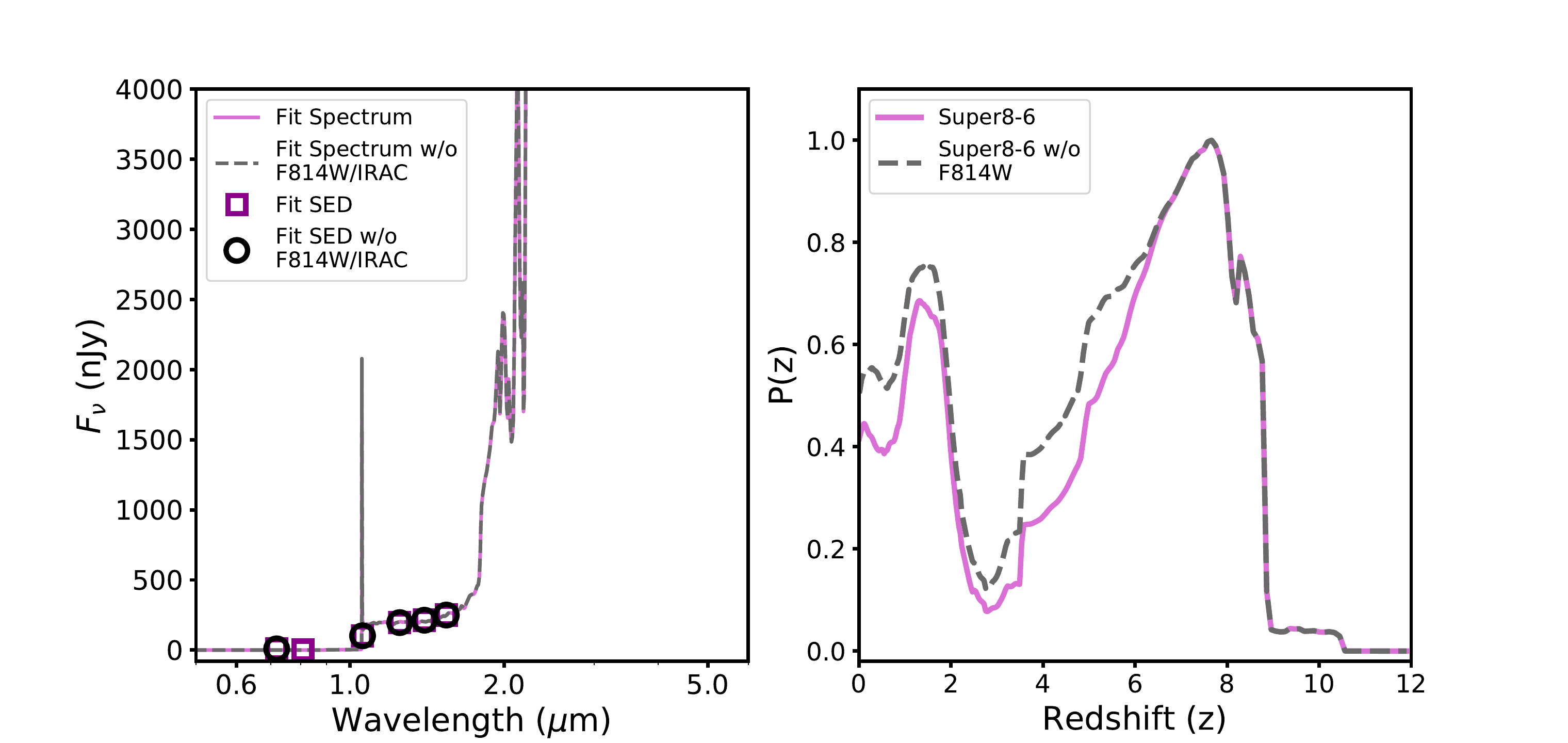}}
\end{subfigure}
\label{sed_compare1}
\caption{Same as Figure~15 but for Super Eights 4-6.}
\end{figure*}
\clearpage

\begin{figure*}[t!]
\centering
\begin{subfigure}
	\centering
	\scalebox{0.45}
	{\includegraphics{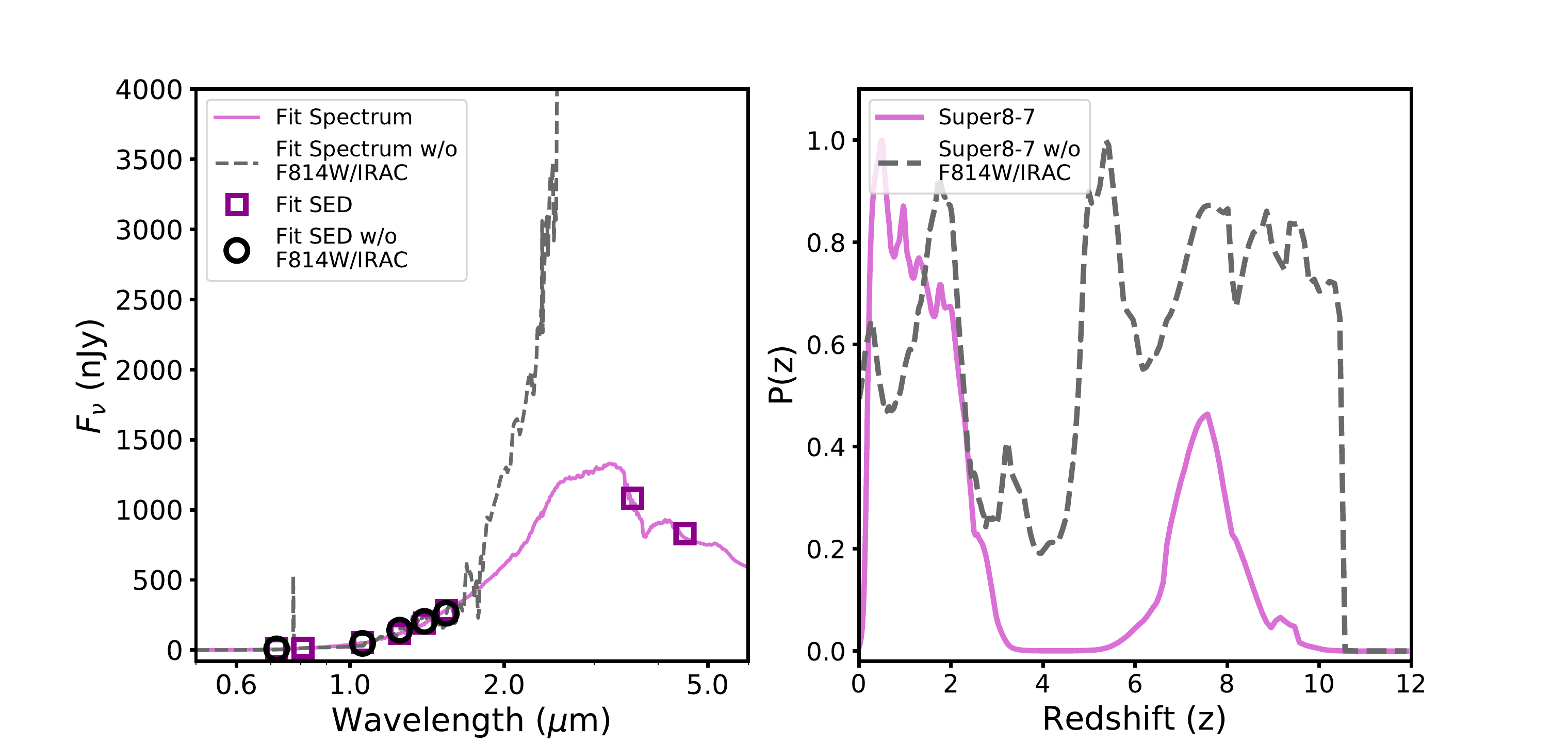}}
\end{subfigure}
\begin{subfigure}
    \centering
	\scalebox{0.45}
	{\includegraphics{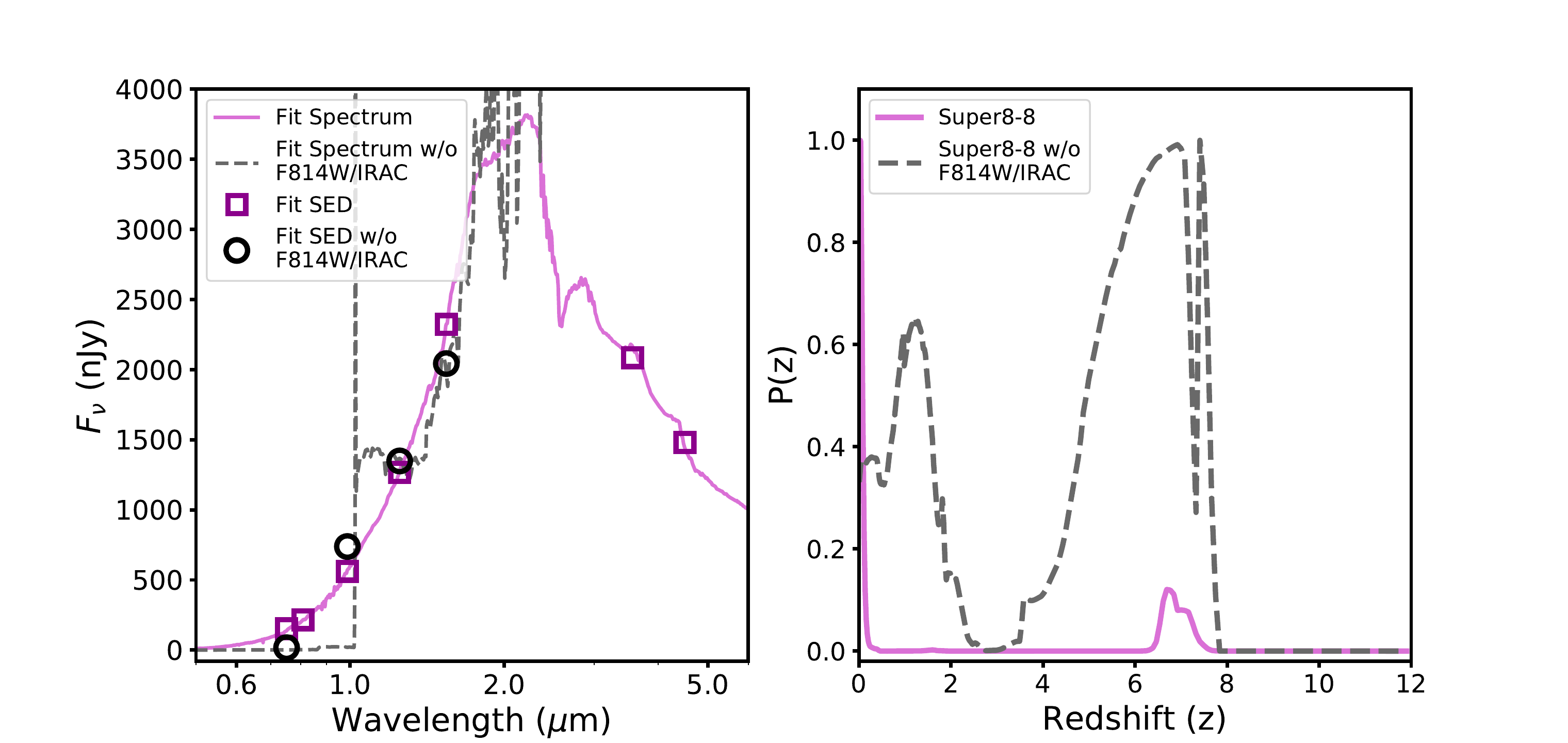}}
\end{subfigure}
\label{sed_compare2}
\caption{Same as Figure~15 but for Super Eight 7 and 8.}
\end{figure*}

\bibliography{SuperEights}

\begin{thebibliography}{}
\expandafter\ifx\csname natexlab\endcsname\relax\def\natexlab#1{#1}\fi
\providecommand{\url}[1]{\href{#1}{#1}}
\providecommand{\dodoi}[1]{doi:~\href{http://doi.org/#1}{\nolinkurl{#1}}}
\providecommand{\doeprint}[1]{\href{http://ascl.net/#1}{\nolinkurl{http://ascl.net/#1}}}
\providecommand{\doarXiv}[1]{\href{https://arxiv.org/abs/#1}{\nolinkurl{https://arxiv.org/abs/#1}}}

\bibitem[{{Astropy Collaboration} {et~al.}(2013){Astropy Collaboration},
  {Robitaille}, {Tollerud}, {Greenfield}, {Droettboom}, {Bray}, {Aldcroft},
  {Davis}, {Ginsburg}, {Price-Whelan}, {Kerzendorf}, {Conley}, {Crighton},
  {Barbary}, {Muna}, {Ferguson}, {Grollier}, {Parikh}, {Nair}, {Unther},
  {Deil}, {Woillez}, {Conseil}, {Kramer}, {Turner}, {Singer}, {Fox}, {Weaver},
  {Zabalza}, {Edwards}, {Azalee Bostroem}, {Burke}, {Casey}, {Crawford},
  {Dencheva}, {Ely}, {Jenness}, {Labrie}, {Lim}, {Pierfederici}, {Pontzen},
  {Ptak}, {Refsdal}, {Servillat}, \& {Streicher}}]{astropy2013}
{Astropy Collaboration}, {Robitaille}, T.~P., {Tollerud}, E.~J., {et~al.} 2013,
  \aap, 558, A33, \dodoi{10.1051/0004-6361/201322068}

\bibitem[{{Atek} {et~al.}(2015){Atek}, {Richard}, {Jauzac}, {Kneib},
  {Natarajan}, {Limousin}, {Schaerer}, {Jullo}, {Ebeling}, \&
  {Egami}}]{atek2015}
{Atek}, H., {Richard}, J., {Jauzac}, M., {et~al.} 2015, \apj, 814, 69,
  \dodoi{10.1088/0004-637X/814/1/69}

\bibitem[{{Ben{\'{\i}}tez}(2000)}]{benitez2000}
{Ben{\'{\i}}tez}, N. 2000, \apj, 536, 571, \dodoi{10.1086/308947}

\bibitem[{{Bernard} {et~al.}(2016){Bernard}, {Carrasco}, {Trenti}, {Oesch},
  {Wu}, {Bradley}, {Schmidt}, {Bouwens}, {Calvi}, {Mason}, {Stiavelli}, \&
  {Treu}}]{bernard2016}
{Bernard}, S.~R., {Carrasco}, D., {Trenti}, M., {et~al.} 2016, \apj, 827, 76,
  \dodoi{10.3847/0004-637X/827/1/76}

\bibitem[{{Bertin} \& {Arnouts}(1996)}]{bertin1996}
{Bertin}, E., \& {Arnouts}, S. 1996, \aaps, 117, 393,
  \dodoi{10.1051/aas:1996164}

\bibitem[{{Bouwens} {et~al.}(2004){Bouwens}, {Illingworth}, {Blakeslee},
  {Broadhurst}, \& {Franx}}]{bouwens2004}
{Bouwens}, R.~J., {Illingworth}, G.~D., {Blakeslee}, J.~P., {Broadhurst},
  T.~J., \& {Franx}, M. 2004, \apjl, 611, L1, \dodoi{10.1086/423786}

\bibitem[{{Bouwens} {et~al.}(2006){Bouwens}, {Illingworth}, {Blakeslee}, \&
  {Franx}}]{bouwens2006}
{Bouwens}, R.~J., {Illingworth}, G.~D., {Blakeslee}, J.~P., \& {Franx}, M.
  2006, \apj, 653, 53, \dodoi{10.1086/498733}

\bibitem[{{Bouwens} {et~al.}(2009){Bouwens}, {Illingworth}, {Franx}, {Chary},
  {Meurer}, {Conselice}, {Ford}, {Giavalisco}, \& {van Dokkum}}]{bouwens2009}
{Bouwens}, R.~J., {Illingworth}, G.~D., {Franx}, M., {et~al.} 2009, \apj, 705,
  936, \dodoi{10.1088/0004-637X/705/1/936}

\bibitem[{{Bouwens} {et~al.}(2011){Bouwens}, {Illingworth}, {Labbe}, {Oesch},
  {Trenti}, {Carollo}, {van Dokkum}, {Franx}, {Stiavelli}, {Gonz{\'a}lez},
  {Magee}, \& {Bradley}}]{bouwens2011}
{Bouwens}, R.~J., {Illingworth}, G.~D., {Labbe}, I., {et~al.} 2011, \nat, 469,
  504, \dodoi{10.1038/nature09717}

\bibitem[{{Bouwens} {et~al.}(2015){Bouwens}, {Illingworth}, {Oesch}, {Trenti},
  {Labb{\'e}}, {Bradley}, {Carollo}, {van Dokkum}, {Gonzalez}, {Holwerda},
  {Franx}, {Spitler}, {Smit}, \& {Magee}}]{bouwens2015}
{Bouwens}, R.~J., {Illingworth}, G.~D., {Oesch}, P.~A., {et~al.} 2015, \apj,
  803, 34, \dodoi{10.1088/0004-637X/803/1/34}

\bibitem[{{Bowler} {et~al.}(2017){Bowler}, {Dunlop}, {McLure}, \&
  {McLeod}}]{bowler2017}
{Bowler}, R.~A.~A., {Dunlop}, J.~S., {McLure}, R.~J., \& {McLeod}, D.~J. 2017,
  \mnras, 466, 3612, \dodoi{10.1093/mnras/stw3296}

\bibitem[{{Bowler} {et~al.}(2014){Bowler}, {Dunlop}, {McLure}, {Rogers},
  {McCracken}, {Milvang-Jensen}, {Furusawa}, {Fynbo}, {Taniguchi}, {Afonso},
  {Bremer}, \& {Le F{\`e}vre}}]{bowler2014}
{Bowler}, R.~A.~A., {Dunlop}, J.~S., {McLure}, R.~J., {et~al.} 2014, \mnras,
  440, 2810, \dodoi{10.1093/mnras/stu449}

\bibitem[{{Bowler} {et~al.}(2015){Bowler}, {Dunlop}, {McLure}, {McCracken},
  {Milvang-Jensen}, {Furusawa}, {Taniguchi}, {Le F{\`e}vre}, {Fynbo}, {Jarvis},
  \& {H{\"a}u{\ss}ler}}]{bowler2015}
---. 2015, \mnras, 452, 1817, \dodoi{10.1093/mnras/stv1403}

\bibitem[{{Bradley} {et~al.}(2012){Bradley}, {Trenti}, {Oesch}, {Stiavelli},
  {Treu}, {Bouwens}, {Shull}, {Holwerda}, \& {Pirzkal}}]{bradley2012}
{Bradley}, L.~D., {Trenti}, M., {Oesch}, P.~A., {et~al.} 2012, \apj, 760, 108,
  \dodoi{10.1088/0004-637X/760/2/108}

\bibitem[{{Bradley} {et~al.}(2014){Bradley}, {Zitrin}, {Coe}, {Bouwens},
  {Postman}, {Balestra}, {Grillo}, {Monna}, {Rosati}, {Seitz}, {Host}, {Lemze},
  {Moustakas}, {Moustakas}, {Shu}, {Zheng}, {Broadhurst}, {Carrasco}, {Jouvel},
  {Koekemoer}, {Medezinski}, {Meneghetti}, {Nonino}, {Smit}, {Umetsu},
  {Bartelmann}, {Ben{\'\i}tez}, {Donahue}, {Ford}, {Infante}, {Jimenez-Teja},
  {Kelson}, {Lahav}, {Maoz}, {Melchior}, {Merten}, \& {Molino}}]{bradley2014}
{Bradley}, L.~D., {Zitrin}, A., {Coe}, D., {et~al.} 2014, \apj, 792, 76,
  \dodoi{10.1088/0004-637X/792/1/76}

\bibitem[{{Brammer} {et~al.}(2008){Brammer}, {van Dokkum}, \&
  {Coppi}}]{brammer2008}
{Brammer}, G.~B., {van Dokkum}, P.~G., \& {Coppi}, P. 2008, \apj, 686, 1503,
  \dodoi{10.1086/591786}

\bibitem[{{Brammer} {et~al.}(2012){Brammer}, {van Dokkum}, {Franx},
  {Fumagalli}, {Patel}, {Rix}, {Skelton}, {Kriek}, {Nelson}, {Schmidt},
  {Bezanson}, {da Cunha}, {Erb}, {Fan}, {F{\"o}rster Schreiber}, {Illingworth},
  {Labb{\'e}}, {Leja}, {Lundgren}, {Magee}, {Marchesini}, {McCarthy},
  {Momcheva}, {Muzzin}, {Quadri}, {Steidel}, {Tal}, {Wake}, {Whitaker}, \&
  {Williams}}]{brammer2012}
{Brammer}, G.~B., {van Dokkum}, P.~G., {Franx}, M., {et~al.} 2012, \apjs, 200,
  13, \dodoi{10.1088/0067-0049/200/2/13}

\bibitem[{{Brinchmann} {et~al.}(2017){Brinchmann}, {Inami}, {Bacon}, {Contini},
  {Maseda}, {Chevallard}, {Bouch{\'e}}, {Boogaard}, {Carollo}, {Charlot},
  {Kollatschny}, {Marino}, {Pello}, {Richard}, {Schaye}, {Verhamme}, \&
  {Wisotzki}}]{brinchmann2017}
{Brinchmann}, J., {Inami}, H., {Bacon}, R., {et~al.} 2017, \aap, 608, A3,
  \dodoi{10.1051/0004-6361/201731351}

\bibitem[{{Bruzual} \& {Charlot}(2003)}]{bruzual2003}
{Bruzual}, G., \& {Charlot}, S. 2003, \mnras, 344, 1000,
  \dodoi{10.1046/j.1365-8711.2003.06897.x}

\bibitem[{{Calvi} {et~al.}(2016){Calvi}, {Trenti}, {Stiavelli}, {Oesch},
  {Bradley}, {Schmidt}, {Coe}, {Brammer}, {Bernard}, {Bouwens}, {Carrasco},
  {Carollo}, {Holwerda}, {MacKenty}, {Mason}, {Shull}, \& {Treu}}]{calvi2016}
{Calvi}, V., {Trenti}, M., {Stiavelli}, M., {et~al.} 2016, \apj, 817, 120,
  \dodoi{10.3847/0004-637X/817/2/120}

\bibitem[{{Calzetti} {et~al.}(2000){Calzetti}, {Armus}, {Bohlin}, {Kinney},
  {Koornneef}, \& {Storchi-Bergmann}}]{calzetti2000}
{Calzetti}, D., {Armus}, L., {Bohlin}, R.~C., {et~al.} 2000, \apj, 533, 682,
  \dodoi{10.1086/308692}

\bibitem[{{Caruana} {et~al.}(2012){Caruana}, {Bunker}, {Wilkins}, {Stanway},
  {Lacy}, {Jarvis}, {Lorenzoni}, \& {Hickey}}]{caruana2012}
{Caruana}, J., {Bunker}, A.~J., {Wilkins}, S.~M., {et~al.} 2012, \mnras, 427,
  3055, \dodoi{10.1111/j.1365-2966.2012.21996.x}

\bibitem[{{Caruana} {et~al.}(2014){Caruana}, {Bunker}, {Wilkins}, {Stanway},
  {Lorenzoni}, {Jarvis}, \& {Ebert}}]{caruana2014}
---. 2014, \mnras, 443, 2831, \dodoi{10.1093/mnras/stu1341}

\bibitem[{{Casertano} {et~al.}(2000){Casertano}, {de Mello}, {Dickinson},
  {Ferguson}, {Fruchter}, {Gonzalez-Lopezlira}, {Heyer}, {Hook}, {Levay},
  {Lucas}, {Mack}, {Makidon}, {Mutchler}, {Smith}, {Stiavelli}, {Wiggs}, \&
  {Williams}}]{casertano2000}
{Casertano}, S., {de Mello}, D., {Dickinson}, M., {et~al.} 2000, \aj, 120,
  2747, \dodoi{10.1086/316851}

\bibitem[{{Chabrier}(2003)}]{chabrier2003}
{Chabrier}, G. 2003, Publications of the Astronomical Society of the Pacific,
  115, 763, \dodoi{10.1086/376392}

\bibitem[{{Charlot} \& {Fall}(2000)}]{charlot2000}
{Charlot}, S., \& {Fall}, S.~M. 2000, \apj, 539, 718, \dodoi{10.1086/309250}

\bibitem[{{Coe} {et~al.}(2013){Coe}, {Zitrin}, {Carrasco}, {Shu}, {Zheng},
  {Postman}, {Bradley}, {Koekemoer}, {Bouwens}, {Broadhurst}, {Monna}, {Host},
  {Moustakas}, {Ford}, {Moustakas}, {van der Wel}, {Donahue}, {Rodney},
  {Ben{\'{\i}}tez}, {Jouvel}, {Seitz}, {Kelson}, \& {Rosati}}]{coe2013}
{Coe}, D., {Zitrin}, A., {Carrasco}, M., {et~al.} 2013, \apj, 762, 32,
  \dodoi{10.1088/0004-637X/762/1/32}

\bibitem[{{Conroy} {et~al.}(2009){Conroy}, {Gunn}, \& {White}}]{conroy2009}
{Conroy}, C., {Gunn}, J.~E., \& {White}, M. 2009, \apj, 699, 486,
  \dodoi{10.1088/0004-637X/699/1/486}

\bibitem[{{Croton} {et~al.}(2006){Croton}, {Springel}, {White}, {De Lucia},
  {Frenk}, {Gao}, {Jenkins}, {Kauffmann}, {Navarro}, \& {Yoshida}}]{croton2006}
{Croton}, D.~J., {Springel}, V., {White}, S. D.~M., {et~al.} 2006, \mnras, 365,
  11, \dodoi{10.1111/j.1365-2966.2005.09675.x}

\bibitem[{{Curtis-Lake} {et~al.}(2016){Curtis-Lake}, {McLure}, {Dunlop},
  {Rogers}, {Targett}, {Dekel}, {Ellis}, {Faber}, {Ferguson}, {Grogin},
  {Kocevski}, {Koekemoer}, {Lai}, {M{\'a}rmol-Queralt{\'o}}, \&
  {Robertson}}]{curtislake2016}
{Curtis-Lake}, E., {McLure}, R.~J., {Dunlop}, J.~S., {et~al.} 2016, \mnras,
  457, 440, \dodoi{10.1093/mnras/stv3017}

\bibitem[{{da Cunha} \& {Charlot}(2011)}]{dacunha2008sc}
{da Cunha}, E., \& {Charlot}, S. 2011, {MAGPHYS: Multi-wavelength Analysis of
  Galaxy Physical Properties}, Astrophysics Source Code Library.
\newblock \doeprint{1106.010}

\bibitem[{{da Cunha} {et~al.}(2008){da Cunha}, {Charlot}, \&
  {Elbaz}}]{dacunha2008}
{da Cunha}, E., {Charlot}, S., \& {Elbaz}, D. 2008, \mnras, 388, 1595,
  \dodoi{10.1111/j.1365-2966.2008.13535.x}

\bibitem[{{da Cunha} {et~al.}(2015){da Cunha}, {Walter}, {Smail}, {Swinbank},
  {Simpson}, {Decarli}, {Hodge}, {Weiss}, {van der Werf}, {Bertoldi},
  {Chapman}, {Cox}, {Danielson}, {Dannerbauer}, {Greve}, {Ivison}, {Karim}, \&
  {Thomson}}]{dacunha2015}
{da Cunha}, E., {Walter}, F., {Smail}, I.~R., {et~al.} 2015, \apj, 806, 110,
  \dodoi{10.1088/0004-637X/806/1/110}

\bibitem[{{Davis} {et~al.}(2007){Davis}, {Guhathakurta}, {Konidaris}, {Newman},
  {Ashby}, {Biggs}, {Barmby}, {Bundy}, {Chapman}, {Coil}, {Conselice},
  {Cooper}, {Croton}, {Eisenhardt}, {Ellis}, {Faber}, {Fang}, {Fazio},
  {Georgakakis}, {Gerke}, {Goss}, {Gwyn}, {Harker}, {Hopkins}, {Huang},
  {Ivison}, {Kassin}, {Kirby}, {Koekemoer}, {Koo}, {Laird}, {Le Floc'h}, {Lin},
  {Lotz}, {Marshall}, {Martin}, {Metevier}, {Moustakas}, {Nandra}, {Noeske},
  {Papovich}, {Phillips}, {Rich}, {Rieke}, {Rigopoulou}, {Salim},
  {Schiminovich}, {Simard}, {Smail}, {Small}, {Weiner}, {Willmer}, {Willner},
  {Wilson}, {Wright}, \& {Yan}}]{davis2007}
{Davis}, M., {Guhathakurta}, P., {Konidaris}, N.~P., {et~al.} 2007, \apjl, 660,
  L1, \dodoi{10.1086/517931}

\bibitem[{{de Barros} {et~al.}(2014){de Barros}, {Schaerer}, \&
  {Stark}}]{debarros2014}
{de Barros}, S., {Schaerer}, D., \& {Stark}, D.~P. 2014, \aap, 563, A81,
  \dodoi{10.1051/0004-6361/201220026}

\bibitem[{{Dijkstra} \& {Wyithe}(2012)}]{dijkstra2012}
{Dijkstra}, M., \& {Wyithe}, J.~S.~B. 2012, \mnras, 419, 3181,
  \dodoi{10.1111/j.1365-2966.2011.19958.x}

\bibitem[{{Eldridge} {et~al.}(2017){Eldridge}, {Stanway}, {Xiao}, {McClelland
  }, {Taylor}, {Ng}, {Greis}, \& {Bray}}]{eldridge2017}
{Eldridge}, J.~J., {Stanway}, E.~R., {Xiao}, L., {et~al.} 2017, Publications of
  the Astronomical Society of Australia, 34, e058, \dodoi{10.1017/pasa.2017.51}

\bibitem[{{Ellis} {et~al.}(2013){Ellis}, {McLure}, {Dunlop}, {Robertson},
  {Ono}, {Schenker}, {Koekemoer}, {Bowler}, {Ouchi}, {Rogers}, {Curtis-Lake},
  {Schneider}, {Charlot}, {Stark}, {Furlanetto}, \& {Cirasuolo}}]{ellis2013}
{Ellis}, R.~S., {McLure}, R.~J., {Dunlop}, J.~S., {et~al.} 2013, \apjl, 763,
  L7, \dodoi{10.1088/2041-8205/763/1/L7}

\bibitem[{{Erb} {et~al.}(2010){Erb}, {Pettini}, {Shapley}, {Steidel}, {Law}, \&
  {Reddy}}]{erb2010}
{Erb}, D.~K., {Pettini}, M., {Shapley}, A.~E., {et~al.} 2010, \apj, 719, 1168,
  \dodoi{10.1088/0004-637X/719/2/1168}

\bibitem[{{Fall} \& {Efstathiou}(1980)}]{fall1980}
{Fall}, S.~M., \& {Efstathiou}, G. 1980, \mnras, 193, 189,
  \dodoi{10.1093/mnras/193.2.189}

\bibitem[{{Fan} {et~al.}(2006){Fan}, {Strauss}, {Becker}, {White}, {Gunn},
  {Knapp}, {Richards}, {Schneider}, {Brinkmann}, \& {Fukugita}}]{fan2006}
{Fan}, X., {Strauss}, M.~A., {Becker}, R.~H., {et~al.} 2006, \aj, 132, 117,
  \dodoi{10.1086/504836}

\bibitem[{{Ferguson} {et~al.}(2004){Ferguson}, {Dickinson}, {Giavalisco},
  {Kretchmer}, {Ravindranath}, {Idzi}, {Taylor}, {Conselice}, {Fall},
  {Gardner}, {Livio}, {Madau}, {Moustakas}, {Papovich}, {Somerville},
  {Spinrad}, \& {Stern}}]{ferguson2004}
{Ferguson}, H.~C., {Dickinson}, M., {Giavalisco}, M., {et~al.} 2004, \apjl,
  600, L107, \dodoi{10.1086/378578}

\bibitem[{{Finkelstein} {et~al.}(2013){Finkelstein}, {Papovich}, {Dickinson},
  {Song}, {Tilvi}, {Koekemoer}, {Finkelstein}, {Mobasher}, {Ferguson},
  {Giavalisco}, {Reddy}, {Ashby}, {Dekel}, {Fazio}, {Fontana}, {Grogin},
  {Huang}, {Kocevski}, {Rafelski}, {Weiner}, \& {Willner}}]{finkelstein2013}
{Finkelstein}, S.~L., {Papovich}, C., {Dickinson}, M., {et~al.} 2013, \nat,
  502, 524, \dodoi{10.1038/nature12657}

\bibitem[{{Finkelstein} {et~al.}(2015){Finkelstein}, {Ryan}, {Papovich},
  {Dickinson}, {Song}, {Somerville}, {Ferguson}, {Salmon}, {Giavalisco},
  {Koekemoer}, {Ashby}, {Behroozi}, {Castellano}, {Dunlop}, {Faber}, {Fazio},
  {Fontana}, {Grogin}, {Hathi}, {Jaacks}, {Kocevski}, {Livermore}, {McLure},
  {Merlin}, {Mobasher}, {Newman}, {Rafelski}, {Tilvi}, \&
  {Willner}}]{finkelstein2015}
{Finkelstein}, S.~L., {Ryan}, Jr., R.~E., {Papovich}, C., {et~al.} 2015, \apj,
  810, 71, \dodoi{10.1088/0004-637X/810/1/71}

\bibitem[{{Fontana} {et~al.}(2010){Fontana}, {Vanzella}, {Pentericci},
  {Castellano}, {Giavalisco}, {Grazian}, {Boutsia}, {Cristiani}, {Dickinson},
  {Giallongo}, {Maiolino}, {Moorwood}, \& {Santini}}]{fontana2010}
{Fontana}, A., {Vanzella}, E., {Pentericci}, L., {et~al.} 2010, \apjl, 725,
  L205, \dodoi{10.1088/2041-8205/725/2/L205}

\bibitem[{{Gonzaga} {et~al.}(2012){Gonzaga}, {Hack}, {Fruchter}, \&
  {Mack}}]{gonzaga2012}
{Gonzaga}, S., {Hack}, W., {Fruchter}, A., \& {Mack}, J. 2012, {The DrizzlePac
  Handbook}

\bibitem[{{Granato} {et~al.}(2004){Granato}, {De Zotti}, {Silva}, {Bressan}, \&
  {Danese}}]{granato2004}
{Granato}, G.~L., {De Zotti}, G., {Silva}, L., {Bressan}, A., \& {Danese}, L.
  2004, \apj, 600, 580, \dodoi{10.1086/379875}

\bibitem[{{Grazian} {et~al.}(2012){Grazian}, {Castellano}, {Fontana},
  {Pentericci}, {Dunlop}, {McLure}, {Koekemoer}, {Dickinson}, {Faber},
  {Ferguson}, {Galametz}, {Giavalisco}, {Grogin}, {Hathi}, {Kocevski}, {Lai},
  {Newman}, \& {Vanzella}}]{grazian2012}
{Grazian}, A., {Castellano}, M., {Fontana}, A., {et~al.} 2012, \aap, 547, A51,
  \dodoi{10.1051/0004-6361/201219669}

\bibitem[{{Grogin} {et~al.}(2011){Grogin}, {Kocevski}, {Faber}, {Ferguson},
  {Koekemoer}, {Riess}, {Acquaviva}, {Alexander}, {Almaini}, {Ashby}, {Barden},
  {Bell}, {Bournaud}, {Brown}, {Caputi}, {Casertano}, {Cassata}, {Castellano},
  {Challis}, {Chary}, {Cheung}, {Cirasuolo}, {Conselice}, {Roshan Cooray},
  {Croton}, {Daddi}, {Dahlen}, {Dav{\'e}}, {de Mello}, {Dekel}, {Dickinson},
  {Dolch}, {Donley}, {Dunlop}, {Dutton}, {Elbaz}, {Fazio}, {Filippenko},
  {Finkelstein}, {Fontana}, {Gardner}, {Garnavich}, {Gawiser}, {Giavalisco},
  {Grazian}, {Guo}, {Hathi}, {H{\"a}ussler}, {Hopkins}, {Huang}, {Huang},
  {Jha}, {Kartaltepe}, {Kirshner}, {Koo}, {Lai}, {Lee}, {Li}, {Lotz}, {Lucas},
  {Madau}, {McCarthy}, {McGrath}, {McIntosh}, {McLure}, {Mobasher},
  {Moustakas}, {Mozena}, {Nandra}, {Newman}, {Niemi}, {Noeske}, {Papovich},
  {Pentericci}, {Pope}, {Primack}, {Rajan}, {Ravindranath}, {Reddy}, {Renzini},
  {Rix}, {Robaina}, {Rodney}, {Rosario}, {Rosati}, {Salimbeni}, {Scarlata},
  {Siana}, {Simard}, {Smidt}, {Somerville}, {Spinrad}, {Straughn}, {Strolger},
  {Telford}, {Teplitz}, {Trump}, {van der Wel}, {Villforth}, {Wechsler},
  {Weiner}, {Wiklind}, {Wild}, {Wilson}, {Wuyts}, {Yan}, \& {Yun}}]{grogin2011}
{Grogin}, N.~A., {Kocevski}, D.~D., {Faber}, S.~M., {et~al.} 2011, \apjs, 197,
  35, \dodoi{10.1088/0067-0049/197/2/35}

\bibitem[{Hashimoto {et~al.}(2018)Hashimoto, Laporte, Mawatari, Ellis, Inoue,
  Zackrisson, Roberts-Borsani, Zheng, Tamura, Bauer, Fletcher, Harikane,
  Hatsukade, Hayatsu, Matsuda, Matsuo, Okamoto, Ouchi, Pell{\'o}, Rydberg,
  Shimizu, Taniguchi, Umehata, \& Yoshida}]{hashimoto2018}
Hashimoto, T., Laporte, N., Mawatari, K., {et~al.} 2018, Nature, 557, 392,
  \dodoi{10.1038/s41586-018-0117-z}

\bibitem[{{Hathi} {et~al.}(2008){Hathi}, {Jansen}, {Windhorst}, {Cohen},
  {Keel}, {Corbin}, \& {Ryan}}]{hathi2008}
{Hathi}, N.~P., {Jansen}, R.~A., {Windhorst}, R.~A., {et~al.} 2008, \aj, 135,
  156, \dodoi{10.1088/0004-6256/135/1/156}

\bibitem[{{Hathi} {et~al.}(2012){Hathi}, {Mobasher}, {Capak}, {Wang}, \&
  {Ferguson}}]{hathi2012}
{Hathi}, N.~P., {Mobasher}, B., {Capak}, P., {Wang}, W.-H., \& {Ferguson},
  H.~C. 2012, \apj, 757, 43, \dodoi{10.1088/0004-637X/757/1/43}

\bibitem[{{Hoag} {et~al.}(2017){Hoag}, {Brada{\v{c}}}, {Trenti}, {Treu},
  {Schmidt}, {Huang}, {Lemaux}, {He}, {Bernard}, {Abramson}, {Mason},
  {Morishita}, {Pentericci}, \& {Schrabback}}]{hoag2017}
{Hoag}, A., {Brada{\v{c}}}, M., {Trenti}, M., {et~al.} 2017, Nature Astronomy,
  1, 0091, \dodoi{10.1038/s41550-017-0091}

\bibitem[{{Holwerda}(2005)}]{holwerda2005}
{Holwerda}, B.~W. 2005, arXiv e-prints, astro.
\newblock \doarXiv{astro-ph/0512139}

\bibitem[{{Holwerda} {et~al.}(2015){Holwerda}, {Bouwens}, {Oesch}, {Smit},
  {Illingworth}, \& {Labbe}}]{holwerda2015}
{Holwerda}, B.~W., {Bouwens}, R., {Oesch}, P., {et~al.} 2015, \apj, 808, 6,
  \dodoi{10.1088/0004-637X/808/1/6}

\bibitem[{{Holwerda} {et~al.}(2014){Holwerda}, {Mu{\~n}oz-Mateos},
  {Comer{\'o}n}, {Meidt}, {Sheth}, {Laine}, {Hinz}, {Regan}, {Gil de Paz},
  {Men{\'e}ndez-Delmestre}, {Seibert}, {Kim}, {Mizusawa}, {Laurikainen},
  {Salo}, {Laine}, {Gadotti}, {Zaritsky}, {Erroz-Ferrer}, {Ho}, {Knapen},
  {Athanassoula}, {Bosma}, \& {Pirzkal}}]{holwerda2014}
{Holwerda}, B.~W., {Mu{\~n}oz-Mateos}, J.-C., {Comer{\'o}n}, S., {et~al.} 2014,
  \apj, 781, 12, \dodoi{10.1088/0004-637X/781/1/12}

\bibitem[{{Huang} {et~al.}(2013){Huang}, {Ferguson}, {Ravindranath}, \&
  {Su}}]{huang2013}
{Huang}, K.-H., {Ferguson}, H.~C., {Ravindranath}, S., \& {Su}, J. 2013, \apj,
  765, 68, \dodoi{10.1088/0004-637X/765/1/68}

\bibitem[{{Huang} {et~al.}(2016){Huang}, {Lemaux}, {Schmidt}, {Hoag}, {Brada{\v
  c}}, {Treu}, {Dijkstra}, {Fontana}, {Henry}, {Malkan}, {Mason}, {Morishita},
  {Pentericci}, {Ryan}, {Trenti}, \& {Wang}}]{huang2016}
{Huang}, K.-H., {Lemaux}, B.~C., {Schmidt}, K.~B., {et~al.} 2016, \apjl, 823,
  L14, \dodoi{10.3847/2041-8205/823/1/L14}

\bibitem[{{Hunter}(2007)}]{hunter2007}
{Hunter}, J.~D. 2007, Computing in Science and Engineering, 9, 90,
  \dodoi{10.1109/MCSE.2007.55}

\bibitem[{{Ilbert} {et~al.}(2009){Ilbert}, {Capak}, {Salvato}, {Aussel},
  {McCracken}, {Sanders}, {Scoville}, {Kartaltepe}, {Arnouts}, {Le Floc'h},
  {Mobasher}, {Taniguchi}, {Lamareille}, {Leauthaud}, {Sasaki}, {Thompson},
  {Zamojski}, {Zamorani}, {Bardelli}, {Bolzonella}, {Bongiorno}, {Brusa},
  {Caputi}, {Carollo}, {Contini}, {Cook}, {Coppa}, {Cucciati}, {de la Torre},
  {de Ravel}, {Franzetti}, {Garilli}, {Hasinger}, {Iovino}, {Kampczyk},
  {Kneib}, {Knobel}, {Kovac}, {Le Borgne}, {Le Brun}, {Le F{\`e}vre}, {Lilly},
  {Looper}, {Maier}, {Mainieri}, {Mellier}, {Mignoli}, {Murayama}, {Pell{\`o}},
  {Peng}, {P{\'e}rez-Montero}, {Renzini}, {Ricciardelli}, {Schiminovich},
  {Scodeggio}, {Shioya}, {Silverman}, {Surace}, {Tanaka}, {Tasca}, {Tresse},
  {Vergani}, \& {Zucca}}]{ilbert2009}
{Ilbert}, O., {Capak}, P., {Salvato}, M., {et~al.} 2009, \apj, 690, 1236,
  \dodoi{10.1088/0004-637X/690/2/1236}

\bibitem[{{Ishigaki} {et~al.}(2018){Ishigaki}, {Kawamata}, {Ouchi}, {Oguri},
  {Shimasaku}, \& {Ono}}]{ishigaki2018}
{Ishigaki}, M., {Kawamata}, R., {Ouchi}, M., {et~al.} 2018, \apj, 854, 73,
  \dodoi{10.3847/1538-4357/aaa544}

\bibitem[{{Jaacks} {et~al.}(2012){Jaacks}, {Choi}, {Nagamine}, {Thompson}, \&
  {Varghese}}]{jaacks2012}
{Jaacks}, J., {Choi}, J.-H., {Nagamine}, K., {Thompson}, R., \& {Varghese}, S.
  2012, \mnras, 420, 1606, \dodoi{10.1111/j.1365-2966.2011.20150.x}

\bibitem[{{Jiang} {et~al.}(2013){Jiang}, {Egami}, {Mechtley}, {Fan}, {Cohen},
  {Windhorst}, {Dav{\'e}}, {Finlator}, {Kashikawa}, {Ouchi}, \&
  {Shimasaku}}]{jiang2013}
{Jiang}, L., {Egami}, E., {Mechtley}, M., {et~al.} 2013, \apj, 772, 99,
  \dodoi{10.1088/0004-637X/772/2/99}

\bibitem[{{Jones} {et~al.}(2001){Jones}, {Oliphant}, {Peterson}, \&
  Others}]{jones2001}
{Jones}, E., {Oliphant}, T., {Peterson}, P., \& Others. 2001, {SciPy}: Open
  source scientific tools for Python.
\newblock \url{http://www.scipy.org/}

\bibitem[{{Kawamata} {et~al.}(2015){Kawamata}, {Ishigaki}, {Shimasaku},
  {Oguri}, \& {Ouchi}}]{kawamata2015}
{Kawamata}, R., {Ishigaki}, M., {Shimasaku}, K., {Oguri}, M., \& {Ouchi}, M.
  2015, \apj, 804, 103, \dodoi{10.1088/0004-637X/804/2/103}

\bibitem[{{Kriek} {et~al.}(2009){Kriek}, {van Dokkum}, {Labb{\'e}}, {Franx},
  {Illingworth}, {Marchesini}, \& {Quadri}}]{kriek2009}
{Kriek}, M., {van Dokkum}, P.~G., {Labb{\'e}}, I., {et~al.} 2009, \apj, 700,
  221, \dodoi{10.1088/0004-637X/700/1/221}

\bibitem[{{Labb{\'e}} {et~al.}(2006){Labb{\'e}}, {Bouwens}, {Illingworth}, \&
  {Franx}}]{labbe2006}
{Labb{\'e}}, I., {Bouwens}, R., {Illingworth}, G.~D., \& {Franx}, M. 2006,
  \apjl, 649, L67, \dodoi{10.1086/508512}

\bibitem[{{Labb{\'e}} {et~al.}(2005){Labb{\'e}}, {Huang}, {Franx}, {Rudnick},
  {Barmby}, {Daddi}, {van Dokkum}, {Fazio}, {Schreiber}, {Moorwood}, {Rix},
  {R{\"o}ttgering}, {Trujillo}, \& {van der Werf}}]{labbe2005}
{Labb{\'e}}, I., {Huang}, J., {Franx}, M., {et~al.} 2005, \apjl, 624, L81,
  \dodoi{10.1086/430700}

\bibitem[{{Labb{\'e}} {et~al.}(2010{\natexlab{a}}){Labb{\'e}}, {Gonz{\'a}lez},
  {Bouwens}, {Illingworth}, {Oesch}, {van Dokkum}, {Carollo}, {Franx},
  {Stiavelli}, {Trenti}, {Magee}, \& {Kriek}}]{labbe2010a}
{Labb{\'e}}, I., {Gonz{\'a}lez}, V., {Bouwens}, R.~J., {et~al.}
  2010{\natexlab{a}}, \apjl, 708, L26, \dodoi{10.1088/2041-8205/708/1/L26}

\bibitem[{{Labb{\'e}} {et~al.}(2010{\natexlab{b}}){Labb{\'e}}, {Gonz{\'a}lez},
  {Bouwens}, {Illingworth}, {Franx}, {Trenti}, {Oesch}, {van Dokkum},
  {Stiavelli}, {Carollo}, {Kriek}, \& {Magee}}]{labbe2010b}
---. 2010{\natexlab{b}}, \apjl, 716, L103, \dodoi{10.1088/2041-8205/716/2/L103}

\bibitem[{{Labb{\'e}} {et~al.}(2013){Labb{\'e}}, {Oesch}, {Bouwens},
  {Illingworth}, {Magee}, {Gonz{\'a}lez}, {Carollo}, {Franx}, {Trenti}, {van
  Dokkum}, \& {Stiavelli}}]{labbe2013}
{Labb{\'e}}, I., {Oesch}, P.~A., {Bouwens}, R.~J., {et~al.} 2013, \apjl, 777,
  L19, \dodoi{10.1088/2041-8205/777/2/L19}

\bibitem[{{Labb{\'e}} {et~al.}(2015){Labb{\'e}}, {Oesch}, {Illingworth}, {van
  Dokkum}, {Bouwens}, {Franx}, {Carollo}, {Trenti}, {Holden}, {Smit},
  {Gonz{\'a}lez}, {Magee}, {Stiavelli}, \& {Stefanon}}]{labbe2015}
{Labb{\'e}}, I., {Oesch}, P.~A., {Illingworth}, G.~D., {et~al.} 2015, \apjs,
  221, 23, \dodoi{10.1088/0067-0049/221/2/23}

\bibitem[{{Laporte} {et~al.}(2014){Laporte}, {Streblyanska}, {Clement},
  {P{\'e}rez-Fournon}, {Schaerer}, {Atek}, {Boone}, {Kneib}, {Egami},
  {Mart{\'{\i}}nez-Navajas}, {Marques-Chaves}, {Pell{\'o}}, \&
  {Richard}}]{laporte2014}
{Laporte}, N., {Streblyanska}, A., {Clement}, B., {et~al.} 2014, \aap, 562, L8,
  \dodoi{10.1051/0004-6361/201323179}

\bibitem[{{Larson} {et~al.}(2018){Larson}, {Finkelstein}, {Pirzkal}, {Ryan},
  {Tilvi}, {Malhotra}, {Rhoads}, {Finkelstein}, {Jung}, {Christensen},
  {Cimatti}, {Ferreras}, {Grogin}, {Koekemoer}, {Hathi}, {O'Connell},
  {{\"O}stlin}, {Pasquali}, {Pharo}, {Rothberg}, {Windhorst}, \& {The FIGS
  Team}}]{larson2018}
{Larson}, R.~L., {Finkelstein}, S.~L., {Pirzkal}, N., {et~al.} 2018, \apj, 858,
  94, \dodoi{10.3847/1538-4357/aab893}

\bibitem[{{Livermore} {et~al.}(2017){Livermore}, {Finkelstein}, \&
  {Lotz}}]{livermore2017}
{Livermore}, R.~C., {Finkelstein}, S.~L., \& {Lotz}, J.~M. 2017, \apj, 835,
  113, \dodoi{10.3847/1538-4357/835/2/113}

\bibitem[{{Livermore} {et~al.}(2018){Livermore}, {Trenti}, {Bradley},
  {Bernard}, {Holwerda}, {Mason}, \& {Treu}}]{livermore2018}
{Livermore}, R.~C., {Trenti}, M., {Bradley}, L.~D., {et~al.} 2018, \apjl, 861,
  L17, \dodoi{10.3847/2041-8213/aacd16}

\bibitem[{{Lorenzoni} {et~al.}(2011){Lorenzoni}, {Bunker}, {Wilkins},
  {Stanway}, {Jarvis}, \& {Caruana}}]{lorenzoni2011}
{Lorenzoni}, S., {Bunker}, A.~J., {Wilkins}, S.~M., {et~al.} 2011, \mnras, 414,
  1455, \dodoi{10.1111/j.1365-2966.2011.18479.x}

\bibitem[{{Malhotra} \& {Rhoads}(2004)}]{malhotra2004}
{Malhotra}, S., \& {Rhoads}, J.~E. 2004, \apjl, 617, L5, \dodoi{10.1086/427182}

\bibitem[{{Maraston}(2005)}]{maraston2005}
{Maraston}, C. 2005, \mnras, 362, 799, \dodoi{10.1111/j.1365-2966.2005.09270.x}

\bibitem[{{Mason} {et~al.}(2015){Mason}, {Trenti}, \& {Treu}}]{mason2015}
{Mason}, C.~A., {Trenti}, M., \& {Treu}, T. 2015, \apj, 813, 21,
  \dodoi{10.1088/0004-637X/813/1/21}

\bibitem[{{Mason} {et~al.}(2018){Mason}, {Treu}, {de Barros}, {Dijkstra},
  {Fontana}, {Mesinger}, {Pentericci}, {Trenti}, \& {Vanzella}}]{mason2018}
{Mason}, C.~A., {Treu}, T., {de Barros}, S., {et~al.} 2018, \apjl, 857, L11,
  \dodoi{10.3847/2041-8213/aabbab}

\bibitem[{{McLure} {et~al.}(2013){McLure}, {Dunlop}, {Bowler}, {Curtis-Lake},
  {Schenker}, {Ellis}, {Robertson}, {Koekemoer}, {Rogers}, {Ono}, {Ouchi},
  {Charlot}, {Wild}, {Stark}, {Furlanetto}, {Cirasuolo}, \&
  {Targett}}]{mclure2013}
{McLure}, R.~J., {Dunlop}, J.~S., {Bowler}, R.~A.~A., {et~al.} 2013, \mnras,
  432, 2696, \dodoi{10.1093/mnras/stt627}

\bibitem[{{McQuinn} {et~al.}(2007){McQuinn}, {Hernquist}, {Zaldarriaga}, \&
  {Dutta}}]{mcquinn2007}
{McQuinn}, M., {Hernquist}, L., {Zaldarriaga}, M., \& {Dutta}, S. 2007, \mnras,
  381, 75, \dodoi{10.1111/j.1365-2966.2007.12085.x}

\bibitem[{{Mesinger} \& {Furlanetto}(2008)}]{mesinger2008}
{Mesinger}, A., \& {Furlanetto}, S.~R. 2008, \mnras, 386, 1990,
  \dodoi{10.1111/j.1365-2966.2008.13039.x}

\bibitem[{{Mo} {et~al.}(1998){Mo}, {Mao}, \& {White}}]{mo1998}
{Mo}, H.~J., {Mao}, S., \& {White}, S.~D.~M. 1998, \mnras, 295, 319,
  \dodoi{10.1046/j.1365-8711.1998.01227.x}

\bibitem[{{Morishita} {et~al.}(2018){Morishita}, {Trenti}, {Stiavelli},
  {Bradley}, {Coe}, {Oesch}, {Mason}, {Bridge}, {Holwerda}, \&
  {Livermore}}]{morishita2018}
{Morishita}, T., {Trenti}, M., {Stiavelli}, M., {et~al.} 2018, \apj, 867, 150,
  \dodoi{10.3847/1538-4357/aae68c}

\bibitem[{{Oesch} {et~al.}(2007){Oesch}, {Stiavelli}, {Carollo}, {Bergeron},
  {Koekemoer}, {Lucas}, {Pavlovsky}, {Trenti}, {Lilly}, {Beckwith}, {Dahlen},
  {Ferguson}, {Gardner}, {Lacey}, {Mobasher}, {Panagia}, \& {Rix}}]{oesch2007}
{Oesch}, P.~A., {Stiavelli}, M., {Carollo}, C.~M., {et~al.} 2007, \apj, 671,
  1212, \dodoi{10.1086/522423}

\bibitem[{{Oesch} {et~al.}(2009){Oesch}, {Carollo}, {Stiavelli}, {Trenti},
  {Bergeron}, {Koekemoer}, {Lucas}, {Pavlovsky}, {Beckwith}, {Dahlen},
  {Ferguson}, {Gardner}, {Lilly}, {Mobasher}, \& {Panagia}}]{oesch2009}
{Oesch}, P.~A., {Carollo}, C.~M., {Stiavelli}, M., {et~al.} 2009, \apj, 690,
  1350, \dodoi{10.1088/0004-637X/690/2/1350}

\bibitem[{{Oesch} {et~al.}(2010){Oesch}, {Bouwens}, {Carollo}, {Illingworth},
  {Trenti}, {Stiavelli}, {Magee}, {Labb{\'e}}, \& {Franx}}]{oesch2010}
{Oesch}, P.~A., {Bouwens}, R.~J., {Carollo}, C.~M., {et~al.} 2010, \apjl, 709,
  L21, \dodoi{10.1088/2041-8205/709/1/L21}

\bibitem[{{Oesch} {et~al.}(2013){Oesch}, {Bouwens}, {Illingworth}, {Labb{\'e}},
  {Franx}, {van Dokkum}, {Trenti}, {Stiavelli}, {Gonzalez}, \&
  {Magee}}]{oesch2013}
{Oesch}, P.~A., {Bouwens}, R.~J., {Illingworth}, G.~D., {et~al.} 2013, \apj,
  773, 75, \dodoi{10.1088/0004-637X/773/1/75}

\bibitem[{{Oesch} {et~al.}(2014){Oesch}, {Bouwens}, {Illingworth}, {Labb{\'e}},
  {Smit}, {Franx}, {van Dokkum}, {Momcheva}, {Ashby}, {Fazio}, {Huang},
  {Willner}, {Gonzalez}, {Magee}, {Trenti}, {Brammer}, {Skelton}, \&
  {Spitler}}]{oesch2014}
---. 2014, \apj, 786, 108, \dodoi{10.1088/0004-637X/786/2/108}

\bibitem[{{Oesch} {et~al.}(2015){Oesch}, {van Dokkum}, {Illingworth},
  {Bouwens}, {Momcheva}, {Holden}, {Roberts-Borsani}, {Smit}, {Franx},
  {Labb{\'e}}, {Gonz{\'a}lez}, \& {Magee}}]{oesch2015}
{Oesch}, P.~A., {van Dokkum}, P.~G., {Illingworth}, G.~D., {et~al.} 2015,
  \apjl, 804, L30, \dodoi{10.1088/2041-8205/804/2/L30}

\bibitem[{{Oesch} {et~al.}(2016){Oesch}, {Brammer}, {van Dokkum},
  {Illingworth}, {Bouwens}, {Labb{\'e}}, {Franx}, {Momcheva}, {Ashby}, {Fazio},
  {Gonzalez}, {Holden}, {Magee}, {Skelton}, {Smit}, {Spitler}, {Trenti}, \&
  {Willner}}]{oesch2016}
{Oesch}, P.~A., {Brammer}, G., {van Dokkum}, P.~G., {et~al.} 2016, \apj, 819,
  129, \dodoi{10.3847/0004-637X/819/2/129}

\bibitem[{{Oke} \& {Gunn}(1983)}]{oke1983}
{Oke}, J.~B., \& {Gunn}, J.~E. 1983, \apj, 266, 713, \dodoi{10.1086/160817}

\bibitem[{{Ono} {et~al.}(2012){Ono}, {Ouchi}, {Mobasher}, {Dickinson},
  {Penner}, {Shimasaku}, {Weiner}, {Kartaltepe}, {Nakajima}, {Nayyeri},
  {Stern}, {Kashikawa}, \& {Spinrad}}]{ono2012}
{Ono}, Y., {Ouchi}, M., {Mobasher}, B., {et~al.} 2012, \apj, 744, 83,
  \dodoi{10.1088/0004-637X/744/2/83}

\bibitem[{{Ono} {et~al.}(2013){Ono}, {Ouchi}, {Curtis-Lake}, {Schenker},
  {Ellis}, {McLure}, {Dunlop}, {Robertson}, {Koekemoer}, {Bowler}, {Rogers},
  {Schneider}, {Charlot}, {Stark}, {Shimasaku}, {Furlanetto}, \&
  {Cirasuolo}}]{ono2013}
{Ono}, Y., {Ouchi}, M., {Curtis-Lake}, E., {et~al.} 2013, \apj, 777, 155,
  \dodoi{10.1088/0004-637X/777/2/155}

\bibitem[{{Ono} {et~al.}(2018){Ono}, {Ouchi}, {Harikane}, {Toshikawa}, {Rauch},
  {Yuma}, {Sawicki}, {Shibuya}, {Shimasaku}, {Oguri}, {Willott}, {Akhlaghi},
  {Akiyama}, {Coupon}, {Kashikawa}, {Komiyama}, {Konno}, {Lin}, {Matsuoka},
  {Miyazaki}, {Nagao}, {Nakajima}, {Silverman}, {Tanaka}, {Taniguchi}, \&
  {Wang}}]{ono2018}
{Ono}, Y., {Ouchi}, M., {Harikane}, Y., {et~al.} 2018, \pasj, 70, S10,
  \dodoi{10.1093/pasj/psx103}

\bibitem[{{Peng} {et~al.}(2002){Peng}, {Ho}, {Impey}, \& {Rix}}]{peng2002}
{Peng}, C.~Y., {Ho}, L.~C., {Impey}, C.~D., \& {Rix}, H.-W. 2002, \aj, 124,
  266, \dodoi{10.1086/340952}

\bibitem[{{Peng} {et~al.}(2010){Peng}, {Ho}, {Impey}, \& {Rix}}]{peng2010}
---. 2010, \aj, 139, 2097, \dodoi{10.1088/0004-6256/139/6/2097}

\bibitem[{{Pentericci} {et~al.}(2011){Pentericci}, {Fontana}, {Vanzella},
  {Castellano}, {Grazian}, {Dijkstra}, {Boutsia}, {Cristiani}, {Dickinson},
  {Giallongo}, {Giavalisco}, {Maiolino}, {Moorwood}, {Paris}, \&
  {Santini}}]{pentericci2011}
{Pentericci}, L., {Fontana}, A., {Vanzella}, E., {et~al.} 2011, \apj, 743, 132,
  \dodoi{10.1088/0004-637X/743/2/132}

\bibitem[{{Pentericci} {et~al.}(2014){Pentericci}, {Vanzella}, {Fontana},
  {Castellano}, {Treu}, {Mesinger}, {Dijkstra}, {Grazian}, {Brada{\v c}},
  {Conselice}, {Cristiani}, {Dunlop}, {Galametz}, {Giavalisco}, {Giallongo},
  {Koekemoer}, {McLure}, {Maiolino}, {Paris}, \& {Santini}}]{pentericci2014}
{Pentericci}, L., {Vanzella}, E., {Fontana}, A., {et~al.} 2014, \apj, 793, 113,
  \dodoi{10.1088/0004-637X/793/2/113}

\bibitem[{{Planck Collaboration} {et~al.}(2016){Planck Collaboration}, {Adam},
  {Aghanim}, {Ashdown}, {Aumont}, {Baccigalupi}, {Ballardini}, {Banday},
  {Barreiro}, {Bartolo}, {Basak}, {Battye}, {Benabed}, {Bernard}, {Bersanelli},
  {Bielewicz}, {Bock}, {Bonaldi}, {Bonavera}, {Bond}, {Borrill}, {Bouchet},
  {Boulanger}, {Bucher}, {Burigana}, {Calabrese}, {Cardoso}, {Carron},
  {Chiang}, {Colombo}, {Combet}, {Comis}, {Couchot}, {Coulais}, {Crill},
  {Curto}, {Cuttaia}, {Davis}, {de Bernardis}, {de Rosa}, {de Zotti},
  {Delabrouille}, {Di Valentino}, {Dickinson}, {Diego}, {Dor{\'e}}, {Douspis},
  {Ducout}, {Dupac}, {Elsner}, {En{\ss}lin}, {Eriksen}, {Falgarone}, {Fantaye},
  {Finelli}, {Forastieri}, {Frailis}, {Fraisse}, {Franceschi}, {Frolov},
  {Galeotta}, {Galli}, {Ganga}, {G{\'e}nova-Santos}, {Gerbino}, {Ghosh},
  {Gonz{\'a}lez-Nuevo}, {G{\'o}rski}, {Gruppuso}, {Gudmundsson}, {Hansen},
  {Helou}, {Henrot-Versill{\'e}}, {Herranz}, {Hivon}, {Huang}, {Ili{\'c}},
  {Jaffe}, {Jones}, {Keih{\"a}nen}, {Keskitalo}, {Kisner}, {Knox},
  {Krachmalnicoff}, {Kunz}, {Kurki-Suonio}, {Lagache}, {L{\"a}hteenm{\"a}ki},
  {Lamarre}, {Langer}, {Lasenby}, {Lattanzi}, {Lawrence}, {Le Jeune},
  {Levrier}, {Lewis}, {Liguori}, {Lilje}, {L{\'o}pez-Caniego}, {Ma},
  {Mac{\'{\i}}as-P{\'e}rez}, {Maggio}, {Mangilli}, {Maris}, {Martin},
  {Mart{\'{\i}}nez-Gonz{\'a}lez}, {Matarrese}, {Mauri}, {McEwen}, {Meinhold},
  {Melchiorri}, {Mennella}, {Migliaccio}, {Miville-Desch{\^e}nes}, {Molinari},
  {Moneti}, {Montier}, {Morgante}, {Moss}, {Naselsky}, {Natoli}, {Oxborrow},
  {Pagano}, {Paoletti}, {Partridge}, {Patanchon}, {Patrizii}, {Perdereau},
  {Perotto}, {Pettorino}, {Piacentini}, {Plaszczynski}, {Polastri}, {Polenta},
  {Puget}, {Rachen}, {Racine}, {Reinecke}, {Remazeilles}, {Renzi}, {Rocha},
  {Rossetti}, {Roudier}, {Rubi{\~n}o-Mart{\'{\i}}n}, {Ruiz-Granados},
  {Salvati}, {Sandri}, {Savelainen}, {Scott}, {Sirri}, {Sunyaev}, {Suur-Uski},
  {Tauber}, {Tenti}, {Toffolatti}, {Tomasi}, {Tristram}, {Trombetti},
  {Valiviita}, {Van Tent}, {Vielva}, {Villa}, {Vittorio}, {Wandelt}, {Wehus},
  {White}, {Zacchei}, \& {Zonca}}]{planck2016}
{Planck Collaboration}, {Adam}, R., {Aghanim}, N., {et~al.} 2016, \aap, 596,
  A108, \dodoi{10.1051/0004-6361/201628897}

\bibitem[{{Reddy} \& {Steidel}(2009)}]{reddy2009}
{Reddy}, N.~A., \& {Steidel}, C.~C. 2009, \apj, 692, 778,
  \dodoi{10.1088/0004-637X/692/1/778}

\bibitem[{{Roberts-Borsani} {et~al.}(2016){Roberts-Borsani}, {Bouwens},
  {Oesch}, {Labbe}, {Smit}, {Illingworth}, {van Dokkum}, {Holden}, {Gonzalez},
  {Stefanon}, {Holwerda}, \& {Wilkins}}]{robertsborsani2016}
{Roberts-Borsani}, G.~W., {Bouwens}, R.~J., {Oesch}, P.~A., {et~al.} 2016,
  \apj, 823, 143, \dodoi{10.3847/0004-637X/823/2/143}

\bibitem[{{Salmon} {et~al.}(2017){Salmon}, {Coe}, {Bradley}, {Bouwens},
  {Bradac}, {Huang}, {Oesch}, {Stark}, {Sharon}, {Trenti}, {Avila}, {Ogaz},
  {Andrade-Santos}, {Carrasco}, {Cerny}, {Dawson}, {Frye}, {Hoag}, {Johnson},
  {Jones}, {Lam}, {Lovisari}, {Mainali}, {Past}, {Paterno-Mahler}, {Peterson},
  {Reiss}, {Rodney}, {Ryan}, {Sendra-Server}, {Strolger}, {Umetsu}, {Vulcani},
  \& {Zitrin}}]{salmon2017}
{Salmon}, B., {Coe}, D., {Bradley}, L., {et~al.} 2017, ArXiv e-prints.
\newblock \doarXiv{1710.08930}

\bibitem[{{Santos}(2004)}]{santos2004}
{Santos}, M.~R. 2004, PhD thesis, California Institute of Technology

\bibitem[{{Schaerer} \& {de Barros}(2009)}]{schaerer2009}
{Schaerer}, D., \& {de Barros}, S. 2009, \aap, 502, 423,
  \dodoi{10.1051/0004-6361/200911781}

\bibitem[{{Schechter}(1976)}]{schechter1976}
{Schechter}, P. 1976, \apj, 203, 297, \dodoi{10.1086/154079}

\bibitem[{{Schenker} {et~al.}(2014){Schenker}, {Ellis}, {Konidaris}, \&
  {Stark}}]{schenker2014}
{Schenker}, M.~A., {Ellis}, R.~S., {Konidaris}, N.~P., \& {Stark}, D.~P. 2014,
  \apj, 795, 20, \dodoi{10.1088/0004-637X/795/1/20}

\bibitem[{{Schenker} {et~al.}(2012){Schenker}, {Stark}, {Ellis}, {Robertson},
  {Dunlop}, {McLure}, {Kneib}, \& {Richard}}]{schenker2012}
{Schenker}, M.~A., {Stark}, D.~P., {Ellis}, R.~S., {et~al.} 2012, \apj, 744,
  179, \dodoi{10.1088/0004-637X/744/2/179}

\bibitem[{{Schlafly} \& {Finkbeiner}(2011)}]{schlafly2011}
{Schlafly}, E.~F., \& {Finkbeiner}, D.~P. 2011, \apj, 737, 103,
  \dodoi{10.1088/0004-637X/737/2/103}

\bibitem[{{Schmidt} {et~al.}(2014){Schmidt}, {Treu}, {Trenti}, {Bradley},
  {Kelly}, {Oesch}, {Holwerda}, {Shull}, \& {Stiavelli}}]{schmidt2014}
{Schmidt}, K.~B., {Treu}, T., {Trenti}, M., {et~al.} 2014, \apj, 786, 57,
  \dodoi{10.1088/0004-637X/786/1/57}

\bibitem[{{Scoville} {et~al.}(2007){Scoville}, {Aussel}, {Brusa}, {Capak},
  {Carollo}, {Elvis}, {Giavalisco}, {Guzzo}, {Hasinger}, {Impey}, {Kneib},
  {LeFevre}, {Lilly}, {Mobasher}, {Renzini}, {Rich}, {Sanders}, {Schinnerer},
  {Schminovich}, {Shopbell}, {Taniguchi}, \& {Tyson}}]{scoville2007}
{Scoville}, N., {Aussel}, H., {Brusa}, M., {et~al.} 2007, \apjs, 172, 1,
  \dodoi{10.1086/516585}

\bibitem[{{Shibuya} {et~al.}(2012){Shibuya}, {Kashikawa}, {Ota}, {Iye},
  {Ouchi}, {Furusawa}, {Shimasaku}, \& {Hattori}}]{shibuya2012}
{Shibuya}, T., {Kashikawa}, N., {Ota}, K., {et~al.} 2012, \apj, 752, 114,
  \dodoi{10.1088/0004-637X/752/2/114}

\bibitem[{{Shibuya} {et~al.}(2015){Shibuya}, {Ouchi}, \&
  {Harikane}}]{shibuya2015}
{Shibuya}, T., {Ouchi}, M., \& {Harikane}, Y. 2015, \apjs, 219, 15,
  \dodoi{10.1088/0067-0049/219/2/15}

\bibitem[{{Shim} {et~al.}(2011){Shim}, {Chary}, {Dickinson}, {Lin}, {Spinrad},
  {Stern}, \& {Yan}}]{shim2011}
{Shim}, H., {Chary}, R.-R., {Dickinson}, M., {et~al.} 2011, \apj, 738, 69,
  \dodoi{10.1088/0004-637X/738/1/69}

\bibitem[{{Smit} {et~al.}(2014){Smit}, {Bouwens}, {Labb{\'e}}, {Zheng},
  {Bradley}, {Donahue}, {Lemze}, {Moustakas}, {Umetsu}, {Zitrin}, {Coe},
  {Postman}, {Gonzalez}, {Bartelmann}, {Ben{\'{\i}}tez}, {Broadhurst}, {Ford},
  {Grillo}, {Infante}, {Jimenez-Teja}, {Jouvel}, {Kelson}, {Lahav}, {Maoz},
  {Medezinski}, {Melchior}, {Meneghetti}, {Merten}, {Molino}, {Moustakas},
  {Nonino}, {Rosati}, \& {Seitz}}]{smit2014}
{Smit}, R., {Bouwens}, R.~J., {Labb{\'e}}, I., {et~al.} 2014, \apj, 784, 58,
  \dodoi{10.1088/0004-637X/784/1/58}

\bibitem[{{Smit} {et~al.}(2015){Smit}, {Bouwens}, {Franx}, {Oesch}, {Ashby},
  {Willner}, {Labb{\'e}}, {Holwerda}, {Fazio}, \& {Huang}}]{smit2015}
{Smit}, R., {Bouwens}, R.~J., {Franx}, M., {et~al.} 2015, \apj, 801, 122,
  \dodoi{10.1088/0004-637X/801/2/122}

\bibitem[{{Smit} {et~al.}(2018){Smit}, {Bouwens}, {Carniani}, {Oesch},
  {Labb{\'e}}, {Illingworth}, {van der Werf}, {Bradley}, {Gonzalez}, {Hodge},
  {Holwerda}, {Maiolino}, \& {Zheng}}]{smit2018}
{Smit}, R., {Bouwens}, R.~J., {Carniani}, S., {et~al.} 2018, \nat, 553, 178,
  \dodoi{10.1038/nature24631}

\bibitem[{{Somerville} {et~al.}(2008){Somerville}, {Barden}, {Rix}, {Bell},
  {Beckwith}, {Borch}, {Caldwell}, {H{\"a}u{\ss}ler}, {Heymans}, {Jahnke},
  {Jogee}, {McIntosh}, {Meisenheimer}, {Peng}, {S{\'a}nchez}, {Wisotzki}, \&
  {Wolf}}]{somerville2008}
{Somerville}, R.~S., {Barden}, M., {Rix}, H.-W., {et~al.} 2008, \apj, 672, 776,
  \dodoi{10.1086/523661}

\bibitem[{{Song} {et~al.}(2016){Song}, {Finkelstein}, {Livermore}, {Capak},
  {Dickinson}, \& {Fontana}}]{song2016}
{Song}, M., {Finkelstein}, S.~L., {Livermore}, R.~C., {et~al.} 2016, \apj, 826,
  113, \dodoi{10.3847/0004-637X/826/2/113}

\bibitem[{{Stark} {et~al.}(2010){Stark}, {Ellis}, {Chiu}, {Ouchi}, \&
  {Bunker}}]{stark2010}
{Stark}, D.~P., {Ellis}, R.~S., {Chiu}, K., {Ouchi}, M., \& {Bunker}, A. 2010,
  \mnras, 408, 1628, \dodoi{10.1111/j.1365-2966.2010.17227.x}

\bibitem[{{Stark} {et~al.}(2011){Stark}, {Ellis}, \& {Ouchi}}]{stark2011}
{Stark}, D.~P., {Ellis}, R.~S., \& {Ouchi}, M. 2011, \apjl, 728, L2,
  \dodoi{10.1088/2041-8205/728/1/L2}

\bibitem[{{Stark} {et~al.}(2013){Stark}, {Schenker}, {Ellis}, {Robertson},
  {McLure}, \& {Dunlop}}]{stark2013}
{Stark}, D.~P., {Schenker}, M.~A., {Ellis}, R., {et~al.} 2013, \apj, 763, 129,
  \dodoi{10.1088/0004-637X/763/2/129}

\bibitem[{{Stark} {et~al.}(2017){Stark}, {Ellis}, {Charlot}, {Chevallard},
  {Tang}, {Belli}, {Zitrin}, {Mainali}, {Gutkin}, {Vidal-Garc{\'{\i}}a},
  {Bouwens}, \& {Oesch}}]{stark2017}
{Stark}, D.~P., {Ellis}, R.~S., {Charlot}, S., {et~al.} 2017, \mnras, 464, 469,
  \dodoi{10.1093/mnras/stw2233}

\bibitem[{{Stefanon} {et~al.}(2017){Stefanon}, {Labb{\'e}}, {Bouwens},
  {Brammer}, {Oesch}, {Franx}, {Fynbo}, {Milvang-Jensen}, {Muzzin},
  {Illingworth}, {Le F{\`e}vre}, {Caputi}, {Holwerda}, {McCracken}, {Smit}, \&
  {Magee}}]{stefanon2017}
{Stefanon}, M., {Labb{\'e}}, I., {Bouwens}, R.~J., {et~al.} 2017, \apj, 851,
  43, \dodoi{10.3847/1538-4357/aa9a40}

\bibitem[{{Steidel} {et~al.}(1996){Steidel}, {Giavalisco}, {Pettini},
  {Dickinson}, \& {Adelberger}}]{steidel1996}
{Steidel}, C.~C., {Giavalisco}, M., {Pettini}, M., {Dickinson}, M., \&
  {Adelberger}, K.~L. 1996, \apj, 462, L17, \dodoi{10.1086/310029}

\bibitem[{{Stringer} {et~al.}(2014){Stringer}, {Shankar}, {Novak},
  {Huertas-Company}, {Combes}, \& {Moster}}]{stringer2014}
{Stringer}, M.~J., {Shankar}, F., {Novak}, G.~S., {et~al.} 2014, \mnras, 441,
  1570, \dodoi{10.1093/mnras/stu645}

\bibitem[{{The Astropy Collaboration} {et~al.}(2018){The Astropy
  Collaboration}, {Price-Whelan}, {Sip{\H o}cz}, {G{\"u}nther}, {Lim},
  {Crawford}, {Conseil}, {Shupe}, {Craig}, {Dencheva}, {Ginsburg},
  {VanderPlas}, {Bradley}, {P{\'e}rez-Su{\'a}rez}, {de Val-Borro}, {Aldcroft},
  {Cruz}, {Robitaille}, {Tollerud}, {Ardelean}, {Babej}, {Bachetti}, {Bakanov},
  {Bamford}, {Barentsen}, {Barmby}, {Baumbach}, {Berry}, {Biscani}, {Boquien},
  {Bostroem}, {Bouma}, {Brammer}, {Bray}, {Breytenbach}, {Buddelmeijer},
  {Burke}, {Calderone}, {Cano Rodr{\'{\i}}guez}, {Cara}, {Cardoso},
  {Cheedella}, {Copin}, {Crichton}, {D{\'A}vella}, {Deil}, {Depagne},
  {Dietrich}, {Donath}, {Droettboom}, {Earl}, {Erben}, {Fabbro}, {Ferreira},
  {Finethy}, {Fox}, {Garrison}, {Gibbons}, {Goldstein}, {Gommers}, {Greco},
  {Greenfield}, {Groener}, {Grollier}, {Hagen}, {Hirst}, {Homeier}, {Horton},
  {Hosseinzadeh}, {Hu}, {Hunkeler}, {Ivezi{\'c}}, {Jain}, {Jenness}, {Kanarek},
  {Kendrew}, {Kern}, {Kerzendorf}, {Khvalko}, {King}, {Kirkby}, {Kulkarni},
  {Kumar}, {Lee}, {Lenz}, {Littlefair}, {Ma}, {Macleod}, {Mastropietro},
  {McCully}, {Montagnac}, {Morris}, {Mueller}, {Mumford}, {Muna}, {Murphy},
  {Nelson}, {Nguyen}, {Ninan}, {N{\"o}the}, {Ogaz}, {Oh}, {Parejko}, {Parley},
  {Pascual}, {Patil}, {Patil}, {Plunkett}, {Prochaska}, {Rastogi}, {Reddy
  Janga}, {Sabater}, {Sakurikar}, {Seifert}, {Sherbert}, {Sherwood-Taylor},
  {Shih}, {Sick}, {Silbiger}, {Singanamalla}, {Singer}, {Sladen}, {Sooley},
  {Sornarajah}, {Streicher}, {Teuben}, {Thomas}, {Tremblay}, {Turner},
  {Terr{\'o}n}, {van Kerkwijk}, {de la Vega}, {Watkins}, {Weaver}, {Whitmore},
  {Woillez}, \& {Zabalza}}]{astropy2018}
{The Astropy Collaboration}, {Price-Whelan}, A.~M., {Sip{\H o}cz}, B.~M.,
  {et~al.} 2018, ArXiv e-prints.
\newblock \doarXiv{1801.02634}

\bibitem[{{Tilvi} {et~al.}(2013){Tilvi}, {Papovich}, {Tran}, {Labb{\'e}},
  {Spitler}, {Straatman}, {Persson}, {Monson}, {Glazebrook}, \&
  {Quadri}}]{tilvi2013}
{Tilvi}, V., {Papovich}, C., {Tran}, K. V.~H., {et~al.} 2013, \apj, 768, 56,
  \dodoi{10.1088/0004-637X/768/1/56}

\bibitem[{{Tilvi} {et~al.}(2016){Tilvi}, {Pirzkal}, {Malhotra}, {Finkelstein},
  {Rhoads}, {Windhorst}, {Grogin}, {Koekemoer}, {Zakamska}, \&
  {Ryan}}]{tilvi2016}
{Tilvi}, V., {Pirzkal}, N., {Malhotra}, S., {et~al.} 2016, \apjl, 827, L14,
  \dodoi{10.3847/2041-8205/827/1/L14}

\bibitem[{{Trenti} \& {Stiavelli}(2008)}]{trenti2008}
{Trenti}, M., \& {Stiavelli}, M. 2008, \apj, 676, 767, \dodoi{10.1086/528674}

\bibitem[{{Trenti} {et~al.}(2010){Trenti}, {Stiavelli}, {Bouwens}, {Oesch},
  {Shull}, {Illingworth}, {Bradley}, \& {Carollo}}]{trenti2010}
{Trenti}, M., {Stiavelli}, M., {Bouwens}, R.~J., {et~al.} 2010, \apj, 714,
  L202, \dodoi{10.1088/2041-8205/714/2/L202}

\bibitem[{{Trenti} {et~al.}(2011){Trenti}, {Bradley}, {Stiavelli}, {Oesch},
  {Treu}, {Bouwens}, {Shull}, {MacKenty}, {Carollo}, \&
  {Illingworth}}]{trenti2011}
{Trenti}, M., {Bradley}, L.~D., {Stiavelli}, M., {et~al.} 2011, \apjl, 727,
  L39, \dodoi{10.1088/2041-8205/727/2/L39}

\bibitem[{{Trenti} {et~al.}(2012){Trenti}, {Bradley}, {Stiavelli}, {Shull},
  {Oesch}, {Bouwens}, {Mu{\~n}oz}, {Romano-Diaz}, {Treu}, {Shlosman}, \&
  {Carollo}}]{trenti2012}
---. 2012, \apj, 746, 55, \dodoi{10.1088/0004-637X/746/1/55}

\bibitem[{{Treu} {et~al.}(2013){Treu}, {Schmidt}, {Trenti}, {Bradley}, \&
  {Stiavelli}}]{treu2013}
{Treu}, T., {Schmidt}, K.~B., {Trenti}, M., {Bradley}, L.~D., \& {Stiavelli},
  M. 2013, \apjl, 775, L29, \dodoi{10.1088/2041-8205/775/1/L29}

\bibitem[{{van der Burg} {et~al.}(2010){van der Burg}, {Hildebrandt}, \&
  {Erben}}]{vanderburg2010}
{van der Burg}, R.~F.~J., {Hildebrandt}, H., \& {Erben}, T. 2010, \aap, 523,
  A74, \dodoi{10.1051/0004-6361/200913812}

\bibitem[{van~der Walt {et~al.}(2011)van~der Walt, Colbert, \&
  Varoquaux}]{vanderwalt2011}
van~der Walt, S., Colbert, S.~C., \& Varoquaux, G. 2011, Computing in Science
  \& Engineering, 13, 22, \dodoi{10.1109/MCSE.2011.37}

\bibitem[{{Vanzella} {et~al.}(2011){Vanzella}, {Pentericci}, {Fontana},
  {Grazian}, {Castellano}, {Boutsia}, {Cristiani}, {Dickinson}, {Gallozzi},
  {Giallongo}, {Giavalisco}, {Maiolino}, {Moorwood}, {Paris}, \&
  {Santini}}]{vanzella2011}
{Vanzella}, E., {Pentericci}, L., {Fontana}, A., {et~al.} 2011, \apjl, 730,
  L35, \dodoi{10.1088/2041-8205/730/2/L35}

\bibitem[{{Windhorst} {et~al.}(2011){Windhorst}, {Cohen}, {Hathi}, {McCarthy},
  {Ryan}, {Yan}, {Baldry}, {Driver}, {Frogel}, {Hill}, {Kelvin}, {Koekemoer},
  {Mechtley}, {O'Connell}, {Robotham}, {Rutkowski}, {Seibert}, {Straughn},
  {Tuffs}, {Balick}, {Bond}, {Bushouse}, {Calzetti}, {Crockett}, {Disney},
  {Dopita}, {Hall}, {Holtzman}, {Kaviraj}, {Kimble}, {MacKenty}, {Mutchler},
  {Paresce}, {Saha}, {Silk}, {Trauger}, {Walker}, {Whitmore}, \&
  {Young}}]{windhorst2011}
{Windhorst}, R.~A., {Cohen}, S.~H., {Hathi}, N.~P., {et~al.} 2011, The
  Astrophysical Journal Supplement Series, 193, 27,
  \dodoi{10.1088/0067-0049/193/2/27}

\bibitem[{{Wyithe} \& {Loeb}(2011)}]{wyithe2011}
{Wyithe}, J.~S.~B., \& {Loeb}, A. 2011, \mnras, 413, L38,
  \dodoi{10.1111/j.1745-3933.2011.01027.x}

\bibitem[{{Zitrin} {et~al.}(2015){Zitrin}, {Labb{\'e}}, {Belli}, {Bouwens},
  {Ellis}, {Roberts-Borsani}, {Stark}, {Oesch}, \& {Smit}}]{zitrin2015}
{Zitrin}, A., {Labb{\'e}}, I., {Belli}, S., {et~al.} 2015, \apjl, 810, L12,
  \dodoi{10.1088/2041-8205/810/1/L12}

\end{thebibliography}
\end{document}